\documentclass[onecollarge]{svjour2}       
\smartqed  
\usepackage{graphicx}

\usepackage{amssymb}
\usepackage{amsmath}
\usepackage{subfigure}

\newcommand{\Lm}{\mbox{$(+,\, -,\, -,\, -)$}}
\newcommand{\aLm}{\mbox{$(-,\, +,\, +,\, +)$}}
\newcommand{\aEm}{\mbox{$(-,\, -,\, -)$}}

\newcommand{\lmax}{\mbox{$\ell_{\textrm {max}}$}}
\newcommand{\lmin}{\mbox{$\ell_{\textrm {min}}$}}

\newcommand{\that}{\mbox{$\mathrm {\mathbf{\hat{t}}}$}}
\newcommand{\xhat}{\mbox{$\mathrm {\mathbf{\hat{x}}}$}}
\newcommand{\yhat}{\mbox{$\mathrm {\mathbf{\hat{y}}}$}}
\newcommand{\zhat}{\mbox{$\mathrm {\mathbf{\hat{z}}}$}}
\newcommand{\ihat}{\mbox{$\mathrm {\mathbf{\hat{i}}}$}}
\newcommand{\jhat}{\mbox{$\mathrm {\mathbf{\hat{j}}}$}}
\newcommand{\khat}{\mbox{$\mathrm {\mathbf{\hat{k}}}$}}

\newcommand{\Exp}{\mbox{$\mathrm e$}}
\newcommand{\sihat}{\mathrm {\mathbf{\hat{i}}}}

\newcommand{\skhat}{\mathrm {\mathbf{\hat{k}}}} 

\newcommand{\vhat}{\mbox{$\mathrm {\mathbf{\hat{v}}}$}}

\newcommand{\ahat}{\mbox{$\mathrm {\mathbf{\hat{a}}}$}}
\newcommand{\bhat}{\mbox{$\mathrm {\mathbf{\hat{b}}}$}}

\newcommand{\Fhat}{\mbox{$\mathrm {\mathbf{\hat{F}}}$}}

\newcommand{\bfa}{\mbox{$\mathrm {\mathbf{a}}$}}
\newcommand{\bfA}{\mbox{$\mathrm {\mathbf{A}}$}}
\newcommand{\bfb}{\mbox{$\mathrm {\mathbf{b}}$}}
\newcommand{\bfab}{\mbox{$\mathrm {\mathbf{ab}}$}}
\newcommand{\bfc}{\mbox{$\mathrm {\mathbf{c}}$}}
\newcommand{\bfd}{\mbox{$\mathrm {\mathbf{d}}$}}
\newcommand{\bfcd}{\mbox{$\mathrm {\mathbf{cd}}$}}
\newcommand{\bfr}{\mbox{$\mathrm {\mathbf{r}}$}}
\newcommand{\bft}{\mbox{$\mathrm {\mathbf{t}}$}}
\newcommand{\bfm}{\mbox{$\mathrm {\mathbf{m}}$}}

\newcommand{\bbC}{\mbox{$\mathrm {\mathbb{C}}$}}
\newcommand{\bbQ}{\mbox{$\mathrm {\mathbb{Q}}$}}

\newcommand{\bbR}{\mbox{$\mathrm {\mathbb{R}}$}}
\newcommand{\bbF}{\mbox{$\mathrm {\mathbb{F}}$}}
\newcommand{\bbV}{\mbox{$\mathrm {\mathbb{V}}$}}
\newcommand{\bbI}{\mbox{$\mathrm {\mathbb{I}}$}}

\renewcommand{\mod}[1]{\mbox{$\left\|#1\right\|$}}

\newcommand{\trans}[2]{\mbox{{\sc Trans}$(#1\rightarrow#2)$}}
\newcommand{\rot}[1]{\mbox{{\sc Rot}$(#1)$}}

\begin{document}

\title{Assumptions and Axioms: Mathematical Structures to Describe the Physics of Rigid Bodies}

\titlerunning{Assumptions and Axioms}        

\author{Philip H. Butler         \and
        Niels G. Gresnigt	\and
        Peter F. Renaud
}


\institute{Philip H. Butler \at
              Department of Physics and Astronomy\\
							University of Canterbury,\\
							Private Bag 4800, Christchurch 8140,\\
							New Zealand \\
              \email{phil.butler@canterbury.ac.nz}           
           \and
           Niels G. Gresnigt \at
           		Department of Physics and Astronomy\\
							University of Canterbury,\\
							Private Bag 4800, Christchurch 8140,\\
							New Zealand \\
              \email{niels.gresnigt@canterbury.ac.nz}
           \and
           Peter F. Renaud	\at
           		Department of Mathematics and Statistics\\
							University of Canterbury,\\
							Private Bag 4800, Christchurch 8140,\\
							New Zealand \\
           \email{peter.renaud@canterbury.ac.nz}
}
\date{Received: date \today / Accepted: date}

\maketitle
\begin{abstract}

This paper challenges some of the common assumptions underlying the mathematics used to describe the physical world. We start by reviewing many of the assumptions underlying the concepts of real, physical, rigid bodies and the translational and rotational properties of such rigid bodies. Nearly all elementary and advanced texts make physical assumptions that are subtly different from ours, and as a result we develop a mathematical description that is subtly different from the standard mathematical structure.

Using the homogeneity and isotropy of space, we investigate the translational and rotational features of rigid bodies in two and three dimensions. We find that the concept of rigid bodies and the concept of the homogeneity of space are intrinsically linked. The geometric study of rotations of rigid objects leads to a geometric product relationship for lines and vectors. By requiring this product to be both associative and to satisfy Pythagoras' theorem, we obtain a choice of Clifford algebras.

We extend our arguments from space to include time. By assuming that $c\delta t = \delta \ell$ and rewriting this in Lorentz invariant form as $c^2t^2-x^2-y^2-z^2 =0$ we obtain a generalization of Pythagoras to spacetime. This leads us directly to establishing that the Clifford algebra $C\ell(1,3)$ is an appropriate mathematical structure to describe spacetime.

Clifford algebras are not division algebras. We show that the existence of non-invertible elements in the algebra is not a limitation of the usefulness to physics of the algebra but rather that it reflects accurately the spacetime properties of physical systems.

\end{abstract}
\section{Preface}\label{sec:pre}

In recent years three well known theoretical physicists have written books challenging the string theory community to reconsider their focus on high-dimensional theories of fundamental physics, especially string theory and its derivatives \cite{smolin2007tpr,woit2006new,penrose2004rrc}. Each of these authors expresses their frustration with the progress of the past 40 years, and argues the case that there needs to be changes to one or more of the current understandings of special relativity, quantum mechanics, quantum field theory, the standard model of particle physics and general relativity.

Penrose \cite{penrose2004rrc} ends his case (page 1045) with: 
\begin{quote}
[T]here are [many] deeply mysterious issues about which we have very little comprehension. It is quite likely that the 21st century will reveal even more wonderful insights than those we have been blessed with in the 20th. But for this to happen, we shall need powerful new ideas, which will take us in directions significantly different from those currently being pursued. Perhaps what we mainly need is some subtle change in perspective---something we have all missed...
\end{quote} 

The aim of this paper is to review the basic assumptions made about physical space, in particular its geometry. From these assumptions we concentrate on developing the most appropriate mathematical framework within which to describe physical phenomena. For maximum clarity, we focus on everyday sized objects. We invite the reader to follow our arguments. We try to be upfront and clearly state all important assumptions. What we find is that the first changes that we wish to make to the physics, and to the mathematics we use to describe the physics, are changes at the geometric foundations. 

We introduce the concept of reference frames from the idea of rigid material objects, made of real atoms. One dimensional rigid rods are, for us, not an abstraction, but like three dimensional rigid bodies, an approximation. The real world is the place where we do measurements, and real measurements do not return exact answers. We endeavour to set up an idealised mathematical world that is a good model of the physical world. Our approach throughout is akin to the axiomatic approach typically found in an introductory mathematics text on vector algebra.

In section \ref{AaAI2010} we set up the concept of a `reference frame' and the concept of a `straight line' by taking the concept of a real, physical, rigid body in a 2-dimensional space and looking at translational properties. We find that the concept of a rigid body and the concept of homogeneity of space are linked. We then set up the mathematical concept of vectors as elements of a vector space over the field of rational numbers, \bbQ. The operations of the vector space are linked to three separate operations on the points of rigid bodies: drawing lines between points; moving a rigid body with respect to another; and transforming from one reference frame to another. These mathematical and physical operations define for us the concept of straight lines used in the expression of Newton's First law. They are intrinsically derived from the property of space known as the `homogeneity of space'.

In section \ref{AaAII2010} we extend the ideas from this section \ref{AaAI2010} and use the isotropy of space to develop the rotational properties of 2D space. We shall use the isotropy of space to introduce the concept of a `right angle'. By asking for a product operation that describes rotations and matches Pythagoras' theorem we are led from a vector space over \bbQ\ to an algebra over \bbQ. The algebra we derive is an example of a Clifford algebra.

Section \ref{AaAIII2010} extends the considerations of homogeneity and isotropy from two spatial dimensions to three. The isotropy of 3D space and the rotation properties of rigid objects lead to a richer set of properties and an eight dimensional Clifford algebra. We demonstrate that the maintenance of cyclic structures of sets of basis lines and sets of basis planes, namely the parity conservation properties of allowable physical movements of rigid bodies, requires the use of the Clifford algebra $C\ell(0,3)$ and not the Clifford algebra $C\ell(3, 0)$. 

Section \ref{AaAIV2010}  studies time as a fourth dimension in a vector space over \bbQ. The observation that the speed of light measured with respect to any inertial rigid body is independent of where and in what direction the light is traveling gives us a generalization of Pythagoras to spacetime. This directly leads us to establish that the 16 dimensional Clifford algebra $C\ell(1,3)$ is an appropriate mathematical structure to describe measurements of the motion of rigid bodies and of light in the reference frame defined by a single rigid body. Our derivation requires us to assume that our rigid body frame is inertial. Finally we show that all this implies that $C\ell(1,3)$ is the appropriate mathematics to describe Lorentz and Poincar\'{e} transformations between rigid body reference frames and is also the appropriate mathematics to describe translations, rotations, and boosts of rigid bodies.

Finally, section \ref{AaAV2010} looks at the algebraic structures of Clifford algebras. In particular we will discuss the differences between various Clifford algebras and seek the matrix representations of them over the reals, \bbR, or its subfield, the rational numbers \bbQ.

\section{Rigid Bodies to Reference Frames, and Homogeneity to Vectors}\label{AaAI2010}

The subject of physics deals with a huge range of scales, from well below the size of the proton, $<10^{-15}$m, to the size of the universe, $> 10$ billion light years or some $10^{26}$m. Even the scales of the objects experienced by people in their daily lives range over some eight orders of magnitude, from fractions of a millimeter to tens of kilometers. Science has learnt ways to observe and measure objects from well below the scale of the proton to the scale of the universe. However in this paper we concentrate on understanding everyday sized objects, and developing the most appropriate mathematical framework within which to describe such objects. In general we expect the mathematical framework in which we work to be larger than our physical space, in the sense that not every mathematical construction or operation has a meaningful physical counterpart. However, we do want the converse to hold; every physically allowed operation can be represented in our mathematical framework. 

The place we shall begin is to seek  to understand what is meant by the geometric content embedded in the usual statements of Newton's First Law, for example given by Serway and Jewett \cite{serway2008psa} as:
\begin{quote}
In the absence of external forces, when viewed from an inertial reference frame, an object at rest remains at rest, and an object in motion continues in motion with a constant velocity (that is, with a constant speed in a straight line).
\end{quote}
It is worth emphasising that it has taken Serway and Jewett some 114 pages of preliminaries to get to that statement, not surprising as this quotation contains some dozen words that have a specific physics meaning.

In this section we shall set up the concept of a `reference frame' and the concept of a `straight line' by taking the concept of a rigid body in a 2-dimensional `toy world'. (A `toy world' is one in which we can study certain processes in a simple way without being distracted by the full richness and complexity of the natural world in which we find ourselves.) The first set of rigid bodies we shall consider are a desktop, and a few transparent sheets of paper which we can move about on the desktop. In addition to the parameters to describe the position and orientation of the pieces of paper in the 2-dimensional world of these material items, we will need another parameter to describe when the pieces of paper are in their different positions as we move them about. 

We then set up the mathematical concept of vectors as elements of a vector space, where the operations of the vector space are linked to three separate operations on the points of rigid bodies: drawing lines between points; moving a rigid body with respect to another; and transforming from one rigid body reference frame to another. These physical operations define for us the concept of straight lines used in the expression of Newton's First law. They are intrinsically derived from the property of space known as the `homogeneity of space'.

Much of the argument presented in this section forms part of those 114 pages of our physics text \cite{serway2008psa}, prior to Newton's First law, although our approach is more akin to the axiomatic approach of an introductory mathematics course on vector algebra than to an introductory physics course.

We finish this section by comparing and contrasting our conclusions with those of standard treatments (such as the introductory text above) and with the arguments in other recent research papers.

\subsection{Points, lines and areas of a rigid body}\label{sec:points}

Let us define various physical idealizations, in particular points and lines, but starting from the concept of a 2D rigid body. There is a logical difficulty lurking here and we do not propose getting into a philosophers' discussion about evidence for, or the nature of, the `objective reality' of philosophers. So we ignore the circularity issues that arise from our trying to describe a `rigid object' before we know how to define `rigid' or `object'. The next subsection will address the first of the properties that allow us to test whether or not we have a `rigid object'.

Consider a 2D rigid object formed by a desktop. 
Mark a set of $n$ `points' $A$, $B$, $C$, \ldots\ on the desktop. We may take these points to be special 2D (rigid) objects idealized as being of negligible or zero size in each of the two dimensions of our toy world.
Now join these points up to form `lines', see figure 1. 
 Again, we need a workable concept of a line. Let us assume we have `strings' or rigid rods that are a special kind of rigid object that we can idealize  to be of finite length but of negligible or zero width. 

There are
 $n^2$ possible lines $AA$, $AB$, $AC$,\ldots, $BB$, $BA$, $BC$,\ldots. Some authors would call our lines  `directed line segments', but we have no need for the (non-physical) concept of lines of infinite length. 
 All our `lines' are directed line segments (or `points' if they are of zero length).
 The line $AB$ is from $A$ to $B$, where we say that $A$ is the `tail' of $AB$ and $B$ the `head' of $AB$. A line $PQ$ for the case where $P = Q$ is a special case in that it has zero length and no direction.
\begin{figure}[ht]
	\centering
	\includegraphics[width=0.60\textwidth]{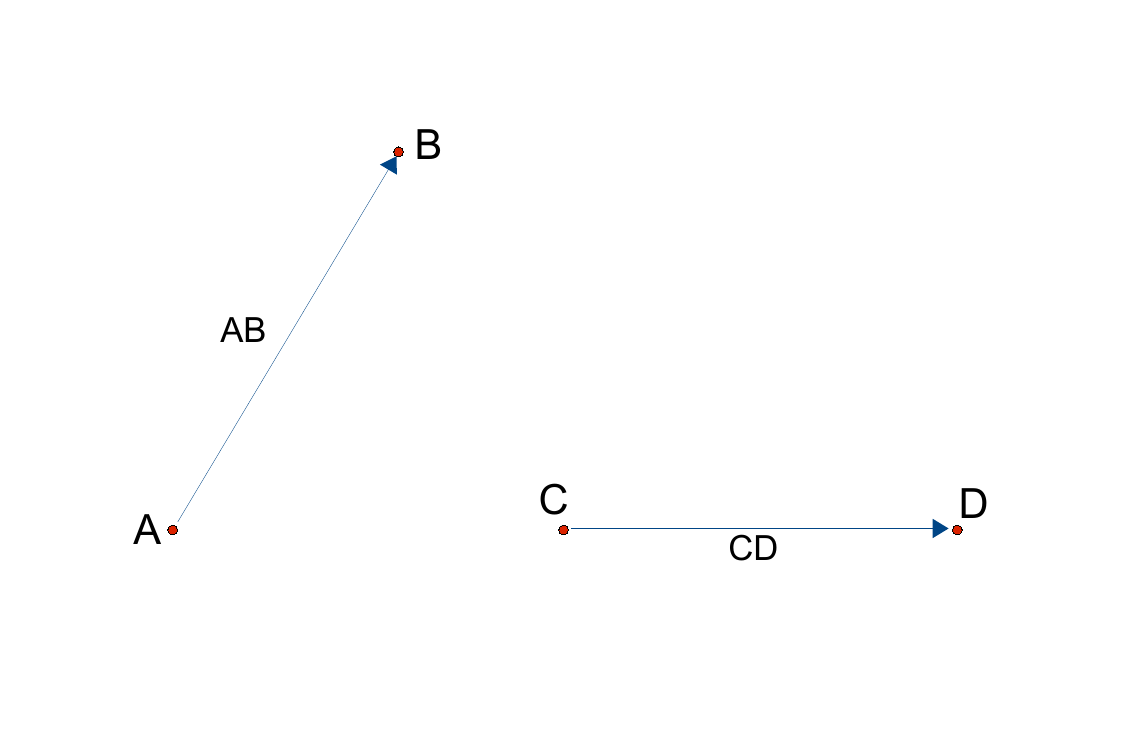}
	\caption{A unique line exists between any two points on the desktop. The line $AB$ is from point $A$ to $B$. Given $n$ points in the space, the total number of possible lines is equal to $n^2$.}
	\label{fig:thedesktopfig1}
\end{figure}
In a natural geometric sense we can define the `passive addition' of lines to lines
\begin{figure}[ht]
	\centering
	\includegraphics[width=0.60\textwidth]{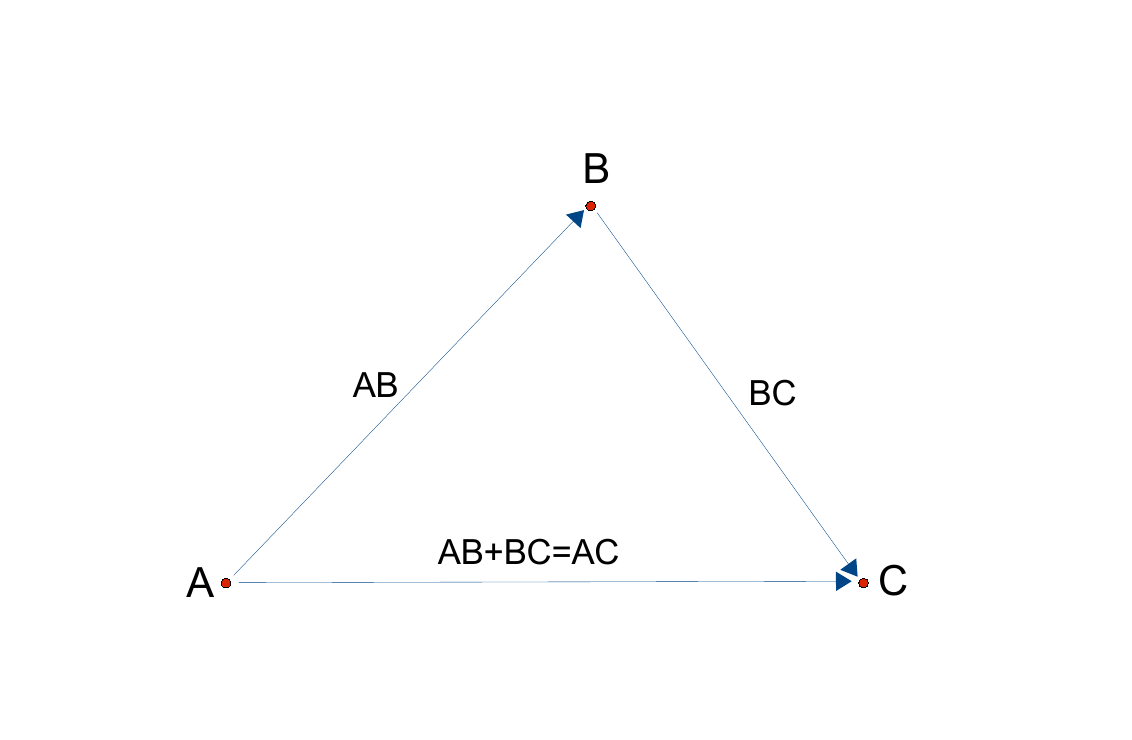}
	\caption{The `passive' operation of joining lines to lines consists of geometrically joining the head of one line to the tail of another line.}
	\label{fig:linejoin}
\end{figure} 
on the desktop and give geometric meaning to expressions such as:
\begin{eqnarray}
	AB + BC &=& AC  \\
	(AB + BC) + CD &=& AB + (BC + CD) = AD
\end{eqnarray}	
This passive addition is just a matter of joining lines, head of the first to tail of the second, see figure 2. 

Likewise in a natural geometric sense we can define the passive addition 
\begin{figure}[ht]
	\centering
	\includegraphics[width=0.90\textwidth]{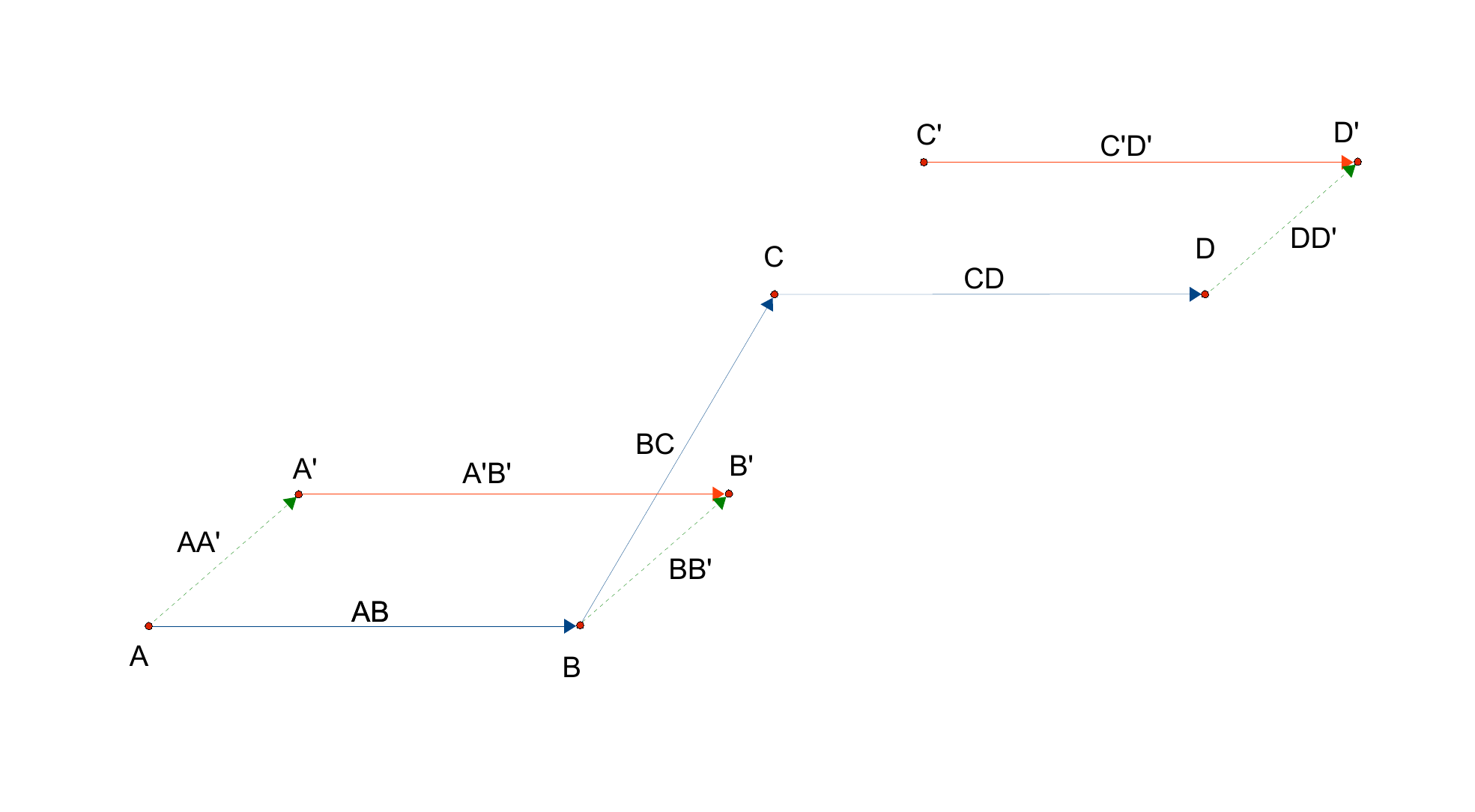}
	\caption{Active addition is the movement of a point from the tail to the head of a line. This figure shows both active and passive translations. Active translations can be used to move points; $A$ is moved to $A'$ by the line $AA'$, or entire lines; the line $AB$ can be moved to the parallel line $A'B'$ using any one of a number of active translations such as $AA'$ and $CC'$.}
	\label{fig:movepoints}
\end{figure}
of a line to a point and  give meanings to expressions such as:
\begin{eqnarray}
	A + AB = B, 	\\
	(A + AB) + BC = C
\end{eqnarray}	
This addition is passive as there is no movement of the object, rather the point $B$ may be considered as a relabeling of point $A$.

It is important to note that passive addition does not contain any concept of translation or equivalence. Therefore we are limited to adding lines where the head of the first line coincides with the tail of the second line, such as $AB$ and $BC$ in figure \ref{fig:linejoin}. It makes no physical sense to add the lines $AB$ and $CD$ of figure \ref{fig:thedesktopfig1} together. Likewise we cannot add a line $AB$ to a point $P$ unless the tail of $AB$ coincides with $P$; that is $A=P$. 

Passive additions are of very limited use and much more useful are active additions which we discuss in the next subsection. Introducing the concept of translations and equivalence we can translate lines through space keeping their length and orientation the same. we can use active translations to move points; $A$ is moved to $A'$ by the line $AA'$, or entire lines; the line $AB$ can be moved to the parallel line $A'B'$ using an active translation such as $AA'$, see figure \ref{fig:movepoints}, and discussion later in this section.

For completeness, we note that three points $A, B, C$, define a triangular area. We return to this concept in more detail in the next section.

\subsection{Rigid body translations and the homogeneity of space}\label{translations}

By considering the motion of several rigid bodies we are led to the `active' addition of points and lines,  then to the concept of the `homogeneity of space'  associated with active motion.

Consider having a 2D rigid transparent object, e.g. a sheet of transparent  paper,  which we can slide about on the desktop. Mark the points $A'$, $B'$, $C'$,\ldots\ on the paper directly above the corresponding points on the desktop $A$, $B$, $C$, \ldots. At this initial position we have $A = A'$, $B = B'$, $C = C'$, \ldots. After a `translation' of the paper, \trans{A}{A'}, the lines $AA'$, $BB'$, $CC'$, \ldots\ are parallel to each other, and of equal length. A translation is defined here to be a movement of a rigid object that is compatible with the ordinary English meaning of translation that is `movement in the absence of rotation'. Mathematically, we say that the lines $AA'$, $BB'$, $CC'$, \ldots\ are equivalent to each other. Any line $AA'$ on the desktop is equivalent to a whole class of parallel lines of the same length on the desktop. We write $[AA']$ to denote this set of lines called an equivalence class.

Alternatively we may use as our definition of translation the observation that, after the motion of the paper relative to the desktop, the lines $AA'$, $BB'$, $CC'$, \ldots\ are parallel and of equal length. The translation can be equally well described by \trans{A}{A'}, or \trans{B}{B'}, or \trans{C}{C'}, \dots, because the lines $AA'$, $BB'$, $CC'$, \ldots\ belong to the same equivalence class. These properties can be tested in our physical world by having a third rigid object, say another piece of paper, on which we mark $X$ at $A$ and $Y$ at $A'$ and slide it about, without rotation, to compare the separation of the other pairs of points, $B$ and $B'$, $C$ and $C'$, \ldots. To compare the length of $BB'$ to $AA'$ we need only translate the second piece of paper $X$ to $B$, whereby $Y$ will be at $B'$\ and no rotation is required. 

As an aside, we observe that the translational motion described above, that has the lines $AA'$, $BB'$, $CC'$, \ldots\  parallel, needs to be changed if we change from our flat desktop to a curved 2D surface, such as the surface of the earth. When sliding objects around curved surfaces it is necessary to generalize to a process known as `parallel transport'. 

The homogeneity of space is the name we give the above geometrical behaviour of rigid body translational motion on a flat surface. 
The concept of a rigid body and the concept of the homogeneity of space are linked. Both require the concept of fixed differences between points, which can be tested for self consistency by our pieces of paper. In the rigid body that is the desktop, we can test the constancy of the length of each one of the lines $AB$, $AC$, $AD$, \ldots, $BC$, $BD$, \ldots\ by repeatedly using our first piece of paper on which we have marked the points $A'$, $B'$, $C'$, \ldots. We can do the same with each one of the lines $A'B'$, $A'C'$, $A'D'$,  \ldots\ on the first piece of paper by matching the points $A'$, $B'$, $C'$, \ldots\ to points $A''$, $B''$, $C''$, \ldots on the second piece of paper.  The combination of the rigidity of the objects and the homogeneity of space, requires that the lengths of the various lines on the various objects do not change as the objects are moved relative to each other. Finally we can verify that the lines  $AA'$, $BB'$, $CC'$, \ldots\  are all the one length and parallel to each other. We know experimentally if a surface is curved by observing that at least some of the lines  $AA'$, $BB'$, $CC'$, \ldots\  have different lengths after the translation $\trans{A}{A'}$.

The existence of rigid bodies in a homogeneous space means that we can extend the passive and active addition rules above, and expand the notation to use the equivalence, under translation,  of the various sets of lines. First note the equivalence of the lines $XY$ (on the second piece of paper) and $AA'$, $BB'$, $CC'$, \ldots\  (between the points on the desktop and on the first piece of paper). Second, note the equivalence of the lines on the desktop to the lines on the first piece of paper ---  the line $AB$ is equivalent to $A'B'$, $AC$ is equivalent to $A'C'$, etc. 

We can say that $AA'$ moves the paper with respect to the desktop by the line $AA'$, and write that all points $P$ on the desktop are moved by \trans{A}{A'}\ to the corresponding points $P'$ on the paper
\begin{eqnarray}
	\trans{A}{A'}(P)= P + AA' = P'
\end{eqnarray} 
Likewise for lines, we can use $AA'$ to  move any line $BC$ on the desktop to its position $B'C'$ on the paper.
\begin{eqnarray}
	\trans{A}{A'}(BC) = BC + AA' = B'C
\end{eqnarray}	
In this notation, which we will use henceforth, $BC+AA'$ does not refer to the passive adding of lines in the sense discussed in the previous subsection, but rather to the active translation of $BC$ by the line $AA'$. We note that $AA'+BC$ is not equal to $BC+AA'$ as they refer to different active translations.

Consider now marking on the desktop the points  $A'$, $B'$, $C'$,\ldots\ that are directly under the corresponding points on the paper (in its moved position).  We now have $2 n$ points on  the desktop. The equations above may now be re-interpreted as actions on points on the desktop.
In particular any line $PQ$ on the desktop may be added to (in the active sense) any other line $AB$ on the desktop, or be used to translate (in the active sense of movement) point $A$ or line $AB$.

As a further subtlety, the action of translation comes in two physical senses. In the first sense we have been considering moving either or both of our two sheets of paper by $AA'$ while leaving the desktop unmoved. In a second sense we can move the desktop by $A'A$ while leaving the pieces of paper unmoved. We have
\begin{eqnarray}
\trans{A}{A'}(paper)=\trans{A}{A'}^{-1}(desktop)=\trans{A'}{A}(desktop)
\end{eqnarray}
The assumption of homogeneity of space says these two senses cannot be distinguished in the physical world. Relative motion is all that can be observed (or measured).

\subsection{Lines constructed by successive additions}\label{sec:additions}

In preparation for deducing that we need a vector space over a field of numbers, we choose to find the smallest field satisfying simple assumptions about the measurement process. We find the field of rational numbers, \bbQ, suffices, although it is usual to use  the field of reals, \bbR.

Starting from a line $AB$, we can form the lines 
\begin{eqnarray}
	2AB = AB +AB\\
	3AB = 2AB +AB
\end{eqnarray}	
giving a natural meaning for the symbol `$nAB$'. (We define $nAB$ to be the line on the rigid body that starts at point $A$.)
This notation incorporates the property of integers
\begin{eqnarray}
	nAB +mAB =(n)AB +(m)AB \\
	= (n+m)AB
	 \label{eq:integeradd}
\end{eqnarray}	

For negative integers, we start from the notion that $BA$ moves the points and lines on any rigid body in the opposite direction to the line $AB$, so that it is natural to write
\begin{eqnarray}
	-AB = BA
\end{eqnarray}	
In general, 
\begin{eqnarray}
	(-n)AB = -(nAB)
\end{eqnarray}
for integer $n$. Thus equation (\ref{eq:integeradd}) applies to all integers small enough so that $nAB$, $mAB$ and  $(n+m)AB$ belong to the rigid body. Henceforth we consider only the cases where this condition is satisfied.

Now, let us use the notation $\mod{AB}$ for the length of $AB$, and use $|n|$ for the absolute value of the integer $n$. It is a property of lines on a  rigid body that 
\begin{eqnarray}
	\mod{nAB} = |n|\ \mod{AB}
\end{eqnarray}
Let us use these notations to compare lengths of parallel lines. (The next section sets up procedures for comparing lengths of non-parallel lines.) Consider only lines that are parallel to the line $AB$ and begin by translating these to have the same tail $A$. Choose a  line $AX$ that is much shorter than  $AB$ as our `short-measuring stick' in the $AB$ 
direction.
 Translate  $AX$ end-to-end  $p$ times (where p is a positive integer) until $pAX$ reaches approximately the point $B$.  (To be precise we say that $pAX$ is approximately at $B$ if $pAX$ is less than or equal to $AB$ and that $(p+1)AX$ is greater than $AB$.) Now translate $AX$ end-to-end $q$ times (where $q$ might be positive or negative) until $qAX$ is approximately at $D'$. 
We conclude that:
\begin{eqnarray}
	\mod{CD} = |q/p|\  \mod{AB}
\label{eq:ratlength}
\end{eqnarray}
to the  accuracy  defined by the length of the chosen `short measuring stick'.

Extending our notation above to rational numbers, we may define the line $rAB$ to be the line at point $A$, parallel to $AB$ of length $|r|\  \mod{AB}$. If $r$ is negative then $rAB$ is sometimes said to be anti-parallel to $AB$. Note that for each given line $AB$, we have imposed a physical limit to the value of $r$. For a rigid body there is a lower limit \lmin\ so that $rAB$ is no smaller that the shortest measurements of length (the shortest measurable line in the direction $AB$) on the rigid body, and an upper limit \lmax\ so that $rAB$ is no longer than the longest measureable length in the $AB$ direction.
(Aside: $r$ is called a rational number, not because it is sensible, but because it is a ratio of integers.)

Having the above definitions and procedures allows us to use any line (and not only a `short measuring stick') as the measuring stick for its direction, but we need the concept of rotations (and the isotropy of space, taken here as the invariance of rigid bodies under rotations), to compare line lengths in non-parallel directions, see the next section. 

\subsection{Discreteness and Continuity}\label{discreteness}

We have seen how the notion of rigid bodies and the homogeneity of physical space are closely related. We have not yet made any assumptions and statements regarding the continuous or discrete nature of physical space.  

Whether our field of numbers is chosen to be the reals or the rationals, there is an underlying assumption which can be expressed in a number of different ways, perhaps the clearest being that between any two numbers, we can find a third. Mathematically we say that the real number field and the rational number field are both dense. On the other hand, if space and time are quantized, this underlying assumption needs to be examined.

An argument is presented by Isham \cite{isham1995lqt} to show that the normal quantum mechanical framework together with the two assumptions;
\begin{itemize}
\item physical space is homogeneous, 
\item  any spatial distance $r$ can be divided
in to two equal parts, $r=r/2+r/2$,
\end{itemize}
leads inevitably to the Heisenberg algebra. The authors of \cite{Ahluwalia1994,doplicher1994sqi} have argued that the Heisenberg algebra, in particular the commutator $\left[{x}_j,{p}_k\right]$, must be modified once gravitational effects associated with the quantum measurement process are accounted for. The appropriate kinematical algebra for this scenario is the Stabilised Poincar\'e Heisenberg algebra (SPHA for short) \cite{chryssomalakos2004:1}, which does feature a modified Heisenberg algebra.

Any modifications to the Heisenberg algebra necessarily induces an associated change in the underlying geometry of physical space, with either the homogeneity or the continuity of space (with the assumption that any spatial distance can be divided into two equal parts) being lost. The authors of \cite{ahluwalia2008ppa} have argued that it is the underlying homogeneity of space that is lost in this case and furthermore that the induced inhomogeneities  may serve as seeds for structure formation in an earlier epoch of the universe (at the present epoch of the universe the modifications to the Heisenberg algebra are very small and hence one would not expect to observe any inhomogeneities today).

In contrast, the authors of the present paper have on an earlier occasion shown that the Clifford algebra $C\ell(1,3)$ generates the SPHA under the action of the familiar Lie bracket \cite{gresnigt2007sph}. We show in this paper that this Clifford algebra necessarily follows from the homogeneity and isotropy of physical space together with Pythagoras theorem (and the generalization to spacetime). In this derivation of $C\ell(1,3)$ physical space is considered to be homogeneous, however no assumption needs to be made about the continuous or discrete nature of physical space. We therefore argue that the spacetime underlying the SPHA is homogeneous and therefore not continuous. The usual alternative is to argue that spacetime is discrete or quantized. 

Perhaps the simplest formulation of a discrete spacetime is given by Meessen \cite{meessen2005stq} who proposes the following basic postulate of spacetime:
\begin{quote}
An ideally exact distance measurement along any direction in any inertial reference frame can only yield integer multiples of the same universally constant quantum of length $a$.
\end{quote}
In Meessen's formulation, spacetime is a lattice of points with minimum length and minimum time scale and furthermore these minimum values are the same for all equivalent observers. In a discrete spacetime a given spatial distance can not always be divided into two equal parts. A direct consequence of this is that there must exist some indivisible minimum unit of length. However this is not satisfactory either. Other ways of quantizing spacetime have been considered, such as quantizing space and time via a random `sprinkling' of points onto a manifold as is done in causal set theory, \cite{bombelli1987stc}. In such an approach the distances between points vary.

It seems to us therefore, that to assume either spacetime is continuous or spacetime is discrete is unjustified.  Some third alternative to quantizing spacetime is required. This issue appears to be a deep problem. Therefore, for the time being, rather than make an unjustified assumption we park the issue and avoid being distracted by it. 

\subsection{Coordinate systems and choices of an origin}\label{sec:coordinates}

Let us first consider the linear independence of translations and then let us derive the relationship between the geometric operations that we have been considering and the axioms of a vector space.

Our statement of linear independence of lines in our 2D toy world of the desktop is as follows. If two lines $AB$ and $CD$ are not parallel, then geometry says that given any two points $P$ and $Q$ (or any line $PQ$) we can find a unique numbers $r$ and $s$ so that
\begin{eqnarray}
	Q = P + rAB + sCD
\end{eqnarray}
where the equality is up to the sizes of the short measuring sticks in the $AB$ and $CD$ directions.
The fact that we need precisely two non-parallel lines and the two numbers is why we say we are working in a 2D (two-dimensional) world.

This expression can be rewritten in the form of three lines
\begin{eqnarray}
	PQ = r\,AB + s\,CD
	\label{eq:linindep}
\end{eqnarray}
and we say that the three lines $AB$, $CD$, and $PQ$ are linearly dependent. Conversely, two lines $AB$ and $CD$ are linearly independent if and only if they are non-parallel.

Taking an arbitrary but fixed point $O$, which we shall call an origin, allows us to associate a unique line $OP$ for every point $P$ on the desktop. Choosing two more points $A$ and $B$, we may rewrite the linear dependence equation, eq(\ref{eq:linindep}), in terms of $P, A$ and $B$, or the lines $OA, OB$ and $OP$, as
$$ OP=r\,OA+s\,OB$$
for any point $P$. With these choices we say that for origin $O$, the lines $OA$ and $OB$ are a `basis' choice for the lines on the desktop, and $(r,s)$ are the coordinates of $P$ (or $OP$) in this basis.
\subsection{Vectors and unit vectors}\label{sec:vectors}

Let us now carry out an abstraction process, where we construct an algebraic system called a vector space. The vector space replicates many of the addition properties of points and lines. A vector space $\bbV$  over a field $\bbF$ is an abstract mathematical construct that is defined by a set of axioms that describe the addition of the elements of the vector space (the `vectors') and the product of vectors with the elements of the field (the `scalars' or the `numbers'). The key axioms are the Abelian properties and the associative properties of the sums of vectors and products of scalars with vectors.

The axioms lead to the concept of linear independence, which in turn leads to the concept of the dimension of the space and the ability to choose a set of basis vectors.

The usual vector space of freshman physics is obtained by defining a vector as  the equivalence class $[AB]$ of all lines parallel to $AB$ and of the same length as $AB$.

Defining 
\begin{eqnarray}
	\mathbf a = [AB]\\
	 \mathbf b = [CD]\\
	 \mathbf c = [PQ]
\end{eqnarray}	 
means we can write the linear dependence equation, eq(\ref{eq:linindep}),  for our vectors above, as
\begin{eqnarray}
	\mathbf c = r \mathbf a + s \mathbf b \textrm{ where } r,s \in \bbQ
\end{eqnarray}		
and in particular the equivalence classes for the lines $rAB$ and $AB$ are related by
\begin{eqnarray}
	[rAB] = r [AB]
\end{eqnarray}		
Observe that there is always one line $OA_0$ in the equivalence class $\mathbf a = [AB]$ that has its tail at the origin $O$ of the coordinate system. The vector $\mathbf a$ can then be described by the point $A_0$. It is easy to confuse, or in some instances conflate, the point $A_0$ on the rigid body, with one or more of the lines on the rigid body in the class $[AB]$, or even with the abstract algebraic entity that is the vector $\mathbf a$.

If the line $AB$ is chosen as the measuring stick in the direction of $AB$, and we choose units for length so that $\mod{AB}$ is $1$ unit of length,  then the vector $\mathbf a = [AB]$ is called the unit vector in the direction of $\mathbf a$. We shall usually label unit vectors with a `hat' symbol, as in \ahat.

In general, given vectors $\mathbf a$, $\mathbf b$, \ldots, we may choose unit vectors such that $\mathbf a = a \ahat$, $\mathbf b =  b \bhat$, \ldots. As with lines, when we write $\mathbf a = a \ahat$, we say $a$ is the magnitude, or length, of $\mathbf a$ and we say that \ahat\ is a unit vector in the direction of $\mathbf a$. The length $a$ is either zero or a positive (rational) number.

Observe that vectors are `mathematical objects', being elements of a vector space, $\bbV$. Vectors have uses outside of geometry, and are often introduced in  mathematics course without any connection to geometry. In this paper we have them firmly linked to geometric objects. Vectors can describe operations  on lines (which are passive objects), or  translations (which are active objects that describe the movement of rigid bodies, with their points, lines and areas). By choosing a point as the origin, a vector can describe a point. It is common to confuse these  different, albeit linked, concepts: the mathematical entity that is a vector and that belongs to a vector space, and the physical entities of points, lines, and  translations. It has long been known that many beginning physics and mathematics students can take a long time to grasp vector algebra because of this. If there  are multiple meanings of the new words and new concepts, and this is not pointed out, confusion reigns in the students' minds. We aim to consistently use a notation that keeps the physical objects clearly separated from the mathematical objects.

It is common to generalize the application of vector spaces from geometric space where vectors represent points, lines and translations
to other physical objects, such as forces, velocities, accelerations. 
In typical physics notation, Newton's Second Law \cite{serway2008psa} is written as
\begin{eqnarray}
	\mathbf F = m \mathbf a
\end{eqnarray}  
where $\mathbf F$ is a force of magnitude $F$ in direction \Fhat, $\mathbf a$ is acceleration of magnitude $a$ in direction \ahat. 
Thus in terms of magnitudes 
\begin{eqnarray}
	F = m \ a 
\end{eqnarray}
and in terms of directions
\begin{eqnarray}
	\Fhat = \ahat
	 \label{eq:Fa}
\end{eqnarray}  
since both the unit vectors \Fhat\ and \ahat\ are dimensionless in the sense of having no units (neither newton nor metre/second/second). The only property that unit vectors have, in this formulation, is direction.

As a further example of the linking of the concepts and the typical abuse of notation, note that $\mathbf a$ can represent a displacement by a distance $a$ (of say 4.5\,metre) in direction \ahat\  (of say 7.3 degrees north of east). 
It is usual to say that \ahat\  is a unit vector, when \ahat\ is really a direction.
We shall abuse notation in this way to the extent that if $\mathbf a$ is of length $1$, we write $\mathbf a = \ahat$ rather than $\mathbf a = 1\ahat$.

\subsection{Vector Addition}\label{sec:vector addition}

The addition of vectors follows simply from the correspondences set up above. To diagrammatically represent the addition of two vectors $\mathbf a$ and $\mathbf b$ we choose two appropriate lines (in the equivalence classes of the two vectors) to represent these vectors. The algebraic equation $\mathbf a + \mathbf b = \mathbf c$ can then be related to the geometric picture of joining the tail of the second line (representing $\mathbf b$) to the head of the first line (representing $\mathbf a$), giving a new line (representing the resultant vector $\mathbf c$) from the tail of the first line to the head of the second.

The commutativity of vector addition  
$\mathbf a + \mathbf b = \mathbf b + \mathbf a$   corresponds diagrammatically to  		
\begin{figure}[ht]
	\centering
	\includegraphics[width=0.70\textwidth]{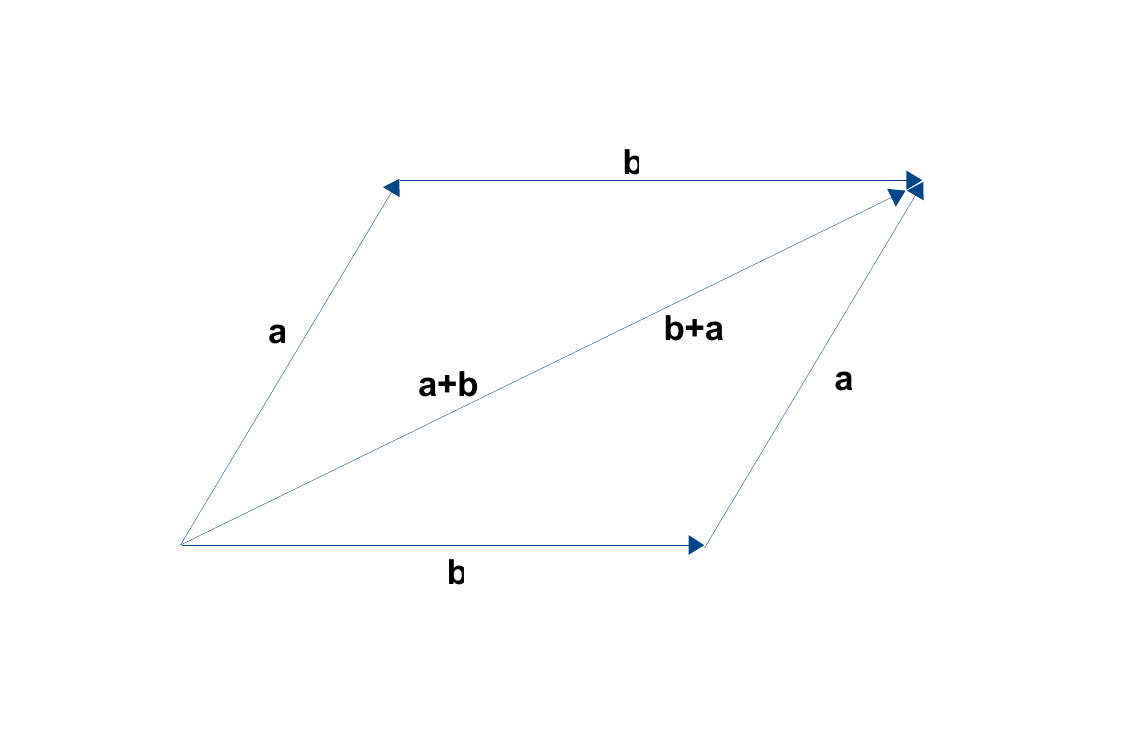}
	\label{fig:abba}
	\caption{By choosing appropriate lines to represent vectors, this figure demonstrates the commutativity of addition of vectors. The vector resulting from adding the vector $\mathbf a$ to the vector $\mathbf b$ is the same as the vector resulting from adding $\mathbf b$ to the vector $\mathbf a$, so that $\mathbf a +\mathbf b =\mathbf b +\mathbf a$.}
\end{figure}                                   
 the parallelogram law, see figure 4.
\begin{figure}[ht]
	\centering
	\includegraphics[width=0.70\textwidth]{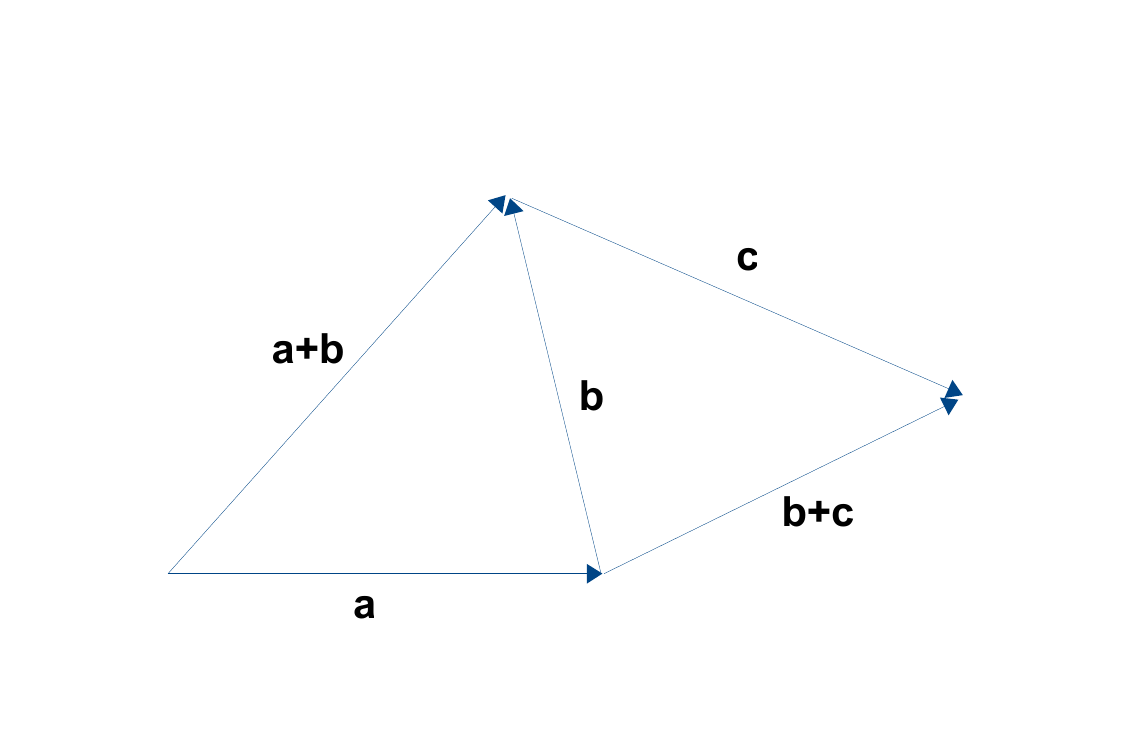}
	\label{fig:abcd}
	\caption{This figure demonstrates the associativity of vector addition $(\mathbf a +\mathbf b )+\mathbf c = \mathbf a +(\mathbf b + \mathbf c )$.}
\end{figure}
We also have associativity  $(\mathbf a + \mathbf b) + \mathbf c = \mathbf a + (\mathbf b + \mathbf c)$, see figure 5.
\subsection{Concluding remarks - What have we achieved?}\label{conclusion}
 
 In this section we have seen that the observed translational features of rigid objects in the geometric space of the 2D physical world led to a set of operations on the points, lines and areas of rigid objects. 
They also led to the abstract construct that is the mathematical structure of a 2~dimensional vector space $\mathbb V$ over the rational number field $\mathbb Q$.
The vector space axioms are chosen so that the mathematical structure of the vector space matches the geometry of space, in particular its homogeneity. However the operations on rigid bodies are described by a subset of those rationals, limited by \lmin\  and \lmax.

Generalising to the full physical world, we can see no physical situation in which we need the number $\infty$ nor lines whose length approaches zero by an infinite (or Cauchy) process. Points, lines and areas are the `observables' of our world of finite sized rigid bodies.

We remind the reader that we avoid making assumptions about the appropriate mathematics except when they have a firm basis in  measurements and observations within the world under consideration. As a particular example we considered the issue of  continuity conditions on the number system that we need to use, and has concluded that we need only the rational numbers. Typically, continuity conditions are assumed. However various theories of spacetime assume a graininess to spacetime or a ``quantum foam''. We will return to such issues later. For the present we ask the reader to not get distracted by these issues and to explore the 2D (and the 3D) world as we find it.

Finally, we also remind the reader that although we insist on being able to represent every physically allowed operation within our mathematical framework, the converse does not hold and there are many mathematical operations which have no physically meaningful counterparts.  

In the next section we extend these ideas to take into consideration the rotational properties of 2D space. We are led from the vector space over \bbQ\ to an algebra over \bbQ. An algebra contains the vector space operations of multiplication of vectors by scalars, and the addition of vectors to vectors. It contains also the operation of multiplication of vectors by vectors. The algebra we derive is an example of a Clifford algebra.

\section{Isotropy plus Pythagoras gives a Clifford Algebra}\label{AaAII2010}

In this section we consider the rotational motion of 2D rigid bodies. This leads to a product operation of vectors with vectors, giving rise an algebra that corresponds closely to the isotropy of physical 2D space and also to the rotational invariance of rigid bodies in this space.

There is a subtlety we have not mentioned: When studying homogeneity by means of translations we talked of lines such as $AB$. Newton's first law talks of `straight lines'. We drew our lines in our toy world as straight lines, but homogeneity would seem to require that lines are merely of constant curvature. However the rotation of the line $AB$ by $\pi$ about its centre, $A+\frac{1}{2}AB$, enables experimental verification that all intermediate points on line $AB$ between $A$ and $B$, also lie on the rotated line $BA$. 

The algebra we obtain in this section is four dimensional. We shall see in section \ref{AaAIII2010} that a $n$-D world naturally leads to an $n$-dimensional vector space to describe the homogeneity and translation properties of the physical space, and to a $2^n$-dimensional algebra to describe its isotropy and rotation properties.

We noted that the physical world did not satisfy all the axioms of the vector space, in particular the physical world is finite in extent, both in the very large and the very small. 
Here we maintain our approach to the assumptions underlying the basic laws of physics: we shall only make the assumptions we need to, and propose mathematical axioms that seem to be required -- absence of evidence is not evidence of absence, nor a reason to make  assumptions to simplify the mathematics.

This section continues our study of our 2D toy world of finite extent (finite both in terms of how small and how large) to deduce some of the geometrical consequences of rotational invariance. The rotational invariance shown by all rigid bodies in the physical world is known as the `isotropy of space'. The consequence of our contemplations is  to find a natural way of comparing lengths of non-parallel lines and to extend the vector space to a Clifford algebra.

\subsection{Isotropy and rotational invariance of rigid objects}\label{sec:isotropy}

Consider our toy world consisting of sheets of paper on the desktop. The most general motion of a sheet of paper relative to the desktop is described by giving the initial ($A$, $B$) and final ($A'$, $B'$) positions on the desktop of two distinct points ($A$, $B$) of the paper. 

We say that we have a rotation about a point $A$ if that point does not move, $A = A'$. If however $A\neq A'$ the motion
can be described either as a translation $A$ to $A'$ followed by a rotation about $A'$, or in some special cases simply as a rotation about some other fixed point. In general if we are given the initial location of two (distinct) points, $A$ and $B$, and the final location of those points, $A'$ and $B'$, then we can describe the motion as a translation $A$ to $A'$ described by the line $AA'$, followed by a rotation about $A'$ where the point $B + AA'$ is rotated to $B'$. Since our world is of finite extent, there are many cases where there is no fixed point. A pure translation is not a rotation about ``the point at infinity'' as that point is not in our physical world, nor in our vector space.

For simplicity let us first consider rotations about a fixed point $A$, so that $A=A'$. 
It is easily confirmed in our toy world that two rotations of a sheet of paper about the same point are equivalent to a single rotation.
There are several special cases of immediate interest:
\begin{itemize}
	\item  The null rotation, $0$\,radian (or $0^\circ$)  where  $A'B' = AB$, for all points $B$.
	\item  The rotation through  $2\pi$ (or $360^\circ$)  where again $A'B' = AB$.
	\item  The rotation through $\pi$ (or $180^\circ$) where $A'B' = - AB$. This  rotation applied twice is equivalent to the null rotation.
	\item The rotation through $\frac{1}{2}\pi$ (or $90^\circ$) where we say that $A'B'$ is orthogonal to $AB$.  This rotation applied twice takes $AB$ to $-AB$.
	\item The rotation through $\frac{3}{2}\pi$ (or $270^\circ$, or $-\frac{1}{2}\pi$, or $-90^\circ$) where $A'B'$ is again orthogonal to  the line $AB$ and again two such rotations take $AB$ to $-AB$. 
\end{itemize}
We observe that when rotating objects in our toy world, then it is a property of the space, and of rigid bodies, that the rotation by any multiple of $2\pi$ is equivalent to the null rotation. A rotation through angle $\theta$ has the same effect on the paper as a rotation through the angle $2\pi + \theta$.

The last two of the rotations in the list above, those through $\frac{1}{2}\pi$ or  $\frac{3}{2}\pi$, are  characterized by the property that applying either of them twice gives a line $AB''$ that is parallel to $BA$ (or equivalently to  $-AB$). This property is so important that lines at an angle of $\pi/2$ (or $90^\circ$) to each other may be described in several ways in English: e.g.\ at right angles, normal, perpendicular, orthogonal. 

In the above we have used two equivalent descriptions of a rotation of the sheet of paper, as an operation on lines \rot{AB\rightarrow A'B'}$(paper)$,  or as an angle about a point \rot{\theta(AB,A'B') \textrm{\ about } A}$(paper)$, where $\theta(AB,A'B')$ is the anti-clockwise angle between the lines $AB$ and $A'B'$. As with translations, rotations of the paper in one direction are equivalent to rotations of the desktop in the opposite direction. For example we have the equalities:
\begin{eqnarray}
\rot{AB\rightarrow A'B'}(paper) 
		&=& \rot{\theta(AB,A'B') \textrm{\ about\ } A}(paper) \nonumber \\
		&=& \rot{-\theta(A'B',AB) \textrm{\ about\ } A}(paper) \nonumber\\
		&=&\rot{A'B'\rightarrow AB}(desktop)		\nonumber\\
			&=& \rot{\theta(A'B',AB) \textrm{about\ } A}(desktop)
\label{eq:rotequiv}
\end{eqnarray}
which all depend on the observation that space is isotropic. The `isotropy of space' is the name we give for the property that a pair of rigid objects do not change their relationship, one to the other, when they are both rotated equal amounts. As with the `homogeneity of space', isotropy is a property that requires the concepts of rigid objects (in our case, at least two pieces of paper and the desktop) and of motion relative to a reference rigid object (in our case, any one of the objects).

We began this subsection by considering several cases of rotations by special angles, $\theta=0,\frac{1}{2}\pi,\pi,\frac{3}{2}\pi,2\pi,$ etc. In a manner similar to the definition of adding and dividing lengths in the previous section, we can define rotations by angles that are rational fractions $r=p/q$ of $\pi$, where $r\in \bbQ$, such that 
\begin{eqnarray}
\rot{r\pi, \textrm{about\ } A}(AB)\cong AB'  \textrm{ to the desired observable accuracy}
\end{eqnarray}

\subsection{Unit measuring sticks and unit vectors}\label{sec:unit vectors}

Another property of our toy world is that any rotations except those through an integer multiple of $\pi$, take $AB$ into a line $AB'$ that is linearly independent of $AB$. 
Both the `short measuring stick', $AX$, and the `measuring stick' $AB$, of the previous section can be represented by  pairs of points on the sheet of paper.  Rotation of the paper from the direction of $AB$ into the direction of another line $CD$ allows the comparison of the length of two sticks in two directions of $AB$ and $CD$.  
Using the translational and rotational invariance of our measuring sticks we can make comparisons of the lengths of all lines in the plane. We therefore conclude that, because of the homogeneity and isotropy of space, only one measuring stick is needed.

A pair of orthogonal lines, $XX'$ and $YY'$, in our toy world gives rise to a pair of orthogonal vectors $\textbf x = [XX']$ and $\textbf y = [YY]$ in our vector space. By choosing the measuring stick to be of unit length (say 1 metre), we can choose the corresponding vectors to be of unit length. We write them as \xhat\ and \yhat. 
Pairs of orthogonal vectors of unit length are said to be orthonormal pairs.

Since our desktop world is 2D, any vectors $\bf a$ and $\bf b$ can be written in terms of the orthonormal vectors  \xhat\ and \yhat.
\begin{eqnarray}
	\mathbf a = a \ahat = a_x\xhat  + a_y\yhat 
	\label{eq:aincoords}
\end{eqnarray}
and
\begin{eqnarray}
	\mathbf b = b \bhat   = b_x \xhat   + b_y \yhat
		\label{eq:bincoords}
\end{eqnarray}					 
It is customary to say that the numbers $a_x$ and $a_y$ are the components of the vector $\bf a$ in the orthonormal basis system (\xhat, \yhat).

The axioms of the vector space allow the addition operation to be written as
\begin{eqnarray}
	{\bf a} + {\bf b} = (a_x + b_x)\xhat + (a_y + b_y)\yhat 
	\label{eq:ab}
\end{eqnarray}
Each of these vector space equations may be carried across to corresponding operations on lines and translations in the plane. A line $AB$ may be written in terms of unit orthogonal lines $XX'$ and $YY'$ as
\begin{eqnarray}
AB=aXX' + bYY'
\end{eqnarray}
and 
\begin{eqnarray}
A+AB=B= A+aXX' + bYY'
\end{eqnarray}

\subsection{Multiplication of lines by lines and vectors by vectors}\label{sec:multiplication}

\begin{figure}[ht]
	\centering
	\includegraphics[width=0.70\textwidth]{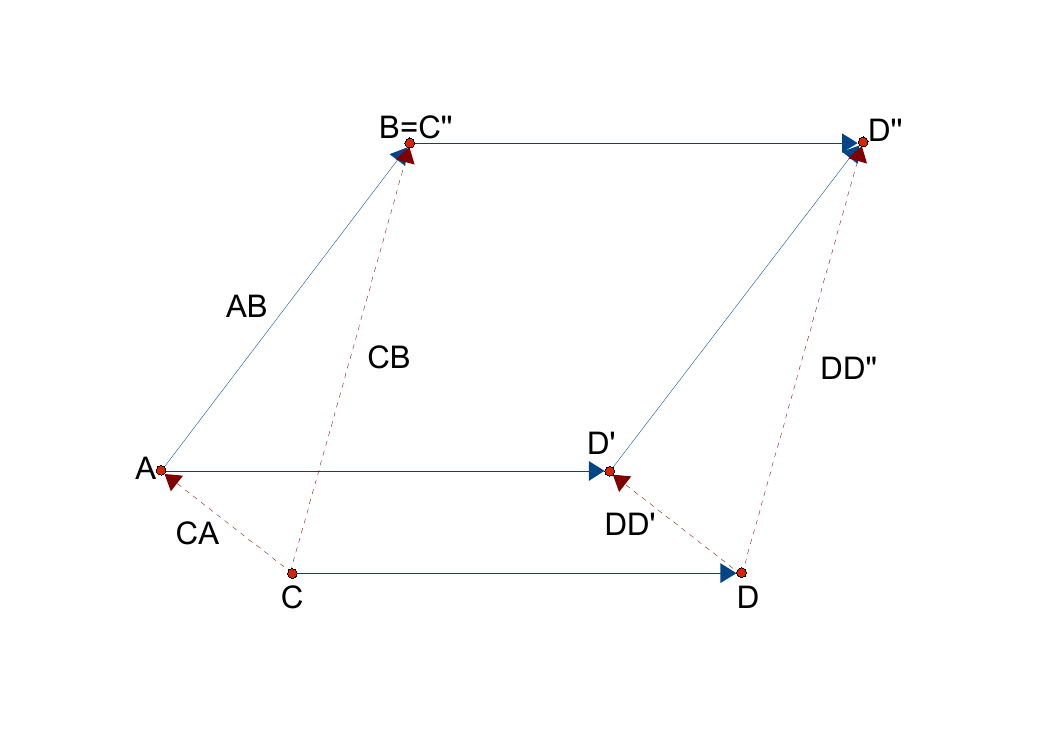}
	
	\caption{The  `product' $(AB,CD)$ of the lines $AB$ and $CD$ is defined as the parallelogram formed by translating the line $CD$ first to $AD'$ then second to $BD''$, and translating $AB$ to $D'D''$}
	\label{fig:linepair}
\end{figure}

We wish to have a geometric definition of the `associative multiplication' or `product' of one line, $AB$, by another, $CD$, which we shall denote by the ordered pair ($AB,CD$).
First translate the line $CD$ so that $C$ moves to $A$. Thus $C' = C +CA = A$ and $D' = D + CA$.
Next  translate the line $CD$ so that $C$ moves to $B$. Thus $C'' = C +CB = D$ and $D'' = D + CB$.
Define the geometric entity  associated with the ordered pair ($AB,CD$) to be the parallelogram $ABD'D''$ as shown in the figure \ref{fig:linepair}. 

Now define  the `multiplication' of one vector, $\bfa=[AA']$, by another, $\bfb = [BB']$,  creating a `bi-vector' denoted by \bfab, as the equivalence class of all products ($AA',BB'$) under appropriate equivalence relations.
\begin{eqnarray}
	\bfab &=& [(AA',BB')]\nonumber
	\label{eq:vectproddef} \\
						&=& \textrm{The equivalence class of all line pairs ($CC',DD'$) that}\nonumber \\
						& & \textrm{are translationally and rotationally equivalent to $(AA',BB')$} 
\end{eqnarray}
The first equivalence relation to use is the translational invariance inherited by the bi-vector from its vectors  
\begin{eqnarray}
	\bfab = [\mathbf a,\,\mathbf b], \textrm{ where } \bfa=[AA'] \textrm{ and } \bfb=[BB']
\end{eqnarray}
We also impose on \bfab, associativity, bi-linearity over the field $\mathbb Q$,  and  rotational invariance. The easiest way to do this is to define \bfab\ in terms of its expression in orthonormal coordinates. (Joyce and Butler  \cite{joyce2002gaa} give a purely geometric argument.)
Define $\mathbf{a b}$ as the term-wise associative expansion (the free product \cite{butler:aem}) of the components written in some orthonormal axis system. 
\begin{eqnarray}
	\mathbf{a b} &=& (a_x \xhat + a_y\yhat)(b_x\xhat  + b_y\yhat) \nonumber \\
     &=& a b\, \ahat \bhat \nonumber \\
     &=& a_x b_x \xhat^2 + a_y b_y \yhat^2 + a_x b_y\xhat\yhat + a_y b_x\yhat\xhat
	\label{eq:freeprod}
\end{eqnarray}
where we have used the property that the components, $a_x$, $a_y$, $b_x$, $b_y$, being numbers, commute with the unit vectors. However  the product of the vectors is {\em not} commutative, as we shall explore in detail in the following.

We now ask that the product incorporate the Euclidean metric and in particular ask that Pythagoras' theorem holds. For the product of vector \bfa\ with itself, we have 
\begin{eqnarray}
	\mathbf{aa} &=& \mathbf a^2 \nonumber \\
							&=& a^2 \ahat^2 \nonumber \\ 
     					&=& a_x^2 \xhat^2 + a_y^2 \yhat^2 + a_x a_y (\xhat\yhat + \yhat\xhat)
	\label{eq:asq}
\end{eqnarray}
If this is to satisfy Pythagoras, 
\begin{eqnarray}
a^2 = a_x^2 + a_y^2
\label{eq:pythag}
\end{eqnarray}
then we must have 
\begin{eqnarray}
	\ahat^2 = \xhat^2 = \yhat^2
	\label{eq:unitssq}
\end{eqnarray}
and
\begin{eqnarray}
	\xhat\yhat + \yhat\xhat =0
	\label{eq:sumzero}
\end{eqnarray}
We want the smallest algebra that contains both $\xhat^2$ and \xhat\yhat\ (and thus also $\yhat^2$, $\ahat^2$ and \yhat\xhat).

We first choose $\ahat^2$  be the rational number $\eta$.  
In the previous subsection we chose unit vectors to have equal length, which we declared to be the length unit, or `standard measuring stick'. The link between unit vectors and the standard measuring stick can be rescaled by any  number in our field \bbQ. However that number appears as a square in eq(\ref{eq:unitssq}),  so we have  two independent cases for $\eta$,  depending on whether $\ahat^2$ is positive ($\eta = +1$) or negative ($\eta = -1$).

The number $\eta$ is known as the metric of the space. The choice of $\eta=-1$ gives $\bfa^2 \leq 0$ for all \bfa. We shall call this choice the `anti--Euclidean metric'. In the next section we compare and contrast the two possible choices of metric, $\eta=\pm 1$. 

The pair of equations, eq(\ref{eq:unitssq}) and eq(\ref{eq:sumzero}), define the Clifford algebras $C\ell(2,0)$ and $C\ell(0,2)$ as $\eta=\pm1$.

The second equality, eq(\ref{eq:sumzero}), introduces a fourth basis element \khat\ (beyond $1$, \xhat\ and \yhat)
\begin{eqnarray}
	\khat = \xhat\yhat = - \yhat\xhat
	\label{eq:defkhat}
\end{eqnarray}   
 into the algebra. We have created an associative algebra of the four basis elements $1$, \xhat, \yhat\ and \khat\, where
\begin{eqnarray}
	\khat^2 &=& (\xhat\yhat)(\xhat\yhat)  \nonumber \\
					&=& - \xhat(\yhat\yhat)\xhat  \nonumber \\
					&=& - \eta\xhat\xhat  \nonumber \\
					&=& - \eta^2 \nonumber \\
					&=& -1
	\label{eq:khat2}
\end{eqnarray}
for both choices for $\eta$.

We shall see that \khat\ is the algebraic unit that describes a unit area in the $xy$-plane, it is not the normal to the plane -- such a normal does not exist in our 2D geometry. Instead, just as the basis vector \xhat\ is the direction of the $x$-axis and is dimensionless, so the basis bi-vector \khat\ is what we may call the `direction' of the $xy$-plane. Being the product of two dimensionless quantities, \khat\ is dimensionless and, as we shall see, is associated with the angle $\frac{1}{2}\pi$ radian.

The four  objects $(1, \xhat, \yhat, \khat)$ together with their negatives $(-1, -\xhat,-\yhat,-\khat)$,  form the eight element Clifford groups, $C\ell^\textrm{ group}(2, 0)$ or $C\ell^\textrm{ group}(0, 2)$, associated with the Clifford algebras $C\ell(2, 0)$ or $C\ell(0, 2)$ as $\eta=\pm1$. The group combination law is the associative product defined above in eq(\ref{eq:freeprod}). The same four objects are also the basis for the four dimensional vector space $\bbQ^4$ over our field, $\mathbb Q$, using the addition operation, with the arbitrary element written 
\begin{eqnarray}
	\textbf A =a+b\xhat+c\yhat+d\khat \qquad\textrm{where } a, b, c, d \in \bbQ
	\label{eq:Aarb}
\end{eqnarray}

An algebra is that mathematical structure that has both the addition and scalar multiplication operations of a vector space, and also the associative multiplication operation of a group. In our case the general elements of the algebra are linear combinations of arbitrary scalars $a$, vectors \bfa, and bi-vectors \bfab. In the above we have derived a four dimensional algebra that is firmly based on the homogeneity and isotropy of our 2D physical toy world, being sheets of rigid paper on the rigid desktop. Let us now explore the vector product, \bfab.

\subsection{The geometric information in the vector product\label{sec:geominfo}}

The vector product $\bfab$ represents both the angle between the lines that correspond to \bfa\ and \bfb, and segments of the plane (parallelograms) spanned by the lines that correspond \bfa\ and \bfb. It has lost the information about the absolute lengths of \bfa\ and \bfb, as can be seen using the bi-linearity of the vector product
\begin{eqnarray}
	\bfab = (\frac{1}{r} \bfa) (r \bfb)
	\label{eq:bilin}
\end{eqnarray}
We remind ourselves that vectors have, in a similar sense, lost the information about the positions of lines, vectors have only length and direction.

The vector \bfa\ represents any vector in the translational equivalence class $\bfa = [AA']$ and similarly for $\bfb = [BB']$. 
 So the equivalence class $\bfab = [([AA'],[BB'])]$ knows neither the start of the lines, nor their length, only the product of their lengths. We shall also see that it  knows only the difference in the directions of the lines, $\bfab$  is  invariant under rotations in the plane of the desktop.

In the next subsection we shall prove  (see eq(\ref{eq:abasrot})) the following. Consider the pair of lines $AA'$ and $BB'$ and form the product ($AA',\, BB'$) with corresponding bi-vector \bfab. Take two other lines $CC'$ and $DD'$ in the desktop, to give the product ($CC',DD'$) and corresponding bi-vector $\bfcd$. Then this second product belongs to the same equivalence class of products as ($AA',BB'$), that is bi-vector $\bfcd$  equals bi-vector $\bfab$, if and only if both $\mod{AA'} \mod{BB'} = \mod{CC'} \mod{DD'}$, and the angle $\theta(AA'\  \mathrm{to}\ BB')$ equals the angle $\theta(CC'\ \mathrm{to}\ DD')$.

Just as the vector \bfa\ represents an equivalence class of lines (passive geometric objects) and also represents an equivalence class of translations (active geometric objects), we shall see that the vector product \bfab\ represents an equivalence class of line pairs (passive geometric objects), and also an equivalence class of rotations (active geometric objects). 

\subsection{Rotations using bi-vectors\label{sec:rotbivector}}

The choice of $\eta= -1$ leads to counter--clockwise  rotations in what follows, while the choice of $\eta= +1$ leads to clockwise  rotations. Often the same effect can be obtained by writing the operator on the right instead of the left. For the remainder of the paper, except where we state otherwise, we choose the value\footnote{The reader is encouraged to work through the equations of the remainder of this section using $\eta$ as a variable or with $\eta=+1$.}
\begin{eqnarray}
	  \eta = -1	\label{eq:negmetric} 
	  \end{eqnarray}
because handedness and parity--conservation arguments in section \ref{AaAIII2010} show that this choice is appropriate for the geometry of the rigid objects of the Universe.

With this choice of $\eta$, we may calculate that the basis bi-vector  \khat\,  when used as an operator acting on the left, rotates \xhat\ into \yhat, and \yhat\ into $-\xhat$, as follows:
\begin{eqnarray}
	  \khat\xhat &=& (\xhat\yhat)\xhat   \nonumber \\
	  	 &=& \xhat(\yhat\xhat)  \nonumber		\\
	  	 &=& -\xhat(\xhat\yhat)   \nonumber		 \\
	  	  &=& -(\xhat\xhat)\yhat  \nonumber	 \\
	  	  &=& -\eta\yhat  \nonumber	 \\
	  	  &=& \yhat  	
\end{eqnarray}
and
\begin{eqnarray}
	  \khat\yhat &=& (\xhat\yhat)\yhat  \nonumber	 \\
	  	 &=& \xhat(\yhat\yhat)  \nonumber	 \\
	  	  &=& \eta\xhat  \nonumber	 \\
	  	  &=& -\xhat  	
\end{eqnarray}
Thus on taking the two vectors \xhat\ and \yhat\ as an ordered pair, (\xhat,\yhat), $\khat$ is the $\frac{1}{2}\pi$ rotation of this pair in the positive sense  
\begin{eqnarray}
\rot{\khat}(\xhat,\yhat) &=& \khat(\xhat,\yhat) \nonumber	 \\
		&=& (\yhat,-\xhat) \nonumber	 \\
			&=&\rot{\frac{1}{2}\pi}(\xhat,\yhat)
\end{eqnarray}
and $-\khat$ is the $\frac{3}{2}\pi$ rotation in the positive sense, the $\frac{1}{2}\pi$ rotation in the negative sense, or the \khat\ operator acting on the right  
\begin{eqnarray}
\rot{-\khat}(\xhat,\yhat) &=& -\khat(\xhat,\yhat) \nonumber	 \\
		 &=& (\xhat,\yhat)\khat \nonumber	 \\
		 &=& (-\yhat,\xhat) \nonumber	 \\
			&=&\rot{-\frac{1}{2}\pi}(\xhat,\yhat)
\end{eqnarray}

Since $\khat^2=-1$, DeMoivres' theorem may be used to write the exponential function $\exp(\theta\khat)$ as the sum of sine and cosine terms. For any object such as \khat\ that squares to $-1$ we have
\begin{eqnarray}
	  \exp(\khat\phi) =
	  \Exp^{\skhat\phi} = 
	  \cos\phi + \khat\sin\phi
	\label{eq:demoivre}
\end{eqnarray}
This result is a generalisation of the result for the complex numbers, where $i=\sqrt{-1}$ and $\exp(i\phi)=\cos\phi+i\sin\phi$.

We may transform the expression for \bfa\  in orthonormal, Galilean coordinates (\xhat, \yhat), eq(\ref{eq:aincoords}), $$\bfa = a_x\xhat + a_y\yhat$$ into circular polar coordinates ($r, \theta$), where $r=a$
\begin{eqnarray}
	 a_x &=& a\cos\theta	\nonumber  \\
	 a_y &=& a\sin\theta	\nonumber  \\
	\bfa &=& a\cos\theta\xhat + a\sin\theta\yhat	
	\label{eq:a_in_polars}
\end{eqnarray}
and so using eq(\ref{eq:demoivre}) and $ \khat\xhat =\yhat$
\begin{eqnarray}
	\bfa &=& a \exp(\khat\theta) \xhat \nonumber \\
	&=& a \, \Exp^{\skhat\theta}\xhat 
		\label{eq:a_in_exp} 
\end{eqnarray}
 
The operator $\exp(\khat\phi)$ of eq(\ref{eq:demoivre}) rotates vector \bfa, when operating on the left, by angle~$\phi$
\begin{eqnarray}
	\Exp^{\skhat\phi}\bfa &=& \Exp^{\skhat\phi}a \, \Exp^{\skhat\theta} \xhat \nonumber \\
	   &=& a \,\Exp^{\skhat(\phi+\theta)} \xhat	\label{eq:arot} 
\end{eqnarray}
However it does not behave this way acting on scalars, $r\in \bbQ$, or on itself. Thus to write a formula for multi-vectors (scalars, vectors and bi-vectors), we require a different form for the operator. This form is as a two--sided operation: If \bfA\ is an arbitrary element of $C\ell(0, 2)$, as in eq(\ref{eq:Aarb}), then
\begin{eqnarray}
	\rot{ \textrm{by }\phi\textrm { in the \khat\ plane}}(\bfA) =
	\Exp^{\skhat\phi/2} \bfA \Exp^{-\skhat\phi/2} 
	\label{eq:rotA}
\end{eqnarray}
because \khat\ commutes with scalars and itself, and anti-commutes with the mono-vectors \xhat\ and \yhat. As we have seen, scalars (that is, numbers) and the bi-vector \khat\ are unchanged by rotations in the $xy$-plane, so $a+d\khat \rightarrow a+d\khat$ while the vector part of \bfA, $b\xhat+c\yhat$, is rotated correctly
\begin{eqnarray}
	\Exp^{\skhat\phi/2} \bfA \Exp^{-\skhat\phi/2} 
	&=& \Exp^{\skhat\phi/2} (a+b\xhat+c\yhat + d\khat) \Exp^{-\skhat\phi/2}
	\nonumber \\
	&=& \Exp^{\skhat\phi/2} \Exp^{-\skhat\phi/2}(a+ d\khat) 
	+ \Exp^{\skhat\phi/2} \Exp^{\skhat\phi/2} (b\xhat+c\yhat) \nonumber \\
	&=&  (a+ d\khat) 
	+ \Exp^{\skhat\phi} (b\xhat+c\yhat)
\end{eqnarray}
where we have used the fact that \xhat\ and \yhat\ anticommute with \khat.

The general bi-vector \bfab, eq(\ref{eq:freeprod}), can be written in terms of scalar and pure bi-vector terms as follows
\begin{eqnarray}
	\bfa &=& a \, \Exp^{\skhat\theta_a} \xhat	\nonumber \\
	\bfb &=& b\, \Exp^{\skhat\theta_b}	\xhat \nonumber \\
	\bfab &=& ab\, \Exp^{\skhat\theta_a}\xhat \Exp^{\skhat\theta_b}  \xhat	\nonumber \\
	 &=& -ab  \Exp^{\skhat(\theta_a-\theta_b)}	\nonumber \\
	 &=& -ab \cos \theta_{ab} + ab\khat \sin \theta_{ab}
	\label{eq:abasrot}
\end{eqnarray}
showing that \bfab\ depends only on the product $ab$ of the lengths of \bfa\ and \bfb, and the angle between them, $\theta_{ab} = \theta_b-\theta_a$. This proves the result stated at the end of subsection \ref{sec:geominfo} above. In subsection \ref{sec:halfangle} below, we shall obtain a simple expression for the bi-vector for half the angle  between lines \bfa\ and \bfb\ as it is needed for rotating general multi-vectors \bfA, as in eq(\ref{eq:rotA}).

In many situations the coordinate free representation of the above results is powerful. Recall that \bfab\ is the equivalence class of all line pairs $[(AA', BB')]$ and where $\bfa = [AA']$ and $\bfb= [BB']$. In general we have that the vector \bfr\ is rotated through the angle from $\mathbf{\hat{a}}$ to $\mathbf{\hat{b}}$, into the vector $\bfr'$, by multiplying on the right by $\mathbf{ab}/\eta ab$ or on the left by $\mathbf{ba}/\eta ab$ as follows
\begin{eqnarray}
	\rot{\bfa\rightarrow\bfb}(\bfr) &=& \bfr' \nonumber \\
	&=& \bfr\bfab/(\eta ab) \nonumber \\
	&=& \bfb\bfa\bfr/(\eta ab)
	\label{eq:rotrab}
\end{eqnarray}
(A metric free version of the above is obtained if \bfa\ and \bfb\ are of the same length, $a=b$, when the rotation operator is simply $\bfab/\bfa^2$).

Because the algebra is associative, this rotation operator has a trivial action on a vector \bfa, acting from the right
\begin{eqnarray}
	\rot{\bfa\rightarrow\bfb}(\bfa) &=& \bfa\bfab/(\eta ab) \nonumber \\
	&=& \eta a^2\bfb/(\eta ab)   \nonumber \\
	&=& \frac{a}{b}\bfb
\end{eqnarray}
which is a line of length $a$ in the direction of \bfb. The corresponding results hold for multiplication on the left
\begin{eqnarray}
	\rot{\bfa\rightarrow\bfb}(\bfa) &=& \bfb\bfa\bfa/(\eta ab) \nonumber \\
	&=& \eta a^2\bfb/(\eta ab)   \nonumber \\
	&=& \frac{a}{b}\bfb
\end{eqnarray}
We note that \bfb\bfa\ is the inverse rotation (the rotation in the opposite sense) to \bfa\bfb, as it rotates \bfb\ into \bfa. $\rot{\bfa\rightarrow\bfb}=\rot{\bfb\rightarrow\bfa}^{-1}$.

\subsection{Half angle rotations\label{sec:halfangle}}
\begin{figure}[ht]
	\centering
		\includegraphics[width=0.70\textwidth]{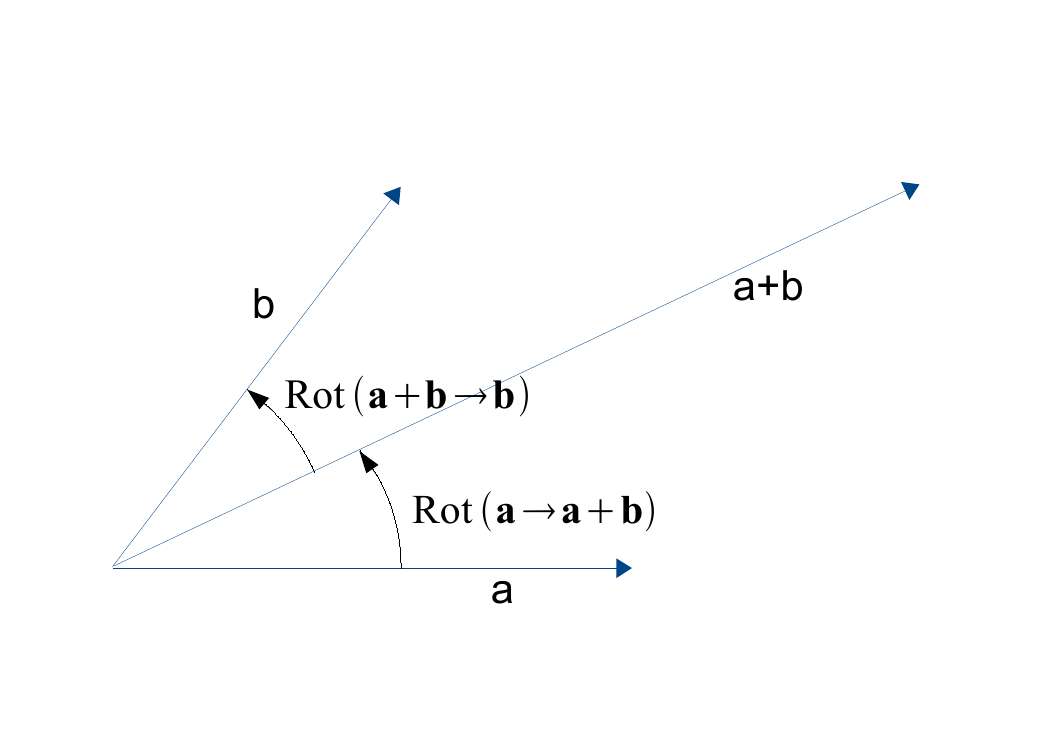}
	\caption{The rotation $\bfa\rightarrow\bfb$ may be composed as the product of two rotations, from \bfa\ to the diagonal $(\bfa+\bfb)$, and then to \bfb.}
	\label{fig:diagpar}
\end{figure}

Figure \ref{fig:diagpar} shows that the rotation \rot{\bfa\rightarrow\bfb} may be composed as the product of two  rotations, first \rot{\bfa\rightarrow\bfa+\bfb}, and then \rot{\bfa+\bfb\rightarrow\bfb}. 
\begin{eqnarray}
	\rot{\bfa\rightarrow\bfb}(\bfA)
	&=&\rot{\bfa+\bfb\rightarrow\bfb}\left(\rot{\bfa\rightarrow\bfa+\bfb}(\bfA)\right)
\end{eqnarray}
If \bfa\ and \bfb\ are of equal length, $a=b$ then the two rotations are through equal angles.

If \bfa\ and \bfb\ are not of equal length, and we wish the steps to be equal, then we need to rescale. We could choose $\bfa'=\ahat$ and $\bfb'=\bhat$, but rather than using unit vectors, let us keep it somewhat more general and define \bfc\ as
\begin{eqnarray}
	\bfc = b\bfa + a\bfb
	\label{eq:diagc}
\end{eqnarray}
The rotation operators for \bfa\ to \bfc, and for \bfc\ to \bfb\ are equal, and are therefore both equal to the square root of the rotation operator for $\bfa\rightarrow\bfb$.
\begin{eqnarray}
	\rot{\bfa\rightarrow\bfc} &=& \rot{\bfc\rightarrow\bfb} \nonumber \\
	&=& \rot{\bfa\rightarrow\bfb}^{\frac{1}{2}} 
	\label{eq:sqrtrotab}
\end{eqnarray}
These rotations can be written in terms of their action on a vector \bfr\ as
\begin{eqnarray}
	\rot{\bfa\rightarrow\bfc}(\bfr) &=& \bfr\bfa\bfc/(\eta ac) \nonumber \\
	= \rot{\bfc\rightarrow\bfb}(\bfr)	&=& \bfr\bfc\bfb/(\eta bc) \nonumber \\
	= \rot{\bfa\rightarrow\bfb}^{\frac{1}{2}} (\bfr)	&=& \bfr(\bfa\bfb/(\eta ab))^{-\frac{1}{2}}  \nonumber \\
					&=& \bfr(\bfb\bfa/(\eta ab))^{\frac{1}{2}}  
	\label{eq:srotaccb}
\end{eqnarray}where the final equality comes from the result that $\rot{\bfa\rightarrow\bfb}=\rot{\bfb\rightarrow\bfa}^{-1}$.
 
The rotation of the general multi-vector \bfA\ is thus given by
\begin{eqnarray}
	\rot{\bfa\rightarrow\bfb}(\bfA)
	&=& (\bfa\bfb/(\eta ab))^{-\frac{1}{2}} \bfA (\bfa\bfb/(\eta ab))^{\frac{1}{2}}\nonumber \\
	&=& ((\bfc\bfa)/(\eta ac)) \bfA (\bfa\bfc/(\eta ac))\nonumber \\
	&=& \bfc\bfa  \bfA \bfa\bfc/(\bfa^2\bfc^2) 
	\label{eq:rotAacb}
\end{eqnarray}
The explicit appearance of the lengths of the various vectors in some of these equations suggest that square roots of  products, such as $\bfa^2$, will need to be taken. However the final result is  square root free, if the original vectors \bfa\ and \bfb\ are of equal length. The generalisation of this result to three or more dimensions is remarkably simpler than the rotation formulas found in standard texts. 

\subsection{Dot and wedge products of vectors}\label{sec:products}

As in freshman algebra, define the dot product, $\bfa\cdot\bfb$ as the symmetric part of the product, $\frac{1}{2}(\bfa\bfb+\bfb\bfa)$. The cross product is not easily defined in 2D, so instead we define a wedge product as the anti-symmetric part of of the product, $\frac{1}{2}(\bfa\bfb-\bfb\bfa)$. Thus  
\begin{eqnarray}
	\bfab = \bfa\cdot\bfb + \bfa\wedge\bfb
	\label{eq:dotwedge}
\end{eqnarray}
where
\begin{eqnarray}
	\bfa\cdot\bfb = \frac{1}{2}(\bfa\bfb+\bfb\bfa)
	\label{eq:2dot}
\end{eqnarray} 
and
\begin{eqnarray}
	\bfa\wedge\bfb = \frac{1}{2}(\bfa\bfb-\bfb\bfa)
	\label{eq:2wedge}
\end{eqnarray} 
Returning to the expression eq(\ref{eq:aincoords}) allows these to be written in coordinates
\begin{eqnarray}
	\bfa\cdot\bfb = \eta(a_x b_x + a_y b_y)
	\label{eq:2dotB}
\end{eqnarray} 
and
\begin{eqnarray}
	\bfa\wedge\bfb = (a_x b_y - a_y b_x)\khat
	\label{eq:2wedgeB}
\end{eqnarray}
Note that  $\bfa\cdot\bfb$ is a scalar (we sometimes say it is a zero-vector) while $\bfa\wedge\bfb$ is a pure bi-vector. Some authors use the term 2-vectors for our term bi-vectors, but this can cause confusion with the name for a vector in 2D space. 
Note that the square of any vector is a scalar, $\bfa^2 = \bfa.\bfa = \eta a^2$, it has no pure bi-vector part.

If the angle from \bfa\ to \bfb\ is $\theta_b-\theta_a$, then we may define  $\alpha\equiv\cos(\theta_b-\theta_a)$  and obtain 
\begin{eqnarray}
	\alpha  &\equiv& \cos(\theta_b-\theta_a)   \nonumber\\
	&=& \bfa\cdot\bfb/ab \nonumber\\
	 &=& (a_x b_x + a_y b_y)/(\eta ab)
\end{eqnarray} 
Likewise by defining 
\begin{eqnarray}
	\beta &\equiv& \sin(\theta_b-\theta_a)   \nonumber\\
	\beta \khat &=&  \bfa\wedge\bfb  \khat  \nonumber\\
	 &=& (a_x b_y - a_y b_x)/(-ab)\khat
\end{eqnarray}
This notation allows us to rewrite eq(\ref{eq:rotrab}) as 
\begin{eqnarray}
	\rot{\bfa\rightarrow\bfb}(\bfr) &=& \bfr' \nonumber \\
	&=& \bfr\exp(-\khat(\theta_b-\theta_a)) \nonumber \\
	&=& r_x'\xhat + r_y'\yhat \nonumber \\
	&=& \bfr(\alpha + \beta\khat) \nonumber \\ 
	&=& (\alpha - \beta\khat) \bfr \nonumber \\ 
	&=& (r_x\alpha - r_y\beta)\xhat + (r_x\beta + r_y\alpha)\yhat 
	\label{eq:trigrotrab}
\end{eqnarray}
or in matrix form, for coordinates written as rows with matrices on the right
\begin{eqnarray}
	(r_x',r_y') &=& (r_x, r_y)
		\left(\begin{array}[pos]{cc}
				\alpha &\beta\\ -\beta &\alpha	
		\end{array}\right) 
	\label{eq:rowmatrot}
\end{eqnarray}while for  coordinates written as columns and with matrices on the left, we have
\begin{eqnarray}
	\left(\begin{array}[pos]{c}
				r_x' \\ r_y'\end{array}\right) 
	&=& \left(\begin{array}[pos]{cc}
			\alpha& -\beta\\ \beta&\alpha	\end{array}\right)
			\left(\begin{array}[pos]{c}
				r_x \\ r_y\end{array}\right)
	\label{eq:colmatrot}
\end{eqnarray}
\subsection{Concluding remarks - What have we achieved?}\label{sec:conclusion}
 
Section \ref{AaAI2010} used homogeneity of 2D space to get the well known properties of vectors in 2D. The only non-standard claims were $(i)$ that because all physical measurements are finite with upper and lower limits \lmax\ and \lmin, not all the mathematical operations of the vector space have physical counterparts, and $(ii)$ that the physics of rigid bodies suggest that $\mathbb{Q}$, the field of rational numbers, is the appropriate field.

In this section we have deduced some old but less familiar consequences of the isotropy of space. Our study of movement, with one point fixed,  of rigid bodies in our 2D toy world of sheets of paper on a desktop, led us to many of the properties of rotations and to an algebra to describe them. 

The concept of a right angle rotation was introduced as a special case of rotations through a rational fraction, $r$ of $2\pi$ (or $360^\circ$). The   angles of $0, \pm\pi, \pm2\pi,\ldots,n\pi,$ are special in that lines $AB$ are rotated into themselves or to their negatives. The right angle rotations $\pm\frac{1}{2}\pi$ or $\pm\frac{3}{2}\pi$, etc., rotate orthogonal pairs of lines $AA'$ and $BB'$ into corresponding pairs $BB'$ and $A'A$, etc. This approach to defining orthogonality from isotropy considerations is not common, usually a metric is defined on the corresponding metric space first.

In our case we introduce the Euclidean metric after defining a product relationship on the lines of the physical space, and a corresponding product on the vectors of the previous section. By requiring the product to be associative, and to incorporate Pythagoras' identity, we obtain a choice of two algebras, each with four basis elements, the scalar, 1, the unit vectors \xhat\ and \yhat, and the less familiar \khat\ describing the plane. One algebra corresponds to the Euclidean metric $(+,\, +)$, and the other to the anti-Euclidean metric $(-,\, -)$. The next section \ref{AaAIII2010} proves that it is the latter metric that describes the geometry of our world.

The product introduced in this section was introduced by Clifford a long time \cite{clifford1878ags} ago in relation to the symmetries of Maxwell's equations, but has been used rarely by physicists. There are however some physicists and computer scientists who have used Clifford algebras, see Hestenes \cite{hestenes2003spg,hestenes1991dla,hestenes1987nfc,hestenes1966sta}, Gull \cite{gull1993inn}, Doran and Lasenby \cite{doran2003gap} and the conference proceedings of the Clifford Society \cite{icca7,icca8}. Our introduction of the Clifford product of two vectors \bfa\ and \bfb\ is by a more geometric route, but one that is rather less common \cite{joyce2002gaa}.  We introduced the bi-vector \bfab\ as representing, in a passive sense, the equivalence class of two sets of lines that lie in the 2D toy world of the plane that is our desktop.
These two sets of lines subtend a fixed angle, $\theta_{ab}$, between each other. The ``angle'' in the algebra is thus an abstraction of the angle between any of the lines in the equivalence class of the vectors \bfa\ and \bfb. Furthermore, the product of the lengths of the lines (or vectors), $ab= \mod{\bfa} \mod{\bfb}$, is fixed.
While all lines $AA'$ belonging to the vector equivalence class, $\bfa=[AA']$, have the same fixed length and direction this is not true for the bi-vector equivalence class. Instead $\bfab = [(AA', BB')]$ can be seen to represent the class of all parallelograms in the desktop, which have the same area and subtend the same angle. All parallelograms that are rotations and translations of the first parallelogram $(AB, CD)$ are in the equivalence class $\bfa\bfb=\left[(AB,CD)\right]$.

Any bi-vector \bfa\bfb\ can be written as a multi-vector with a scalar part, $\eta ab\cos\theta_{ab}$, and a pure bi-vector part, $ab\sin\theta_{ab}\khat$. Although $\khat^2=-1$, $\khat$ is not the complex number $i$. The usual formulation of rotations in a plane use complex numbers, giving the four basis elements $(\xhat,\yhat, i\xhat, i\yhat)$ to use. In the Clifford algebra formulation derived here, the product properties of the basis elements $(1,\xhat,\yhat,\khat)$ are very different to the complex number properties. 

In parallel to the above passive interpretations of bi-vectors $\mathbf{ab}$, the bivectors are, in an active sense, operators that rotate all elements  of the Clifford algebra, $C\ell(0, 2)$.  
Scalars and pure bi-vector elements are unchanged under rotations, while for vectors rotation has a very simple formula: $\rot{\bfa\rightarrow\bfb}(\bfr) = \bfr \bfa\bfb/\bfa^2$. The expression for the rotation of the general element \bfA\ of the algebra is not much more complicated, and is given by eq(\ref{eq:rotAacb}). 
In this active sense bi-vectors correspond to a class of line pairs, $(AB,AC)$ that rotate the points, lines, areas and indeed entire rigid bodies, relative to each other, about the point $A$, being point in common of the lines $AB$ and $AC$. Any of the vector and vector-product results of this section can therefore be written as purely geometric expressions acting on the points, lines and areas of rigid bodies by selecting representative lines and line products for the vectors and vector products.

\section{Motion in 3D and Parity gives $C\ell(0,3)$}\label{AaAIII2010}

The generalization to three spatial dimensions of the results of our 2D toy world of the previous two sections is straightforward. This is particularly true for extending homogeneity considerations of section \ref{AaAI2010}. 
The key result of is that a 2D vector space over the field of rational numbers, \bbQ, describes the homogeneity of the desktop world. The key underlying concept is the invariance of the size and shape of rigid bodies under movement in straight lines. subsection \ref{sec:3trans} will extend the ideas to a three dimensional vector space by considering translational motion of rigid 3D objects relative to each other.

The extension to 3D of the rotational ideas of section \ref{AaAII2010} follows in a similar manner in subsection \ref{sec:3rot}. The algebra has the additional basis vector arising from the translational motion, but there are two additional basis bi-vectors associated with rotations in each of two extra basis planes, and also a new object, a tri-vector that represents volumes. The algebra is thus eight dimensional. The three basis bi-vectors of rotation do not commute among themselves and give rise to the quaternion algebra. 

Subsection \ref{sec:metrics} shows that in the case of the $\eta = -1$ metric choice, there are four sets of basis elements in the algebra that behave as quaternionic sets and maintain a cyclic relationship. Since handedness is preserved for rigid bodies under the physically realized invariances of space, homogeneity and isotropy, we conclude that $C\ell(0,3)$ describes space. Further, we conclude that $C\ell(3,0)$ does not.

Many of the algebraic differences between $C\ell(0,3)$  and $C\ell(3,0)$ are highlighted in section \ref{AaAV2010} where we seek matrix representations of them over the reals, \bbR, or its subfield, the rational numbers \bbQ. 

We end this section in with a few words about transformations between reference frames and by showing that the Clifford algebra is a powerful tool for finding formulas for the rotation between different orientations of rigid bodies. The expressions obtained, unlike Euler angle formulations, do not use complex numbers.

\subsection{Translations in 3D space \label{sec:3trans}}

In section \ref{AaAI2010} we considered the 2D toy world of rigid objects consisting of a desktop and several sheets of transparent paper on it. In such a world we could move the objects relative to each other, by translational motion (sliding) the paper around in `straight line motion', to use the words of Newton's First Law. With transparent paper, any points on one object can be marked on the other objects, and after sliding, the distances between points can be compared directly. In this toy world we were led via the concepts of relative lengths of parallel lines, and via linear independence, to the mathematical concept of the basis vectors of a vector algebra.

In our 2D toy world, we could bring any two parallel lines to coincidence and compare lengths. Extending this process to the 3D world of an office raises an immediate problem. A rigid object, such as a book,  cannot be brought into coincidence with another rigid object. In the 2D world it is possible, in fact we have two ways of doing it. Several sheets of paper can lie on top of each other, as in our toy world we consider only the horizontal position, not the height above the desktop. The parameter `height' can be used to distinguish objects with the same position on the desktop. The second way is by using the parameter `time' to describe the different positions of the points (and lines and parallelograms) of a single sheet of paper on the desktop.

Consider now the example of several books in my office. We can translate the books parallel to the horizontal surfaces (e.g. the desktop, floor or ceiling), parallel to the north-south walls, and parallel to the east-west walls, or any linear combination thereof. What we cannot do is place two books in the same position in the room. We do not have a parameter equivalent to `height', only a time parameter. We will study the time parameter in section \ref{AaAIV2010}.

Another issue is that we cannot compare  the points, lines and surfaces inside one book with the corresponding points and lines of a second book. We need to restrict ourselves to comparisons of only some of the points, lines and surfaces on the surfaces of the books. To look inside we either need to take the rigid object apart, or use some form of remote measurement or remote sensing. 

However for many position measurements, the process for 3D rigid objects is little different from the process with 2D objects. Distances between points on the surfaces of rigid bodies can be measured by direct comparison with points on the surface of another rigid 3D body (using a measuring stick). As in the 2D toy world, all such measurements will be limited by the upper limit \lmax,  and  lower limit \lmin, associated with the relevant scales for measurements with the apparatus. 

Bearing in mind these restrictions however, it is clear that any line $AB$ can be written as a linear combination of three non-parallel lines, $OX, OY, OZ$.
Correspondingly, we need three basis vectors to describe the directions of the passive entities, the lines, and the active entities, the translations. Jumping ahead now to the conclusions of the next subsection, we can choose an origin $O$, and choose these basis lines $OX, OY, OZ$ to be orthogonal so that the corresponding basis vectors are an orthonormal set \xhat, \yhat, \zhat. We may write the vector \bfr\ in this basis as
\begin{eqnarray}
	\bfr = r_x \xhat + r_y\yhat + r_z\zhat
	\label{eq:rtoxyz}
\end{eqnarray}

\subsection{Rotations in 3D \label{sec:3rot}}

With the same provisos as in the subsection above regarding the somewhat indirect measurement process needed for the inside points of a 3D rigid body, the extension to rotational motion of a rigid body from the 2D toy world of section \ref{AaAII2010} to 3D follows simply.

\begin{figure}[htp]
  \begin{center}
    \subfigure[A book lying on our desktop.]{\label{fig:1a}\includegraphics[scale=0.42]{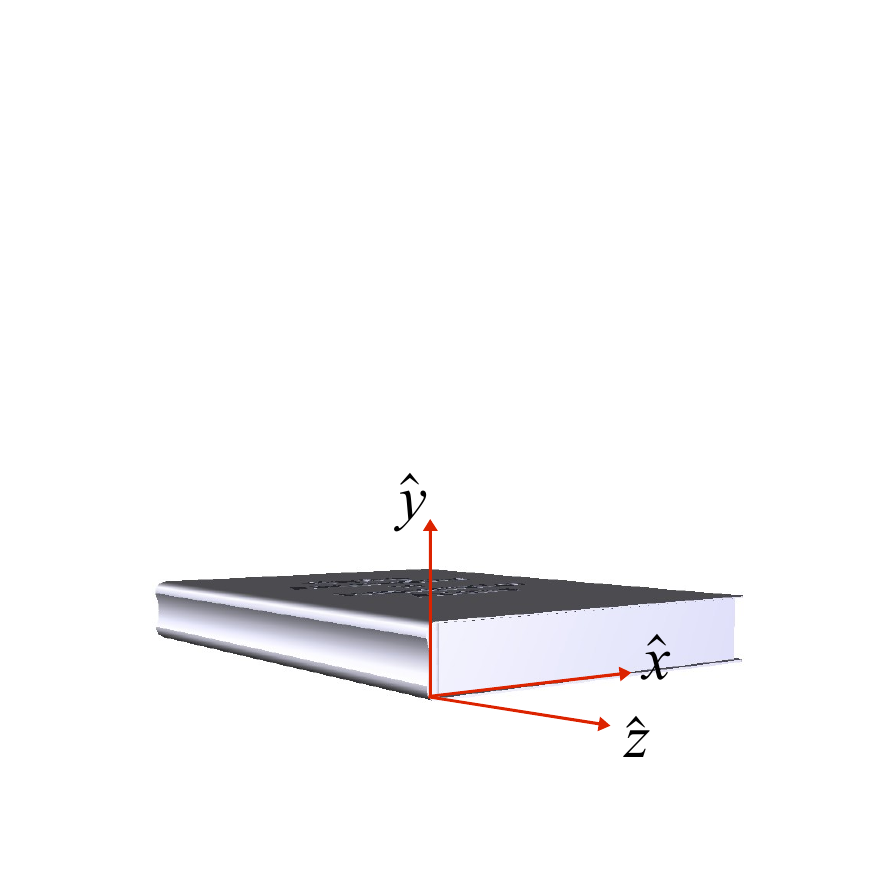}}
    \subfigure[The book after a rotation of $\pi /2$ in the \ihat-plane or about the \xhat-axis.]{\label{fig:1b}\includegraphics[scale=0.42]{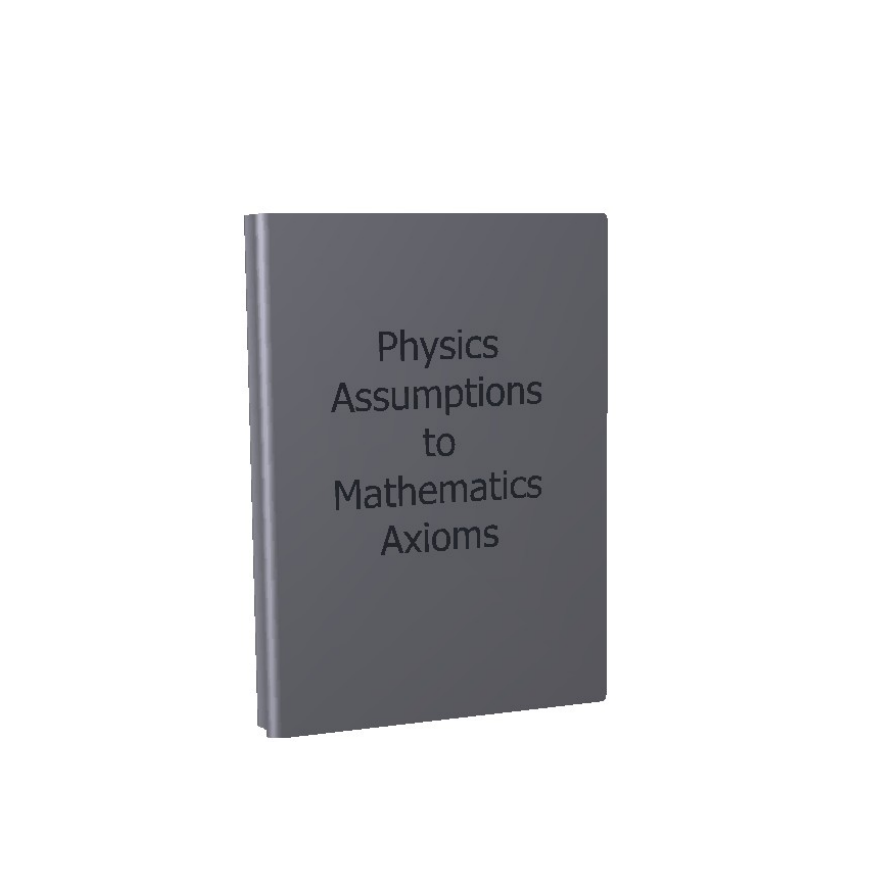}}
    \subfigure[The book after a rotation of $\pi /2$ in the \jhat-plane or about the \yhat-axis.]{\label{fig:1c}\includegraphics[scale=0.42]{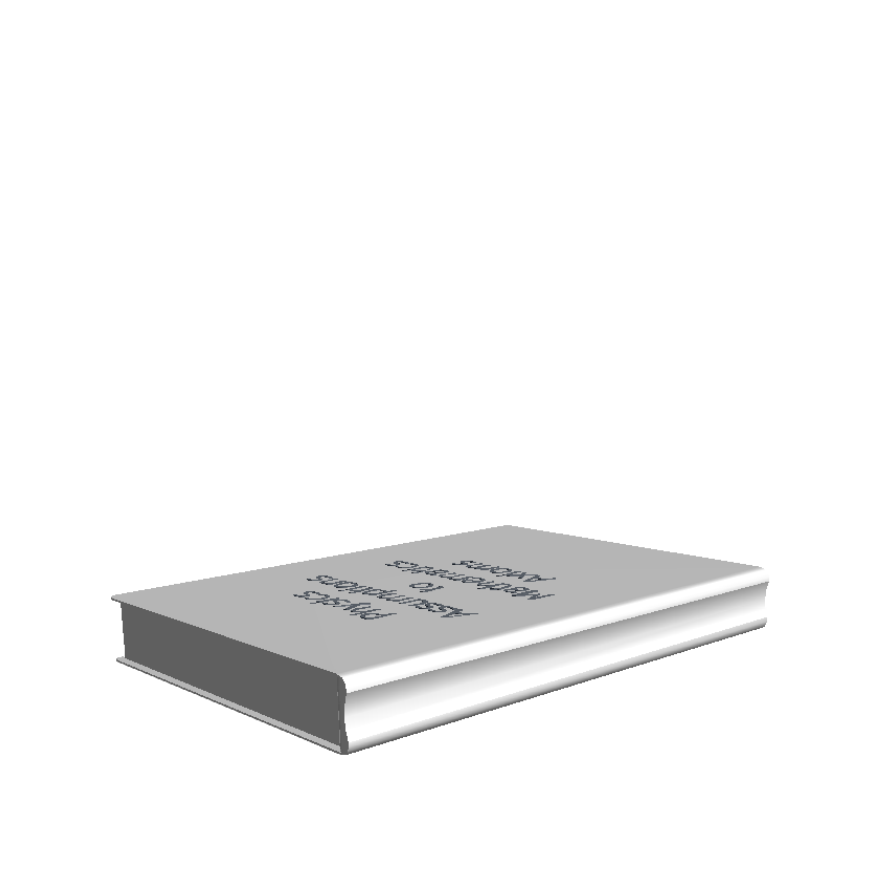}}
    \subfigure[The book after a rotation of $\pi/2$ in the \khat-plane or about the \zhat-axis.]{\label{fig:1d}\includegraphics[scale=0.42]{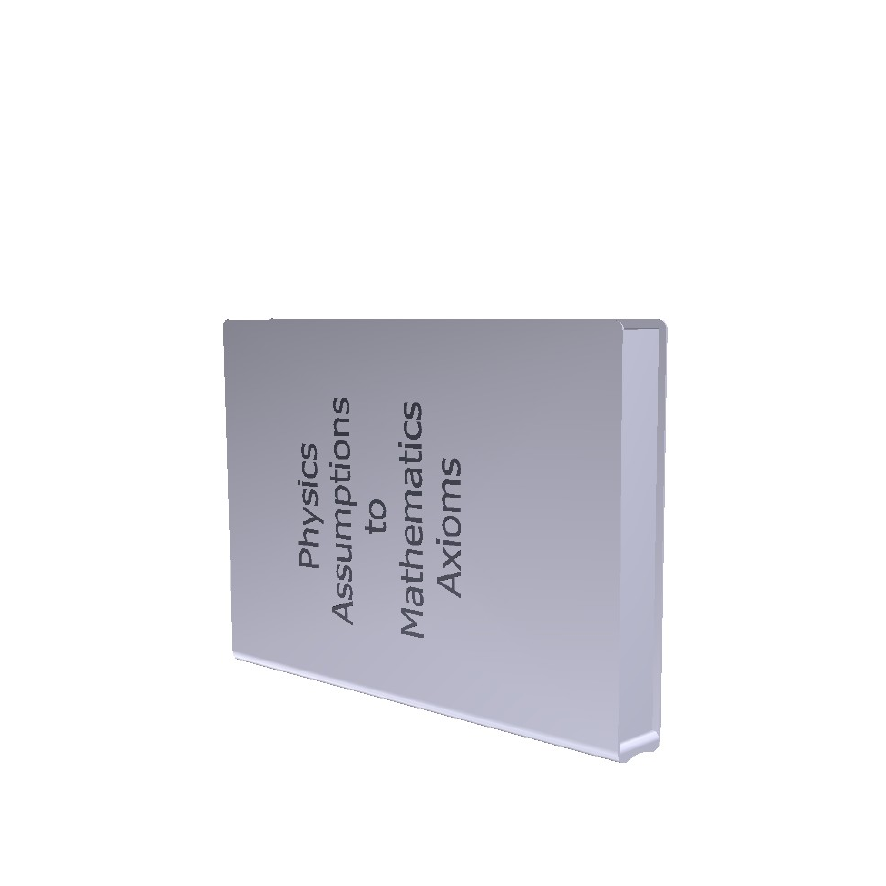}}
  \end{center}
  \caption{A book resting on the desktop, and rotations by $\pi/2$ about the \xhat, \yhat, and \zhat\ axes or equivalently the \ihat, \jhat, and \zhat\ planes respectively.}
  \label{fig:1}
\end{figure}

Consider rotating a book about the corner at the near, bottom, left when it is initially resting on the desktop, as shown in 
figure \ref{fig:1}. Rotations can take place about the \xhat, \yhat, and \zhat\ axes. 

We showed in section \ref{AaAII2010} that isotropy and Pythagoras, when applied to considerations of rotations in the plane of the desktop (the $xy$-plane) led to the requirement that $\xhat^2=\yhat^2=\eta$ and also that $\xhat\yhat=-\yhat\xhat=\khat$. Applying the argument to the two vertical planes ($yz$ and $zx$) leads to
\begin{eqnarray}
	\xhat^2  &=&	\ \ \yhat^2\,\  =\ \zhat^2\ =\ \eta \\
	\xhat\yhat &=&  -\yhat\xhat\ =\ \khat  \\
	\label{eqn4}\yhat\zhat &=&  -\zhat\yhat\ =\ \ihat  \\
	\zhat\xhat &=&  -\xhat\zhat\ =\ \jhat \\
		\label{eqn6}\xhat\yhat\zhat &=& \ \vhat 
	\label{eq:3ddefs}
\end{eqnarray}
where we define \ihat, \jhat\ and \vhat\ as in eqs(\ref{eqn4} to \ref{eqn6}). The rotations of figure 1 can thus be equivalently labeled as being in the \ihat, \jhat, and \khat\ planes respectively, as shown in the figure. The definitions are chosen retain the cyclic order of the basis vectors (\xhat, \yhat, \zhat) when defining the basis bi-vectors (\ihat, \jhat, \khat), and the basis tri-vector \vhat. The eight elements $\{1,\xhat, \yhat, \zhat,\ihat, \jhat, \khat,\vhat\}$ form the basis of the Clifford algebra $C\ell(0,3)$ when $\eta=-1$ and the Clifford algebra $C\ell(3,0)$ when $\eta=+1$.

The above results for (\xhat, \yhat, \zhat) and definitions for (\ihat, \jhat, \khat) lead to the properties 
\begin{eqnarray}
	\ihat^2 &=& (\yhat\zhat)(-\zhat\yhat)\ =\ -\eta^2=-1  \\
	\ihat\jhat &=& (\yhat\zhat)(\zhat\xhat)\ \ =\ -\eta\khat\\
	        &=&\khat \textrm{ if and only if } \eta = -1\\
	 \vhat^2 &=& -\eta
\end{eqnarray}  
Thus if we choose the anti-Euclidean metric, $\eta=-1$, the  basis bi-vectors satisfy
\begin{eqnarray}
	\ihat^2  &=&\	\jhat^2\, \ =\ \khat^2=-1 \label{eq:ii}\\
	\ihat\jhat &=&  -\jhat\ihat\ =\ \khat  \\
	\jhat\khat &=&  -\khat\jhat\ =\ \ihat  \\
	\khat\ihat &=&  -\ihat\khat\ =\ \jhat  \\
	\ihat\jhat\khat &=& -1
	\label{eq:ijkprops}
\end{eqnarray}
These equations are the relations that characterize Hamilton's quaternions \cite{hamilton1843qon}
\begin{eqnarray}
	i^2=j^2=k^2=ijk=-1
	\label{eq:quaternions}
\end{eqnarray}
as the relations of eqs(\ref{eq:ii} to \ref{eq:ijkprops}) can be readily derived from eqs(\ref{eq:quaternions}). 

Observe that there are four sets of  basis elements in $C\ell(0,3)$ that match the quaternion relations, the ordered set of bi-vectors (\ihat, \jhat, \khat), and the ordered sets (\xhat, \yhat, \khat), (\yhat, \zhat, \ihat) and (\zhat, \xhat, \jhat).
Six of the eight basis elements square to $-1$, namely \xhat, \yhat, \zhat, \ihat, \jhat, \khat, while $1$ and \vhat\ square to $+1$.

On the other hand for $C\ell(3,0)$ only one set of the basis elements matches the quaternion relations, namely the set of bi-vectors $(-\ihat, -\jhat, -\khat)$, where the  minus sign is required to retain the cyclic structure. The bi-vectors \ihat, \jhat, \khat\ and the tri-vector \vhat\ are the four of the eight that square to $-1$, while the other four, $1, \xhat, \yhat, \zhat,$ square to $+1$.

\subsection{Rotations do not commute\label{sec:commute}}

It is a fact that rotations in different planes are non-commutative. One of the reasons that the appropriate algebraic structure to describe rotations is an algebra, is that  products in an algebra are not necessarily commutative. In the example below we choose rotations of $\pi/2$ about the basis planes, as these are easiest to draw.

\begin{figure}[htp]
  \begin{center}
    \subfigure[A book lying on our desktop ready to be read]{\label{fig:2a}\includegraphics[scale=0.55]{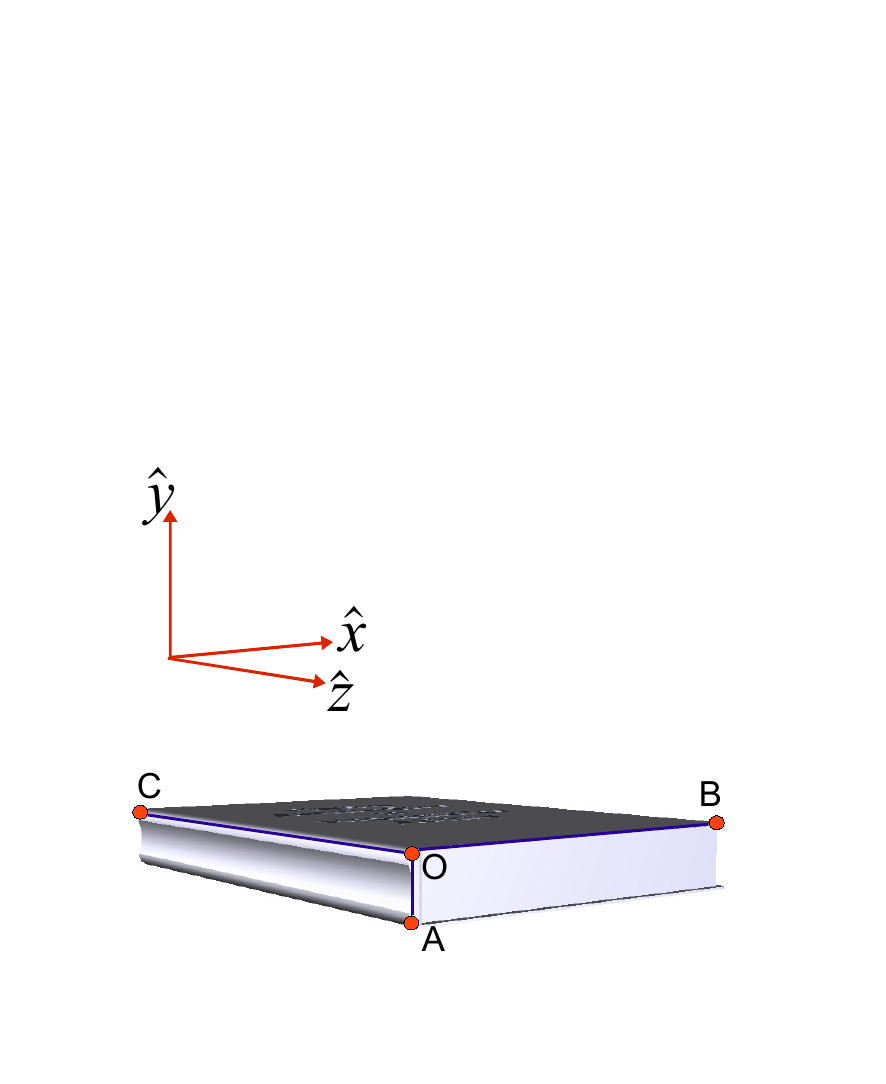}}
    \subfigure[The same book after a rotation of $\pi /2$ in the $xy$-plane]{\label{fig:2b}\includegraphics[scale=0.55]{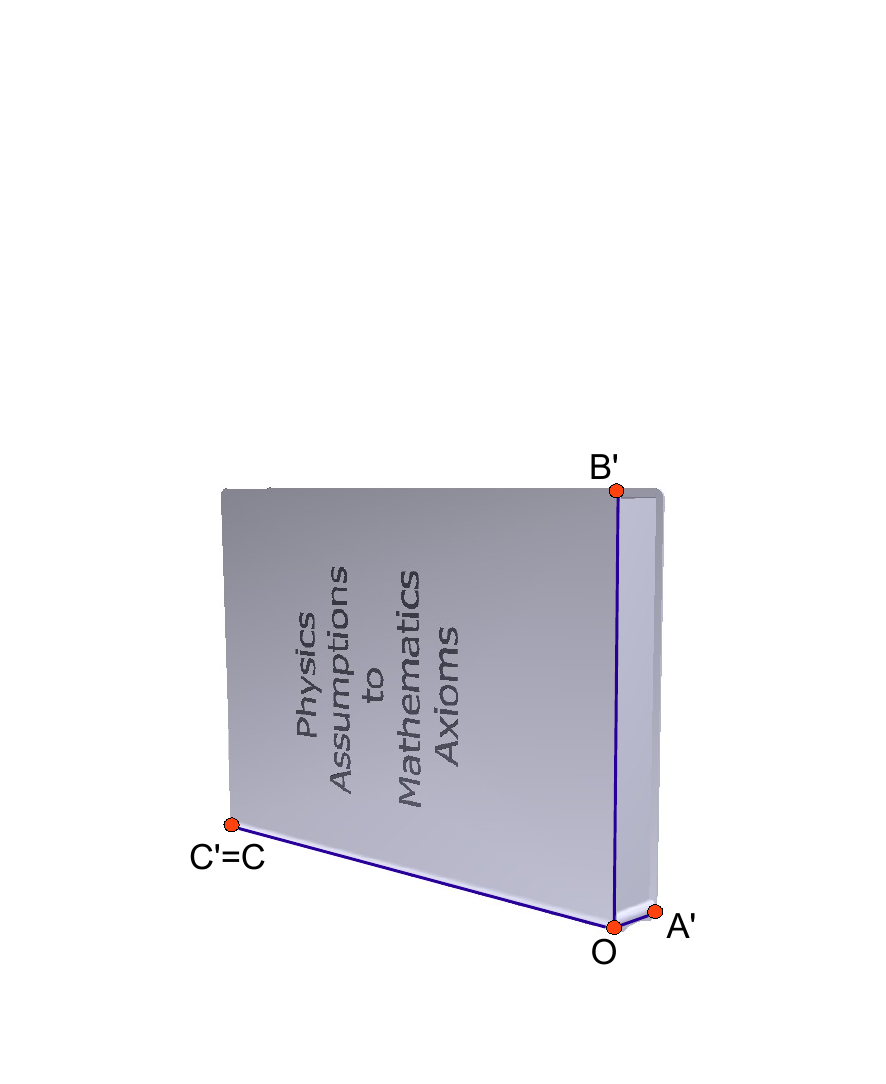}}
    \subfigure[The same book after a second rotation of $\pi /2$ in the $yz$-plane]{\label{fig:2c}\includegraphics[scale=0.47]{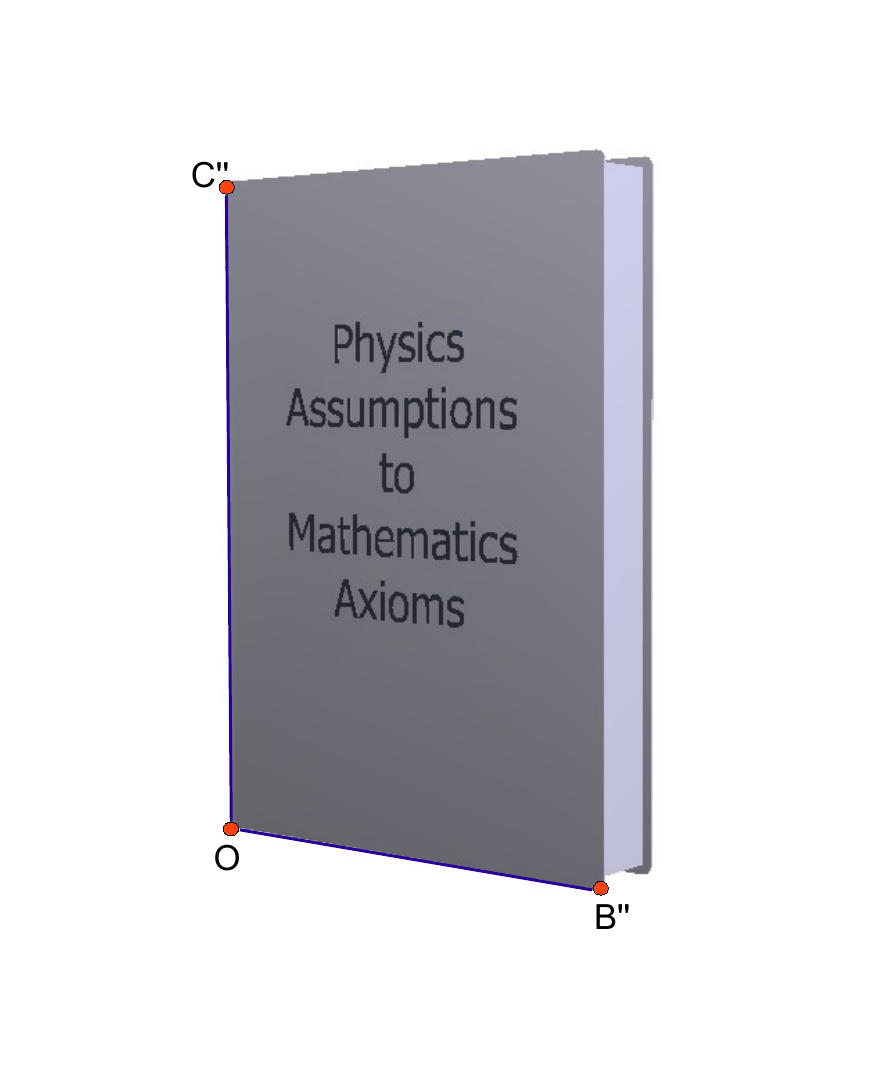}}
  \end{center}
  \caption{A book lying on our desktop, and rotated $\xhat\rightarrow\yhat$ then $\yhat\rightarrow\zhat$}
  \label{fig:2}
\end{figure}

Consider the example of the book of fig \ref{fig:2a} initially lying on the desktop ready to be opened and read. First rotate the book by $\pi/2$ in the $xy$-plane, that is the vertical plane, \khat. 

The line product $({OA}, {OB})$ rotates line $OA$ into line $OA'$ parallel to $OB$, and $OB$ into $OB'$ parallel to $-OA$. Line $OC$ is not moved, $OC=OC'$, see fig \ref{fig:2b}. The Clifford algebra operation that takes the arbitrary element of the book 
\begin{eqnarray}
	\bfA^{(0)} _{\textrm {book}} = a+b\xhat+c\yhat+d\zhat+e\ihat+f\jhat+g\khat+f\vhat
	\label{eq:Ain03}
\end{eqnarray}
through the angle $2\pi$ in the $xy$-plane, to its image $\bfA^{(1)}$, is
\begin{eqnarray} 
	\rot{\xhat\rightarrow\yhat}(\bfA^{(0)}_{\textrm {book}})
		&=& \rot{OA\rightarrow OB}(\bfA^{(0)}_{\textrm {book}})\nonumber \\
       &=&\rot{\textrm{by } \pi/2\textrm{ in plane \khat}} 
       (\bfA^{(0)}_{\textrm {book}})\nonumber \\
	 &=& \bfA^{(1)}_{\textrm {book}} \nonumber \\
       &=& (\xhat+\yhat)\xhat \bfA^{(0)}_{\textrm {book}} 
       \xhat(\xhat+\yhat) /2  \nonumber \\	
	&=& \Exp^{\skhat\pi/4} \bfA^{(0)} _{\textrm {book}} \Exp^{-\skhat\pi/4}
	\label{eq:book1}
\end{eqnarray}
This generalisation of the results of section \ref{AaAII2010} works for all parts of $\bfA^{(0)}_{\textrm {book}}$. For example, the proof for multi-vectors follows from linearity and by inserting the identity operator in the appropriate places, e.g. 
\begin{eqnarray}
	\rot{\pi/2, \khat} (\bfab) 
	&=& \Exp^{\skhat\pi/4} (\bfa\bfb) \Exp^{-\skhat\pi/4} \nonumber \\
	&=& (\Exp^{\skhat\pi/4} \bfa \Exp^{-\skhat\pi/4})( \Exp^{\skhat\pi/4}\bfb \Exp^{-\skhat\pi/4}) \nonumber \\
	&=& \left(\rot{\pi/2, \khat} (\bfa)\right) \left(\rot{\pi/2, \khat} (\bfb)\right)
\end{eqnarray}

Rotate  the book now by $\pi/2$ in the $yz$-plane, as shown in fig \ref{fig:2c}. 

The line product $({OA}, {OC})$ rotates the line $OB'$ into the line $OB''$ parallel to $-OC$, and $OC'$ into $OC''$ parallel to $-OA$. The line $OA'$ is not moved, $OA''=OA'$ which is parallel to $OB$, see fig \ref{fig:2c}.
\begin{eqnarray}
	\rot{\yhat\rightarrow\zhat}(\bfA^{(1)}_{\textrm {book}})
		&=& \rot{OB\rightarrow OC}(\bfA^{(1)}_{\textrm {book}})\nonumber \\
        &=& \rot{\textrm{by }\pi/2\textrm{ in plane }\ihat}
 (\bfA^{(1)}_{\textrm {book}}) \nonumber \\
       &=& \bfA^{(2)}_{\textrm {book}} \nonumber \\
	&=&  \Exp^{\sihat\pi/4} \bfA^{(1)}_{\textrm {book}} \Exp^{-\sihat\pi/4} \nonumber \\
	&=& \Exp^{\sihat\pi/4}  \Exp^{\skhat\pi/4} \bfA^{(0)}_{\textrm {book}} \Exp^{-\skhat\pi/4} \Exp^{-\sihat\pi/4} 
	\label{eq:book2}
\end{eqnarray}
The book is now standing on its lower edge with its front cover closest to us. The three lines $OA, OB, OC$ have moved to lines parallel to $OB, -OC, -OA$ respectively. 

Now do these two rotations in the reverse order, 
\begin{figure}[htp]
  \begin{center}
    \subfigure[A book lying on our desktop ready to be read]{\label{fig:3a}\includegraphics[scale=0.55]{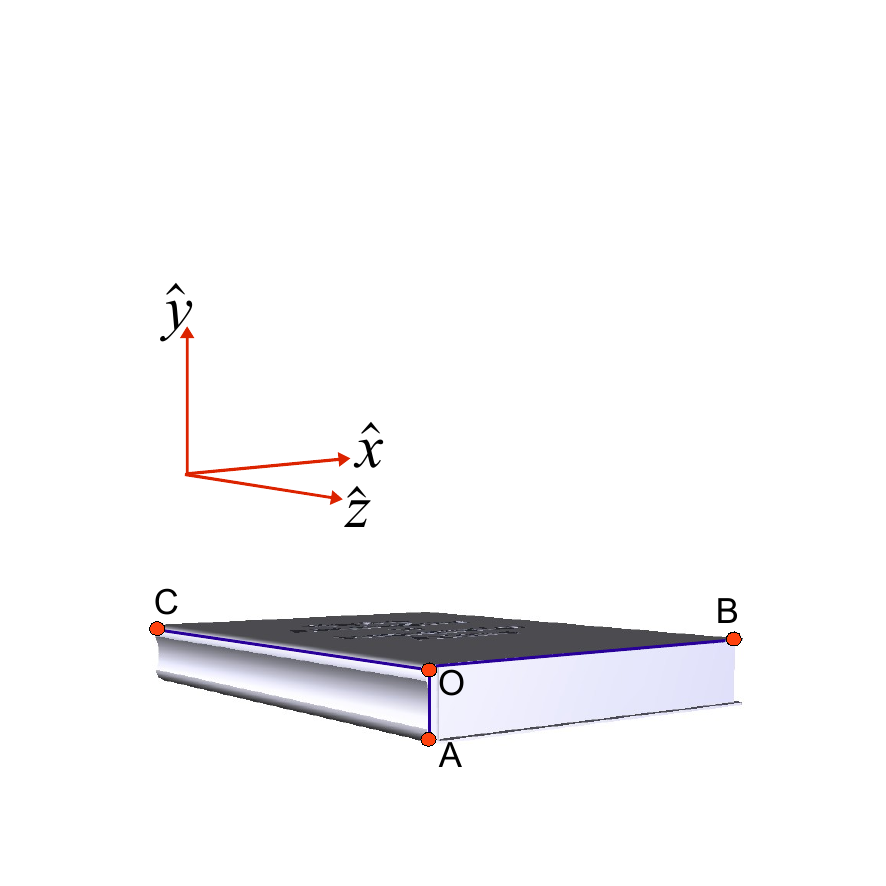}}
    \subfigure[The same book after a rotation of $\pi /2$ in the $yz$-plane]{\label{fig:3b}\includegraphics[scale=0.55]{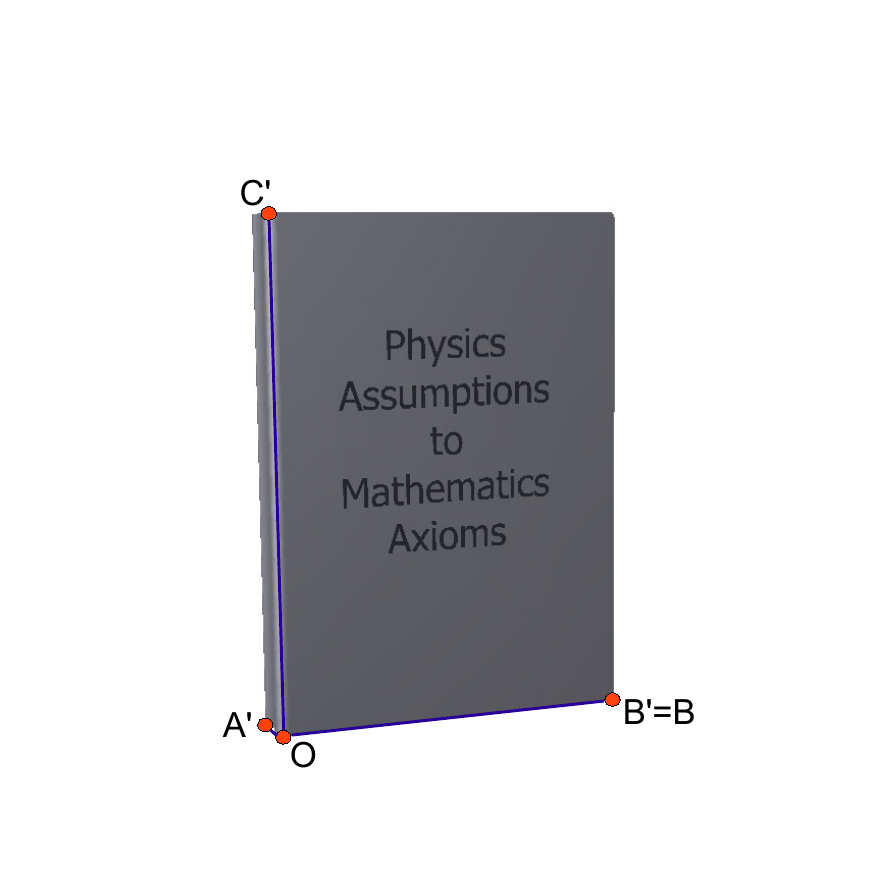}}
    \subfigure[The same book after a second rotation of $\pi /2$ in the $xy$-plane]{\label{fig:3c}\includegraphics[scale=0.55]{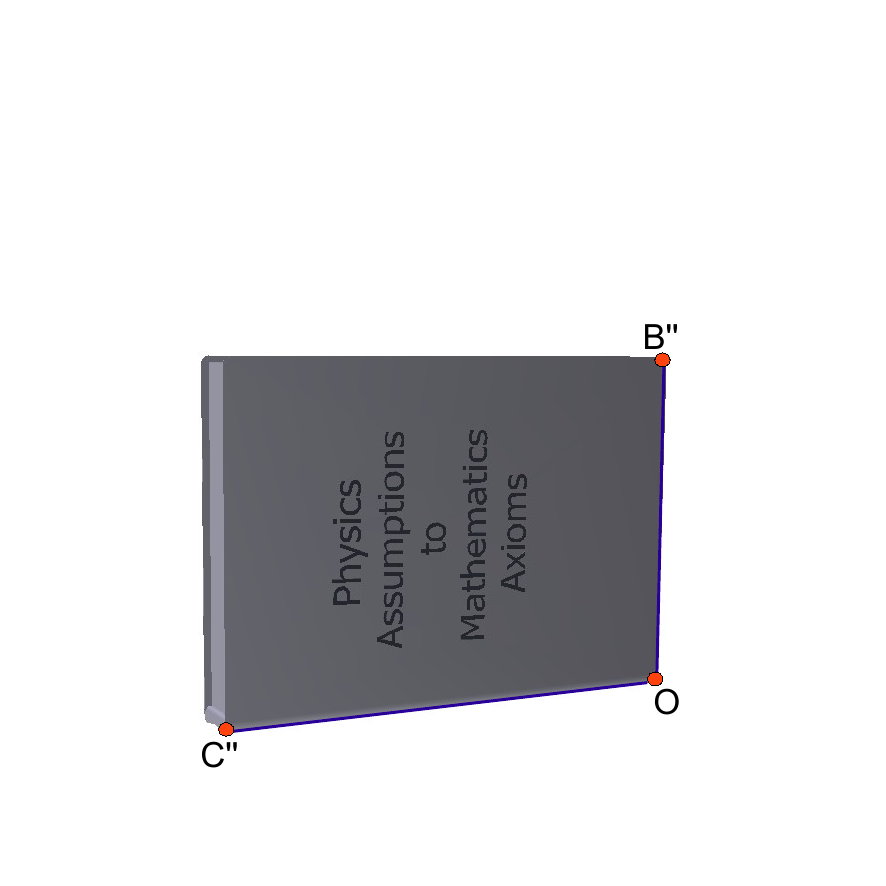}}
  \end{center}
  \caption{A book lying on our desktop, and rotated $\yhat\rightarrow\zhat$ then $\xhat\rightarrow\yhat$}
  \label{fig:3}
\end{figure}
first rotate by $\pi/2$ in the $yz$-plane, as shown in fig \ref{fig:3b}. The line product $({OA}, {OC})$ rotates the line $OA$ into line $OA'$ parallel to $OC$, and $OC$ into $OC'$ parallel to $-OA$. Line $OB$ is not moved, $OB'=OB$.
\begin{eqnarray}
	\rot{\yhat\rightarrow\zhat}(\bfA^{(0)}_{\textrm {book}})
		&=& \rot{OB\rightarrow OC}(\bfA^{(0)}_{\textrm {book}})\nonumber \\
        &=&	\rot{\textrm{by }\pi/2\textrm{ in plane }\ihat}
 (\bfA^{(0)}_{\textrm {book}})  \nonumber \\
       &=& \bfA^{(3)}_{\textrm {book}} \nonumber \\
	&=&  \Exp^{\sihat\pi/4} \bfA^{(0)}_{\textrm {book}} \Exp^{-\sihat\pi/4}
	\label{eq:book3}
\end{eqnarray}
Next rotate the book by $\pi/2$ in the $xy$-plane, as shown in fig \ref{fig:3c}. The line $OB'$ is rotated into line $OB''$ parallel to $-OA$, and $OC'$ into $OC''$ parallel to $-OB$ by the line product $({OA}, {OB})$. Line $OA'$ is not moved, $OA''=OA'$.

\begin{eqnarray}
	\rot{\textrm{by }\pi/2\textrm{ in plane }\khat} (\bfA^{(3)}_{\textrm {book}}) &=& \bfA^{(4)}_{\textrm {book}} \nonumber \\
	&=& \Exp^{\skhat\pi/4} \bfA^{(3)}_{\textrm {book}} \Exp^{-\skhat\pi/4}  \nonumber \\
	&=& \Exp^{\skhat\pi/4} \Exp^{\sihat\pi/4} \bfA^{(0)}_{\textrm {book}} \Exp^{-\sihat\pi/4} \Exp^{-\skhat\pi/4} 
	\label{eq:book4}
\end{eqnarray}
The book is now standing on its spine, front cover facing right. The three lines $OA, OB, OC$ have moved to lines parallel to $-OC, -OA, -OB$, and the two results differ by a rotation by $2\pi/3$ about the line $\xhat-\yhat+\zhat$. 

The exponential expressions for the rotation operator can be rewritten in terms of the vector products, for example for fig \ref{fig:2} we have
\begin{eqnarray}
		\rot{\pi/2, \ihat}\left(\rot{\pi/2, \khat} (\bfA^{(0)}_{\textrm {book}})\right)		 &=& \bfA^{(2)}_{\textrm {book}} \nonumber \\
	&=& \rot{\yhat\rightarrow\zhat}\left( \rot{\xhat\rightarrow\yhat} 
								\bfA^{(0)}_{\textrm {book}}\right) \nonumber \\
	&=& (\yhat+\zhat)\yhat(\xhat+\yhat)\xhat \bfA^{(0)}_{\textrm {book}} \xhat(\xhat+\yhat) \yhat(\yhat+\zhat) /4  \nonumber \\
	&=& (\eta-\ihat)(\eta-\khat) \bfA^{(0)}_{\textrm {book}} (\eta+\khat) (\eta+\ihat)/4 
	\label{eq:book4vec}
\end{eqnarray}
which equals eq(\ref{eq:book2}) when $\eta=-1$ because $\cos(\frac{\pi}{4}) = \sin(\frac{\pi}{4}) = \frac{1}{\sqrt{2}}$.

\subsection{Handedness conservation and choice of metric\label{sec:metrics}}

It is well known that the metric chosen for special relativity is subject to the choice of \Lm\ or \aLm, a choice known by some USA researchers as the East Coast versus West Coast choice --- merely a matter of taste! Certainly the choice seems governed more by the one used by nearby colleagues than physical principles. Indeed, for most situations the choice is irrelevant. One reason for this is that in most physical calculations are done in the complex number field, \bbC. The effect of the sign of the metric disappears. However complexifying the algebra changes the topology, and it is in the topology that we should look for the differences. Our concern with the effects of topology is one of the reasons not to assume the complex number field in our development.

One of us has argued earlier \cite{butler:aem} that to respect the cyclic properties of the triads (\xhat, \yhat, \zhat) and  (\ihat, \jhat, \khat) in $C\ell(0,3)$ we must have the metric \aEm\ to preserve the handedness in the observed world. The first part of arguments regarding these special cyclic properties of the basis vectors of $C\ell(0,3)$ were presented above, at the close of subsection \ref{sec:3rot}. To be explicit, physical operations in our physical 3D space retain their handedness, that is, they retain the cyclic structure in the ordering of axes and planes in rigid bodies. It is only in the Clifford algebra $C\ell(0,3)$, that the sets (\xhat, \yhat, \zhat) and  (\ihat, \jhat, \khat) have a cyclic structure that is respected by the operations of the algebra.
Some operations of $C\ell(3,0)$ take some cyclic orderings to their reverses.

Another explicit example of when the sign of the spatial metric distinguishes the two algebras, is the dual operation defined by left (or right) multiplication of the unit tri-vector \vhat, where \vhat\ was defined in eq(\ref{eq:3ddefs}). (In the usual mathematical study of algebras there is some interest in operations that, when applied twice, are equivalent to the identity operation.  Operators familiar to students in university courses in introductory mathematics include the inverse operation $A^{-1}$, matrix transposition and  complex conjugation.)

Use $A^*$ to denote the dual of an arbitrary element $A$ formed by right multiplication by \vhat, $A^*=A\vhat$. We call this dual the \vhat-dual, or pseudo-scalar-dual, or spatial dual. Using the linearity properties, we need only study the action $^*$ on the basis elements, and obtain
\begin{eqnarray}
	\{1^*, \xhat^*, \yhat^*, \zhat^*, \ihat^*, \jhat^*, \khat^*, \vhat^*\} 
	= \{\vhat, \eta\ihat, \eta\jhat, \eta\khat, -\xhat, -\yhat, -\zhat, 1\}
	\label{eq:3duals}
\end{eqnarray}
The scalar and tri-vector are dual to each other, and each vector is dual to the bi-vector (or plane) to which it is orthogonal. It is only for $\eta=-1$ that we have $(\bfA^*)^* = \bfA$ for this simple definition of a dual.

section \ref{AaAV2010} takes the study of the differences between $C\ell(0,3)$ and $C\ell(0,3)$ a step further by studying matrix representations of the algebras, and their corresponding groups.

\subsection{The conventional vector cross product\label{sec:cross}}

The familiar cross product $\bfa\times \bfb$ of the Heaviside-Gibbs 
algebra is the spatial dual of the anti-symmetric part of \bfab, $\bfa\wedge\bfb = \frac{1}2{}(\bfab-\bfb\bfa)$ 
\begin{eqnarray}
	\bfa\times \bfb &=&  (\bfa\wedge\bfb)^* \nonumber \\ 
				&=& -(\bfa\wedge\bfb)\vhat
	\label{eq:cross}
\end{eqnarray}
Writing  $\bfa\times \bfb$  out in components gives
\begin{eqnarray}
	\bfa\times \bfb &=&  (a_y b_z -a_z b_y)\xhat
			+ (a_z b_x -a_x b_z)\yhat + (a_x b_y -a_y b_x)\zhat 
\end{eqnarray}
where the $z$-componnent of the product is obtained from $a_x$ times $b_y$ and so on, cyclicly. For the wedge product, we get similarly
\begin{eqnarray}
	\bfa\wedge \bfb &=& ( (a_y b_z -a_z b_y)\ihat
			+ (a_z b_x -a_x b_z)\jhat  + (a_x b_y -a_y b_x)\khat 
\end{eqnarray}
so that for example the $i$ component of the wedge product is the $x$ component of the cross product
\begin{eqnarray}
	(\bfa\wedge\bfb)|_i &=& (\bfa\times \bfb)|_x = (a_y b_z -a_z b_y)
\end{eqnarray}


The key result from the above is that $\bfa\wedge\bfb$ is a pure bi-vector that represents an equivalence class of squares of the appropriate area in the plane in which \bfa\ and \bfb\ lie, whereas $\bfa\times \bfb$ is a vector that represents an equivalence class of lines of the appropriate length that are normal to that plane.

\subsection{The arbitrary rotation axis \label{sec:arbaxis}}

Subsection \ref{sec:3rot} and \ref{sec:commute} both used rotations in the plane of the given vectors.  The rotations were expressed either in the form $\rot{\bfa\rightarrow\bfa'}$ which is defined as the rotation in the $\bfa\bfa'$-plane by the angle between \bfa\ and $\bfa'$, $\theta_{a'a}$, or in the alternative form $\rot{\textrm{by }\theta_{a'a}\textrm { in the $\bfa\wedge\bfa'$-plane}}$. In 3D, the line $\bfa\times\bfa'$ is  perpendicular to the $\bfa\wedge\bfa'$-plane, and thus we may write it in the various forms
\begin{eqnarray}
	\rot{\bfa\rightarrow\bfa'} &=& 
	\rot{\textrm{by }\theta_{a'a}\textrm { in the $\bfa\wedge\bfa'$-plane}} \nonumber \\
&=&\rot{\textrm{by }\theta_{a'a}\textrm { about the line }\bfa\times\bfa'}
 \nonumber \\
&=&\rot{\textrm{by }\theta_{a'a}\textrm { about the line }(\bfa\wedge\bfa')^*}
\end{eqnarray}
the last expression follows from the duality between a line (or vector) and the plane (or bi-vector) normal to it. In the above we assume our standard convention of an anti-clockwise rotation from \bfa\ to $\bfa'$. The rotation $\rot{\bfa'\rightarrow\bfa}$ is the inverse to $\rot{\bfa\rightarrow\bfa'}$, but since all our rotations are anti-clockwise, the angle $\theta_{a'a} = 2\pi - \theta_{aa'}$. 

\begin{figure}[ht]
	\centering
		\includegraphics[width=0.70\textwidth]{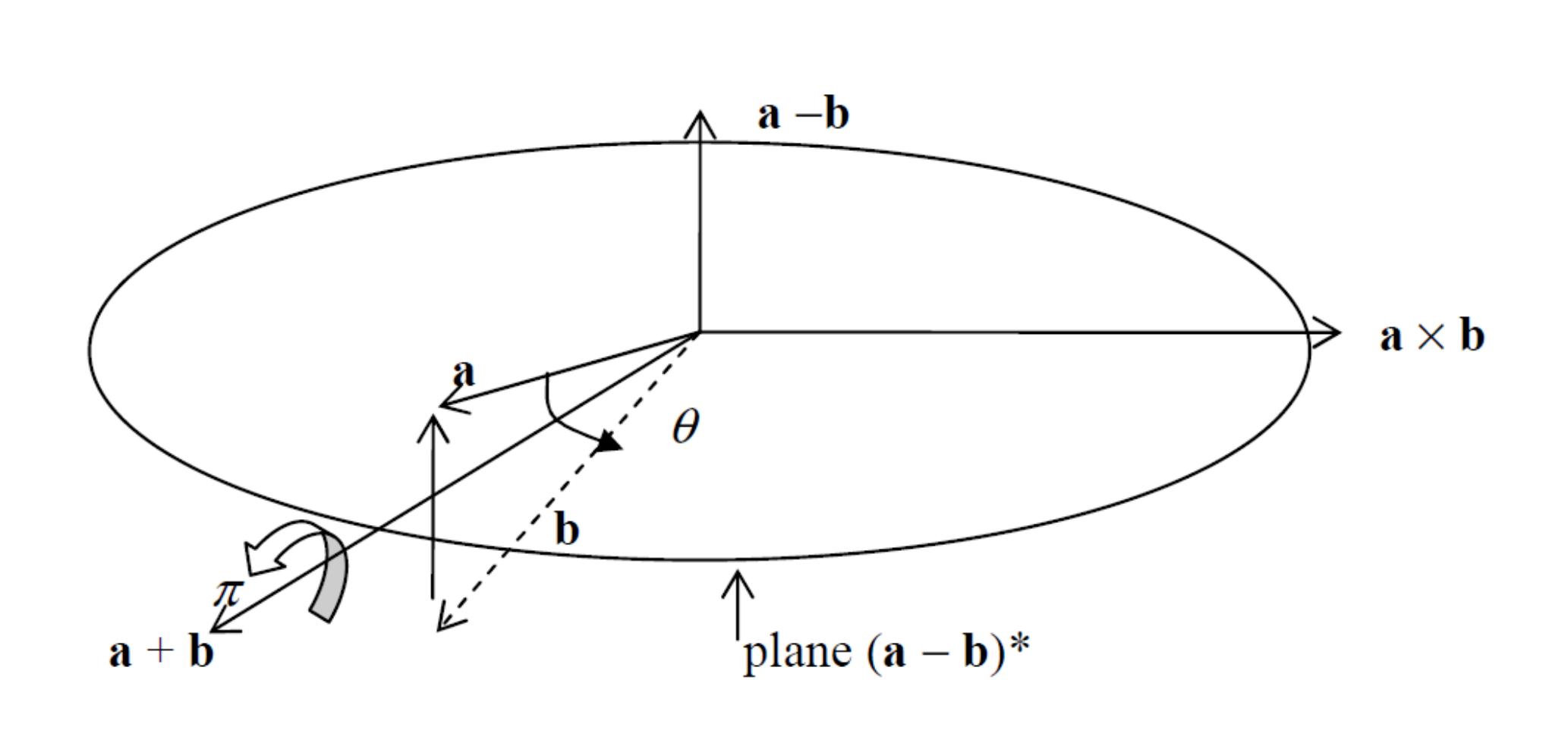}
	\caption{The plane formed from $\bfa\times\bfb$ and $\bfa+\bfb$ is perpendicular to $\bfa-\bfb$. The plane $\bfa\bfb$ is vertical in this diagram and the rotation of $\bfa$ into $\bfb$ in this plane is a rotation through angle $\theta_{ab}$ about the horizontal axis $\bfa\times\bfb$. The rotation through $\pi$ about axis $\bfa+\bfb$ also rotates \bfa\ into $\bfb$}\label{fig:rotaxes}%
\end{figure}

The minimum rotation angle $\theta_{aa'}$ is the smaller of $\theta_{aa'}$ and $\theta_{a'a}$. We have $\cos\theta_{aa'}=\bfa.\bfa'/\bfa^2$ (We assume here that \bfa\ and $\bfa'$ are the same length, $a=a'$) and choose to have $\sin \theta_{aa'}=\mod{\bfa\times\bfa'/\bfa^2}$, so that $0\leq\theta\leq\pi$.
However in 3D there are many rotations, not just $\rot{\bfa'\rightarrow\bfa}$ and $\rot{\bfa\rightarrow\bfa'}$, that take a vector \bfa\ into its image $\bfa'$.  
The rotation by $\pi$ about the vector $\bfa + \bfa'$ that lies halfway between \bfa\ and $\bfa'$ also suffices, as does the rotation by the appropriate angle about any axis lying in the $(\bfa\times\bfa', \bfa+\bfa')$-plane. This plane is perpendicular to the vector $\bfa-\bfa'$, see figure \ref{fig:rotaxes}, and so is the $(\bfa-\bfa')^*$-plane.

Any vector that is a linear combination of the above two vectors
\begin{eqnarray}
	\bfm=p(\bfa+\bfa')+q(\bfa\times\bfa') \ \textrm{ for any } p,q,\in \bbQ
\end{eqnarray}
may be used as the axis of rotation for $\bfa\rightarrow\bfa'$. The rotation angle is the angle between the projections of the vectors \bfa\ and $\bfa'$ onto the plane $\bfm^*$.
\begin{figure}[ht]
	\centering
		\includegraphics[width=0.70\textwidth]{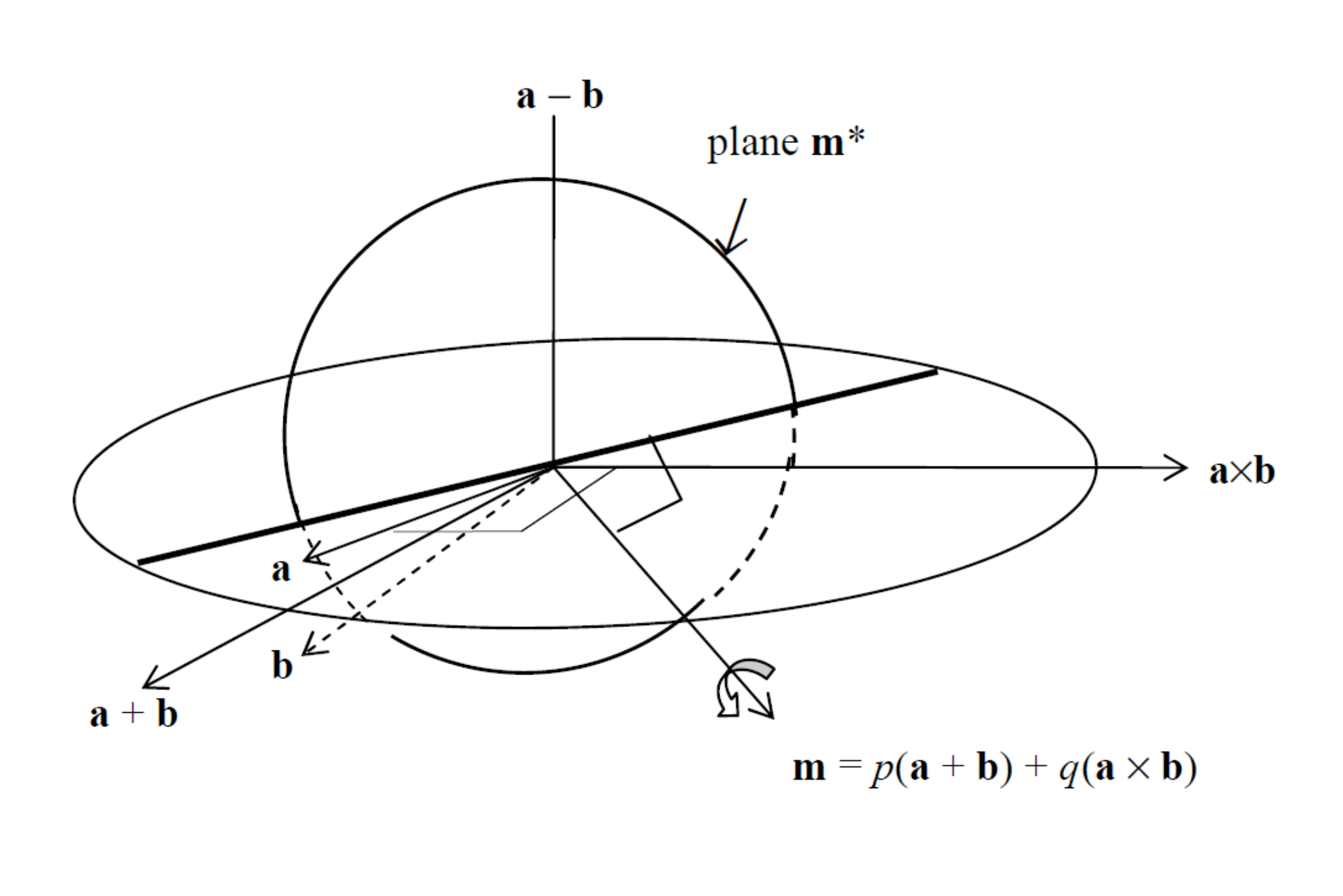}
	\caption{Rotating about the arbitrary axis \bfm\ in the $(\bfa+\bfb,\bfa\times\bfb)$-plane by the appropriate angle will take \bfa\ to $\bfb$. The angle needed is obtained by projecting \bfa\ and $\bfb$ onto the plane $\bfm^*$}
	\label{fig:rotaxis-m}%
\end{figure} 

\subsection{The general rotation of a rigid object \label{sec:3rotate}}

Let us now use the 3D Clifford algebra that we have derived in this section, to derive a simple closed formula for the rotation of a rigid object, from one known position to another.

\begin{figure}[ht]
	\centering
		\includegraphics[width=0.80\textwidth]{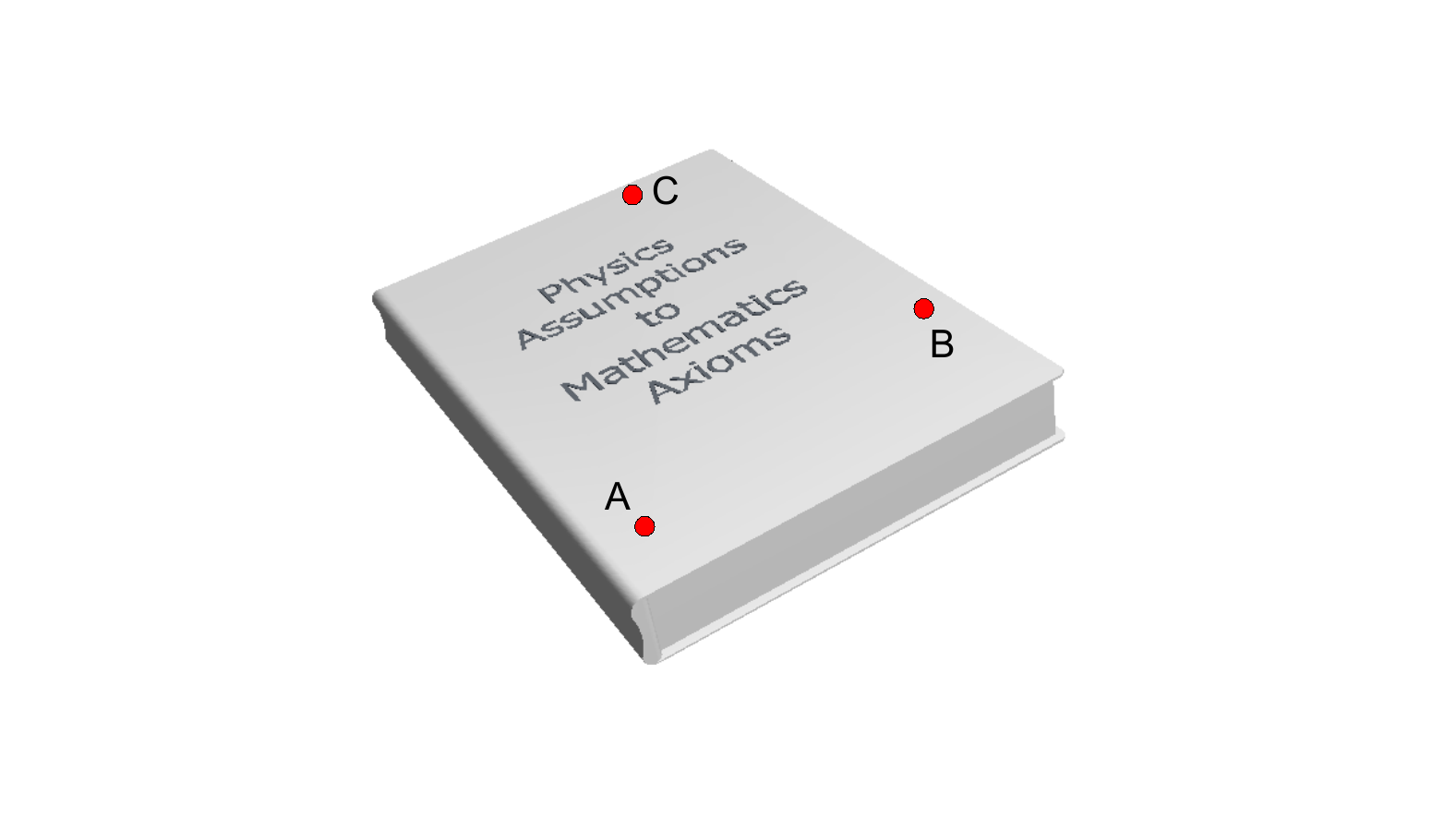}
	\caption{The position of a book is given by giving the location of any three points $A,B,C$ if $A,B,$ and $C$ are not co-linear.}
	\label{fig:Fig6}
\end{figure}

The location and orientation of a rigid body, e.g.\ a book, in 3-space is given by specifying the location of three non-collinear points, e.g.\ $A,B,C$ as in figure \ref{fig:Fig6}. 

As a preliminary, observe that the location of a rigid body can be uniquely specified by the location of precisely three non-collinear points of the body only because axis systems attached to rigid bodies retain their handedness under all those movements that are physically possible.

If the book is moved then the three points will move to new positions $A',B',C'$ relative to the coordinate frame of the observer. 
The problem we wish to find a general solution for, is given the initial points $A,B,C$ and the final points  $A',B',C'$, and that an arbitrary but known point $R$ has moved to $R'$, find $R'$.

\begin{figure}[ht]
	\centering
		\includegraphics[width=1.00\textwidth]{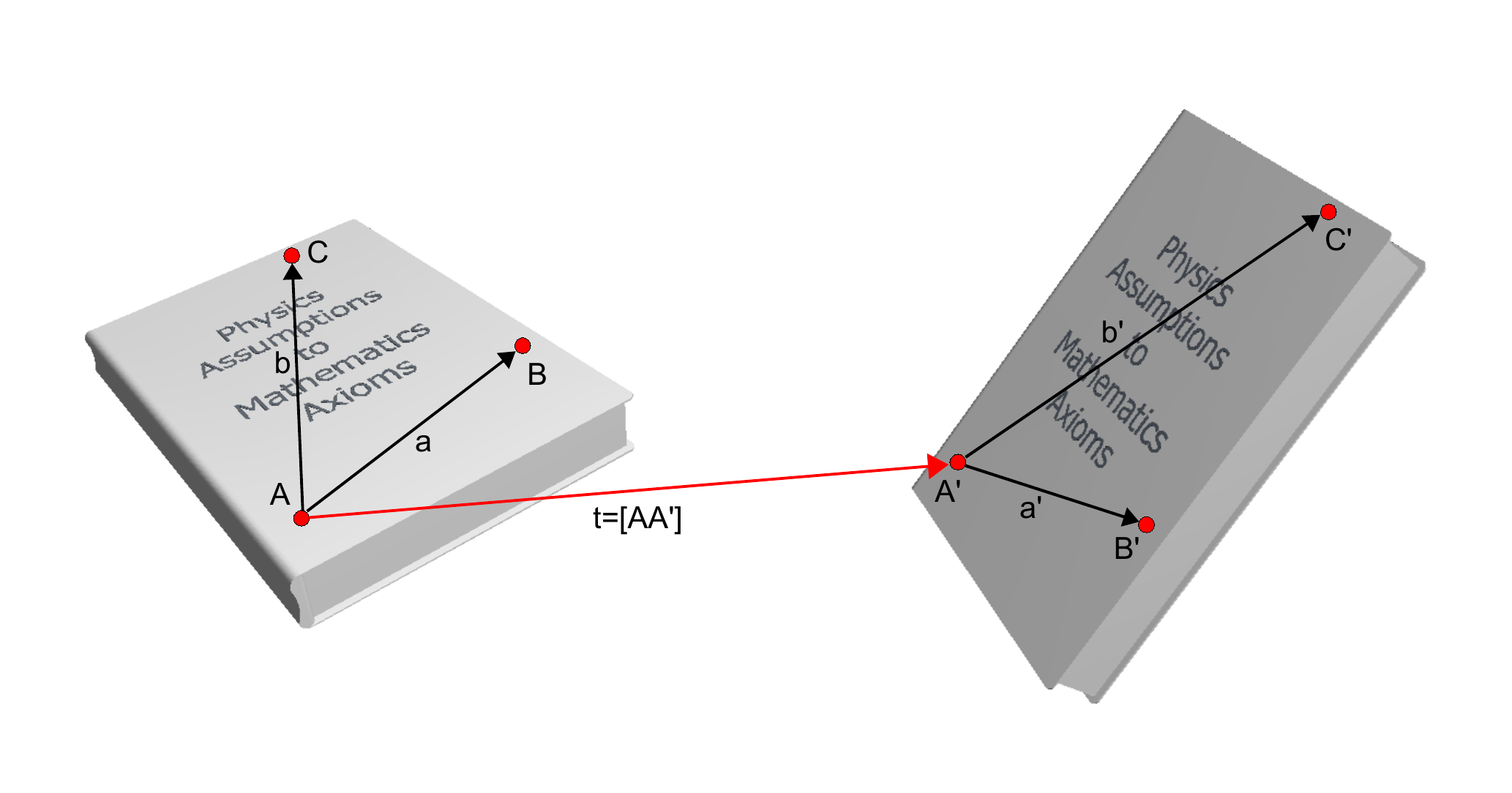}
	\caption{The movement of a book from points $A,B,C$ to $A',B',C'$ is described by the translation vector $\bft=[AA']$ and the initial vectors \bfa\ and \bfb, and final vectors $\bfa'$ and $\bfb'$. }
	\label{fig:Fig7}
\end{figure}

The motion can be described as a translation, followed by a rotation, see figure \ref{fig:Fig7}. It is by assumption a rigid body, so the lengths of the lines $AB, BC, CA$ remain unchanged: $\mod{A'B'}=\mod{AB}, \mod{B'C'}=\mod{BC}$ and $\mod{C'A'}=\mod{CA}$.  We may choose the translation to be specified by the active action of the line $AA'$, that is by the translation $AA'$, and seek to find the rotation about $A'$. Let us find this in terms of the elements of the Clifford algebra with the origin at $A'$. Let \bfa\ correspond to the line $AB$ (strictly \bfa\ is the equivalence class $[AB]$), \bfb\ correspond to $AC$, $\bfa'$ to $A'C'$ and $\bfa'$ to $A'C'$, as in figure \ref{fig:Fig7}. 

The rotation that simultaneously rotates \bfa\ into $\bfa'$ and \bfb\ into $\bfb'$  is about a line that is in both the plane $(\bfa+\bfa',\bfa\times\bfa')$ and the plane $(\bfb+\bfb',\bfb\times\bfb')$. The first plane is the plane $(\bfa-\bfa')^*$, that is the set of lines orthogonal to the line $(\bfa-\bfa')$. The second plane is  the set of lines orthogonal to the line $(\bfb-\bfb')$. In the general case the line \bfm\ that we need is therefore
\begin{eqnarray}
	\bfm= (\bfa-\bfa')\times(\bfb-\bfb')
\end{eqnarray}
and the angle may be found by projecting either \bfa\ and $\bfa'$, or \bfb\ and $\bfb'$ onto the plane $\bfm^*$ orthogonal to the line \bfm. Thus
\begin{eqnarray}
	\bfc=\bfa_{\perp \textrm{\textbf m}}= \bfa-\bfa.\bfm/\bfm^2
\label{eq:vect_c}
\end{eqnarray}
and
\begin{eqnarray}
	\bfc'=\bfa_{\perp \textrm{\textbf m}}= \bfa'-\bfa'.\bfm/\bfm^2
\label{eq:vect_cp}
\end{eqnarray}
are suitable vectors. Observe that $\bfc,\bfc'\in \bfm^*$, and $\bfc\times\bfc'$ is parallel to \bfm. In the special case that $(\bfa-\bfa')$ is parallel to $(\bfb-\bfb')$, then \bfm\ is zero. This corresponds to the rotations \bfa\ to $\bfa'$ and \bfb\ to $\bfb'$ being equal and we can choose $\bfc = \bfa$ and $\bfc'=\bfa'$. 

Thus the operator that rotates the rigid object so that points $A,B,C$ go to points $A'=A, B', C'$ is
\begin{eqnarray}
\rot{A,B,C\rightarrow A'=A, B',C'} &=& \rot{\bfa\rightarrow\bfa', \bfb\rightarrow\bfb'} \nonumber \\
	&=& \rot{\bfc\rightarrow\bfc' \textrm{ in the cc}'\textrm{-plane}} \nonumber \\
	&=& \rot{\bfc\rightarrow\bfc' \textrm{ about the \bfm\ axis}} 
\label{eq:generalrotation}
\end{eqnarray}
where \bfc\ and $\bfc'$ are given by eqs(\ref{eq:vect_c}-\ref{eq:vect_cp}).

Combining all these results gives the final expression for the rotation of a general 3D multi-vector \bfA\ associated with the rigid object as 
\begin{eqnarray}
	\rot{A,B,C\rightarrow A'=A, B', C'}(\bfA)
       &=&\rot{\bfc\rightarrow \bfc'\ \textrm{about\ }\bfm }(\bfA)\\
	&=& \bfd\bfc  \bfA \bfc\bfd/(\bfc^2\bfd^2) 
	\label{eq:rotA3D}
\end{eqnarray}
where \bfd, the bisection of $\bfc$ and $\bfc'$ is given by 
\begin{eqnarray}
	\bfd 	&=& c'\bfc + c\bfc' 
\end{eqnarray}

\subsection{Reference frames\label{sec:frames}}

In earlier sections, and in this section, we showed how to set up an orthonormal coordinate system for each rigid body. Using this coordinate system we can then measure the location (the position and orientation) of other rigid bodies relative to that coordinate system. The operations of the Clifford algebra  $C\ell(0,3)$ gave the mathematical transformation for the location of an object measured relative to one rigid body (one coordinate system) to the location of that object measured in any other coordinate system.

To personalise this in the usual way, the observations of various observers are related to one another by making the appropriate adjustments (the Galilean transformations) to the positions and orientations of the observers' coordinate systems. Using the term `frame of reference': the measurements of one observer, $S_1$, of the locations of the points, lines, planes and volumes of other rigid bodies, using that observer's frame of reference, $(\xhat_1, \yhat_1, \zhat_1)$, may be transformed using the operations of the Clifford algebra  $C\ell(0,3)$ to the locations of the points, etc., as measured by other observers, $S_2, S_3, \ldots$ using their various frames of reference, $(\xhat_i, \yhat_i, \zhat_i)$ where $i=2,3,\dots$. Some of those measurements are the locations of the frames relative to one another. One set of these measurements consists of the vectors $\bfa_{12}, \bfb_{12}$ and $\bfc_{12}$, being the position of three points $A,B,C$ (for example the origin $O_2$, and the ends of the unit lines $O_2X_2$ and $O_2Y_2$) that describe the position of the origin and the orientation of frame $S_2$ relative to the frame $S_1$. The position of the origins of two frames are related via a translation. Their orientations are related via a rotation
\begin{eqnarray}
	\trans{S_2}{S_1}O_2&=&O_1,\\
	\rot{S_2\rightarrow S_1}(\xhat_2,\yhat_2,\zhat_2)&=&(\xhat_1,\yhat_1,\zhat_1)
\end{eqnarray}

The transformation of the location of (say) rigid body $S_3$ measured in frame $S_2$ as $\bfA_{32}$, to frame $S_1$ (measured as $\bfA_{31}$ is thus first a translation of the origin of frame $S_2$ to frame $S_1$, and then a rotation of the form given by eq(\ref{eq:rotA3D}) of the previous subsection. 
\begin{eqnarray}
	\rot{S_2\rightarrow S_1}\trans{S_2}{S_1}\bfA_{32} &=& \bfA_{31} 
\label{eq:changeframe}
\end{eqnarray}

Of particular interest, and fundamental importance, is the reciprocity of this relationship. It follows from the vector space (homogeneity) properties of translations, and the Pythagorean metric (which includes the isotropy of space).
The position of the origin of frame $S_1$ in the frame $S_2$, and the orientation of frame $S_1$ in the frame $S_2$ are the inverses of the above
\begin{eqnarray}
	\trans{S_2}{S_1} &=& \trans{S_1}{S_2}^{-1}  \\
	\rot{S_2\rightarrow S_1} &=& \rot{S_1\rightarrow S_2}^{-1}
\label{eq:changeobserver}
\end{eqnarray}
and consequently the inverse of equation (\ref{eq:changeframe}) is given by
\begin{eqnarray}
\rot{S_2\rightarrow S_1}\trans{S_2}{S_1}\bfA_{32}&=&\left[\rot{S_2\rightarrow S_1}\trans{S_2}{S_1}\right]^{-1}\bfA_{31}\\
&=&\trans{S_2}{S_1}^{-1}\rot{S_2\rightarrow S_1}^{-1}\bfA_{31}
\end{eqnarray}

Thus we see that the Clifford algebra $C\ell(0,3)$ enables the measurements of observer $S_2$ to be transformed to those of observer $S_1$, and conversely.

\subsection{Concluding remarks - What have we achieved?}\label{sec:conclusion}
 
This section has extended the consequences of homogeneity and isotropy from a 2D world to a 3D world. Homogeneity and the properties of translations of rigid objects led to a rather simple extension of the addition properties of lines and the corresponding vector space properties. The isotropy of 3D space, and the rotation properties of rigid objects led to a richer set of properties, properties that are described by the Clifford algebra $C\ell(0,3)$.

As an example of the power of a coordinate-free formulation of the Clifford algebra, we obtained the operator that rotates a rigid object from a known position (specified by the location of three points) to a second position (specified by the new location of these three points). It may be that this result has not been previously found, certainly the authors have not found such an expression in the literature.

We demonstrated that the maintenance of cyclic structures of sets of basis lines and sets of basis planes, namely the parity conservation properties of allowable physical movements of rigid bodies, requires the use of the Clifford algebra $C\ell(0,3)$ and not the Clifford algebra $C\ell(3, 0)$.

Parity conservation in physical 3D space would seem to be a property of homogeneity and isotropy in $n$D space. In 1D, rigid objects can be modelled as beads on a wire or trains on tracks, and while they can be moved (translated) backwards and forwards, they cannot be turned around. To do so would require turning the object over, using a second dimension of physical space.

In 2D, we modelled rigid objects as sheets of section on a desktop. The objects can be translated in two orthogonal directions, represented by the unit vectors \xhat\ and \yhat, and the expressions $-\xhat$ and $-\yhat$ make operational sense as independent translations. But a rigid 2D object cannot be rotated so that only one of a pair of orthogonal lines (say $OA$ and $OB$) becomes its negative, if we have $-OA=AO$ then $-OB=BO$ also. To have only one would require turning the paper over, using the third dimension of physical space. In our 3D space of rigid bodies, there is no evidence of another spatial dimension, and no evidence that parity of rigid objects is not an absolute conservation law. Note: parity does not seem to be conserved in some experiments involving the weak force, however the weak force does not act on rigid bodies. The time parameter adds another dimension to the mathematical description of our world, but it differs from the spatial dimensions in many ways, as we study in the next section.

Hamilton\cite{hamilton1843qon,hamilton1853lq} spent many years seeking a generalization of the algebra of complex numbers that seemed, via the Argand diagram, to give a good mathematical description of the geometry of the plane. Complex numbers gave the mathematics describing translations in 2D, as addition of pairs of numbers for coordinates in the $x$ and $y$ directions. Complex numbers describe rotation by multiplications. However the structure of 3-complexes that he sought does not exist, but he did
find the necessary four--dimensional generalization, which he called the quaternions. 

The story of Hamilton recognizing what was needed is part of the oft quoted folklore of mathematical discovery. He reports that it came to him ``in a flash'' while walking with his wife along a Dublin canal on a Sunday afternoon. The generalization for 3D of the ``ordered pair'' or 2-complex needed for 2D geometry, was to an ``ordered 4-tuple'' or ``quaternion''. The quaternion components are the basis of Hamilton's non-commutative algebra. Hamilton's quaternion algebra was the first formal non-commutative algebra, and represents the non-commutativity of rotations in 3D.

Hamilton was en route to developing the appropriate algebra for describing the geometry of space. In his lectures to the Dublin Royal Society in 1853\cite{hamilton1853lq} Hamilton carefully distinguished between polar vectors (which he called lines) and axial vectors (which he called versors). Polar vectors describe the positions of the points of objects, and also describe translations. Axial vectors describe the orientations of (non-point) objects and also describe rotations. He then proceeded (page 71) to write {\it  both} polar vectors and axial vectors in his axial basis which he labeled by the letters ${\it i, j, k}$. Unfortunately for the development of the subject he fails to maintain this distinction, and writes
\begin{quote}
  And I conceive that we may {\it now} legitimately, and with advantage, avail ourselves of the same analogy, or of the theorem to which it
  corresponds, to {\it dispense} with that {\it symbolic distinction} which has been above observed, between the three quadrantal {\it
    versors} {\it i, j, k}, and the three {\it lines} i, j, k, which have respectively the directions of their three axes.
  [Emphasis as in the original.]
\end{quote}
We presume he made this identification so as to keep his algebra small, had he retained the distinction he would probably been led to the conclusions of Clifford\cite{clifford1878ags}. The appropriate algebra has the three lines i, j, k, which are now called the basis polar vectors and we write as \xhat, \yhat, \zhat. It also has the three versors {\it i, j, k}, which are now called the basis axial vectors and we write as \ihat, \jhat, and \khat. The complete algebra closes with the addition of two more basis elements, the scalar, 1, and an element we represent as \vhat\ which relates to a basis volume element. Thus to describe 3D geometry accurately we need to use the eight dimensional algebra which we label $C\ell(0,3)$.

Hamilton proved that the axial vectors ${\it i, j}$ and ${\it k}$ square to $-1$, and that they form his famous quaternion algebra
$$ i^2=j^2=k^2=ijk=-1$$
However the incorrect identification between the versors ${\it i, j, k}$ and the lines i, j, k continues in the labeling, by many physics texts, of the polar vector basis elements as \ihat, \jhat, and \khat\  where $\ihat\times\jhat=\khat$.  This product is correct for axial vectors but not for polar vectors. This unfortunate identification by Hamilton has to be patched up by ignoring the distinction of polar and axial vectors, or equivalently by identifying lines with planes (or translations with rotations). Put simply, in most approaches since Hamilton a plane is identified with the line that is normal to it.

This conflation of lines (or unit vectors) $\xhat, \yhat, \zhat$ and planes (or unit rotators) $\ihat, \jhat, \khat$ continues to this day. Simon Altmann\cite{altmann1986rqa} gives the fullest history of this mess that we are aware of.
We recommend that readers who wish to pursue some of the history of the
geometric product and Clifford algebras refer to the review by Altmann\cite{altmann1986rqa}, and that they also read the lectures by
Hamilton \cite{hamilton1853lq}.

A further consequence of the polar--axial identification is that the vector algebra is too small to describe the geometry and the physics contained in that geometry. This evidences itself in many ways. The first one we have discussed above --- there is the need in the usual 3D algebra to use complex numbers, effectively a six dimensional space, to describe rotations. The eight dimensional Clifford algebra contains all we need without complex numbers. Second is the so-called proof that quantum mechanics needs complex numbers. In a future paper we plan to review the argument as presented by Sakurai\cite{sakurai1995mqm}, to conclude that yes, you do need more than a three dimensional algebra, but no, complex numbers are not needed if both polar and axial vectors are used. Our argument is, in brief, that the complex number basis $\mathbb{C}^3$ $\xhat, i\xhat, \yhat, i\yhat, \zhat, i\zhat$ can be used for some of the geometry of 3D, if used with care, but the basis $\xhat, \ihat, \yhat, \jhat, \zhat, \khat$ is a better description of the geometry of the physical 3D world. 

\section{Time and the Speed of Light, the Algebra of Spacetime}\label{AaAIV2010}

If the assumptions and axioms up till now are accepted, then their extension for spacetime seems trivial. It was easy to extend homogeneity (section  \ref{AaAI2010}) and isotropy (section \ref{AaAII2010}) from 2D to 3D (section \ref{AaAIII2010}). However, time is different from space in many ways, Our preceding analysis was based on the translational and rotational invariance of rigid bodies. Clocks are not rigid bodies, although most definitions of a clock rely in part on rigid bodies -- the swing of a pendulum, the oscillation of a crystal, the bouncing of light between mirrors held a fixed distance apart.

The translations and rotational properties of rigid bodies, and reference frames defined by rigid bodies, led us to the conclusion that the Clifford algebra $C\ell(0,3)$ over the field of rational numbers is the mathematical structure to transform from one inertial reference frame to another, the measurements of position and orientation of rigid bodies. It may seem that extending the arguments to spacetime would be as trivial as the extension from 2D to 3D space. Rather, the authors have found this section the most difficult, both to understand what we wish to write and also to write it clearly. The reason for this would seem to lie in the fact that the conventional derivation of Lorentz and Poincar\'e transformations assume more than what is needed. Correspondingly, the literature is full of paradoxes and seemingly unsolved problems in special relativity, see for example the collections of papers in the conference proceedings \cite{selleri1998open} or papers by Selleri \cite{selleri2004sagnac,selleri2005zero}. Some, like the twin paradox, rely on confusion between inertial objects (the Earth bound twin) and the accelerated twin. Some argue that paradoxes can only be resolved by considering that the synchronisation of separated clocks is related to the one-way speed of light. In this section we retain the homogeneity and isotropy of space and imbed this into an assumption about the one-way speed of light. By making the minimal assumptions about these matters, this section aims to come to the strongest conclusions about spacetime transformations. We conclude that the algebra to describe them is the Clifford algebra $C\ell(1,3)$. 

The subject of this section is to extend the work of the previous three sections to include time, not as a mere parameter, but as a fourth dimension. In particular we seek to understand how to mathematically describe the motion of rigid bodies in spacetime. In general, the motion of a rigid body in spacetime can be described by a sequence of events. The term `event' is the generalisation to spacetime of the term `point' of space from the previous sections. Saying `event' is an alternative to saying `point in spacetime.' As we shall see, spacetime needs exactly four linearly independent parameters to specify position, up from the three needed for 3D space. In Newton's First Law, the sequence of events is the set of locations of the points on (or in) rigid bodies in space and parameterised by time. 

In sections \ref{AaAI2010} and \ref{AaAII2010} of this series we considered observations of the movement of rigid objects in a 2D toy world. The rigid objects in these sections were sheets of paper, moving about on top of another rigid object, the desktop. The points, lines, and the pieces of paper could be described as being at particular coordinates on the desktop at different values of a third parameter. We had the choice between two options for this third parameter to describe the location of the objects as they moved. We could plot 2D position against values of a time parameter, or against values of a height parameter. However neither the parameter `time', nor the parameter `height' needed any scale for that discussion. All that was needed was some means of characterizing the sequence of the positions of one rigid body relative to another. When extending these arguments to 3D in section \ref{AaAIII2010}, we had only the parameter `time' to characterise the various positions of the book or other 3D objects. Once again though, no scale was attributed to the parameter.

The first task in this section is to discuss the concept of equal times, and to associate a scale to the time parameter by developing a time measuring stick (known as a `clock').  The means to do this is to use natural systems that provide clocks, so \ref{sec:clocks} discusses such natural systems and shows that we may treat distances in the time direction (time intervals) in similar ways that we treat distance in any one of the three linearly independent space directions.
The invariance properties of clocks, that is the fact that many of our world's clocks behave in the same way yesterday, today and tomorrow, corresponds strongly with the homogeneity of space --  rigid bodies do not change when moved in space. There is however an important difference between translations in time and translations in space. While we can move our rigid objects back and forth in space (within the limits imposed by our experiments), we cannot move our rigid objects back and forth in time.

It may help the reader to use the term `timeline' here. Timelines do not always have a scale attached, they often just show the time ordering of events, but clocks and their invariance always allow us to attach a scale. This invariance property allows us to define corresponding algebraic entities, vectors in the one dimensional vector space that is the time direction of spacetime, and also the unit vector for the time direction, \that. 

Einstein \cite{einstein1905elektrodynamik} caused a major revision to the way we view time and revised our view of simultaneity. Einstein showed that we have a choice of assuming that clocks measure the same time intervals for all inertial observers, or that the measured speed of light is the same for all inertial observers. Experiment shows it is the second option that is correct, at least for rigid body frames. This changes the way we interpret what we observe, in particular what we consider simultaneous.  Our second key task is therefore to explore some consequences of the fact that the speed of light in a vacuum is the same for all inertial rigid body observers.

The relationship of the time axis to the space axes is given by the Lorentz metric, just as the relationship of the space axes to each other is given by the Pythagorean metric. The Lorentz metric is shown to follow from the invariance of the speed of light, just as the Pythagorean metric was shown in sections \ref{AaAII2010} and \ref{AaAIII2010} to follow from the isotropy of space, being the invariance of rigid bodies under rotation.

The previous sections were limited to exploring the geometric consequences of the concept of straight line motion of a rigid body as used in Newton's First Law. We have shown that the homogeneity and isotropy of our 3D space are well described by the Clifford algebra $C\ell(0,3)$. However this is only part of the statement of Newton's First Law. Not only does the absence of forces lead to straight line motion (which we have seen is not a simple concept), but the motion has constant speed, or in other words is `uniform.' This section looks at how to extend the Clifford algebra $C\ell(0,3)$ to incorporate our knowledge of the relationship of the time parameter to the three spatial parameters. 

We find that the `distance' between events in spacetime is given by Lorentz' generalisation of  Pythagoras' result.  
Section \ref{AaAIII2010} used the isotropy of space to compare the length of measuring sticks in different directions, and thereby to choose unit vectors to be of the same length $\mod{\xhat} = \mod{\yhat} =\mod{\zhat}$. The concept of `distance' in spacetime arising from the constancy of the speed of light enables us to expand the four-dimensional vector space to an associative algebra of dimension $2^4 = 16$. This algebra describes much more general transformations than the translations described by a vector space. We explore this algebra, the Clifford algebra $C\ell(1,3)$, in subsection \ref{sec:spacetime} and find that it contains the transformations of Lorentz and Poincar\'{e} \cite{weinberg1995}, many of which are transformations between measurements of different observers, and a few of which  describe physical operations on rigid bodies or on clocks.

\subsection{Clocks and the uniformity of time\label{sec:clocks}}

The previous sections have made much use of the homogeneity of space and the concept of a rigid body to define measuring sticks to measure length. The measuring sticks can be translated (and with isotropy, rotated) to compare lengths of objects. Time is rather different, as before the invention of the fob-watch in the sixteenth century and later the wrist watch, we could not `pick up our time' and compare it with some one else's. Human use and understanding of time was based on the day--night, lunar and yearly cycles.  All time measurements shorter than a day were subject to considerable variability and inaccuracy, and more or less unrepeatable. Excepting sun dials, water clocks and swinging chandeliers, and of course one's pulse or heart beat, time was difficult to measure. Prior to mechanical clocks there was nothing for measuring time that was analogous to the rigid objects that provide reproducible measuring sticks for space.

Today it is quite different. Standard wrist watches have an accuracy of better than seconds per day, a few parts in $10^5$ or so, and 
 we are used to computer clock frequencies of gigahertz, not only the 0.2 to 2 hertz of a chandelier or one's heart. Time is now the accurate measurement, defined by the period associated with the cesium atomic clock.  Distance is now defined by wavelength of light, as a product of the (assumed, but well tested) constant speed of light and the frequency of light emitted by the appropriate atoms.

Modern clocks, such as used for laboratory measurements, are based on atomic phenomena which typically have a `tick' of $10^{-15}$\,s and an accuracy of around $1:10^{-20}$. Such short times are beyond the comprehension of the proverbial `person in the street', who is perhaps limited to a minimum time interval, $t_{\textrm{min}}$, of a millisecond, $10^{-3}$\,s. However high energy physics experimentalists are familiar with particle lifetimes as short as  $t_{\textrm{min}}=10^{-24}$\,s. Likewise a child's perception of a longest time interval time, $t_{\textrm{max}}$ is a few years, or $10^{7}$\,s. Cosmologists have their $t_{\textrm{max}}$ as the age of the Universe, $t_{\textrm{Universe}}\approx 10^{17}$\,s. Physicists thus consider the ratio $t_{\textrm{max}}/t_{\textrm{min}}$ of about $10^{41}$

The invariance properties of clocks, that is the fact that many of our world's clocks behave in the same way yesterday, today and tomorrow, corresponds strongly with the homogeneity of space --  rigid bodies do not change when moved in space. It is of course an assumption that clocks will behave the same way tomorrow as they did yesterday, but it is an assumption that can be tested in 24 hours time. We have good records of how clocks behaved in the recent past, perhaps for several hundred years, and indirect evidence that (many) clocks have not changed for billions of years.

\subsection{Time as an axis of 4D spacetime\label{sec:time}}

Although our perception of time is perhaps controlled by internal clocks in the body, such as the beating heart and electrical and biochemical processes in our brain, it is clear that time has many similarities to position. It is reasonable therefore to treat time as a fourth coordinate, that is to treat spacetime as a four dimensional vector space. We can compare the length of time intervals in our laboratory by first choosing an origin $O$ and consequently choosing  our `time measuring stick', a clock with a `tick', $OT$ starting from the origin $O$, of length 1\,s, which in turn is calibrated by our cesium atomic clock. The time coordinate is to be regarded  as a `coordinate', `parameter' or `direction' that is linearly independent of the three space `coordinates', `parameters' or `directions'. We add to the  three orthonormal spatial measuring sticks $(OX, OY,OZ)$ and the three orthonormal unit vectors $(\xhat,\yhat, \zhat)$, the time measuring stick $OT$ and the unit vector \that. Just as \xhat\ is the  class of all lines equivalent by space translation to the unit spatial measuring stick, $OX$, that is $\xhat=[OX]$, so is \that\ the  class of all time intervals equivalent by time translation to the unit time interval, $OT$, namely $\that=[OT]$. There is however an important difference between translations in time and translations in space. While we can move our rigid objects back and forth in space (within the limits imposed by our experiments), we cannot move back and forth in time. The reason for this brings us back to the concluding remarks of the previous section. Just as the handedness (parity) of rigid bodies in space are conserved, so the handedness of time must be conserved also. Because we have only a single time dimension, the direction of time cannot be reversed via any physical operation. Thus both the handedness of space and the handedness of time for all rigid bodies are conserved by all physically allowed operations. An event $a$ in spacetime can now be labelled in terms of the four basis vectors $(\that,\xhat,\yhat,\zhat)$ describing the coordinates of the event relative to the origin of the reference frame, or describing translations of elements of the linear space.

It seems now an appropriate time to introduce the usual notation for unit vectors in special and general relativity, and in the study of Clifford algebras. The initial letter for the number one in German, {\em ein}, is a common choice. The four basis vectors are
\begin{eqnarray}
	e_0 &=& \that \nonumber \\
	e_1 &=& \xhat \nonumber \\
	e_2 &=& \yhat \nonumber \\
	e_3 &=& \zhat 
\label{eq:defe}
\end{eqnarray}
which we index by Greek letters, $0,1,2,3$, and use as usual Latin letters for the spatial indices $1,2,3$. An event $a$ measured in (coordinatised in) frame $S_1$ is thus 
\begin{eqnarray}
	a &=& a_0 \that + a_1\xhat +a_2\yhat +a_3\zhat \nonumber \\
	 &=& a_0e_0 + a_1e_1 +a_2e_2 +a_3e_3 \nonumber\\
	 &=& a_0e_0 + a_i e_i \nonumber\\
	 &=& a_\mu e_\mu
\label{eq:4vect}
\end{eqnarray}
following the usual convention of using non-bold Latin font for 4-vectors (vectors in spacetime). An exception to this is that for the basis vectors $\that, \xhat, \yhat, \zhat,$ we retain the 3D bold-hat notation. In the above, and in the following we use the usual `Einstein summation convention' whereby doubled indices are summed over. However because we are using a Clifford algebra, we do not need to use raised and lowered indices to take into account the metric -- the metric is built into the basis vectors.

The time measuring stick is linearly independent of the three space measuring sticks, and we now have the concepts to fully understand Newton's First Law (see our discussion of this matter in section \ref{AaAI2010}). But how do we determine whether or not the time measuring stick is orthogonal to the spatial ones? Indeed, what does it mean for time to be orthogonal to space, since we cannot rotate space into time? Prior to 1905 time and space were seen as independent, but Einstein showed that there was a way to generalise the isotropy considerations of section \ref{AaAII2010}. He found a way to interpret certain measurements as `rotations' in spacetime, and thus a way to find a spacetime distance measure that was the generalisation of Pythagoras. A modern variant of his argument can be found in introductory physics texts. We review this in \ref{sec:speedlight}. We offer a simpler derivation and explore the result in detail in \ref{sec:spacetime}.

\subsection{Relative speed of rigid bodies \label{sec:worldines}} 

We want to use an inertial rigid body (and its clocks) to define a frame and measure other rigid bodies with respect to this reference frame. In general these other rigid bodies can be moving with uniform velocity, undergoing linear acceleration, or rotating. For the cases where there is any acceleration, either linear or angular, the associated forces acting on the rigid body will have to be considered. Particularly in the case of rotating rigid bodies, centrifugal (and Coriolis) forces arise. 
Although we have set up the mathematics to deal with rotating frames and accelerating frames, in this section we are concerned only with inertial frames. Whereas frames are allowed to be moving at a constant speed and two frames are allowed to be rotated with respect to each other, that is have different orientations, we do not allow a frame to be accelerating or rotating. Such a frame would not be an inertial frame. In future work we plan to use the Clifford algebra to account for any accelerations that are being experienced by observers. This will allow us to deduce the extension of the properties of rigid bodies and clocks to accelerated motion. We further anticipate that this future work will resolve many of the paradoxes, such as the twin paradox, and what many \cite{selleri1998open} regard as open problems in special relativity.

The  speed, $v_{21}$, of  rigid body $S_2$ relative to rigid body $S_1$, is defined as the ratio of distance traveled, $\ell_1$, as measured by $S_1$, to the travel time, $t_1$, also as measured by $S_1$. We have 
\begin{eqnarray}
	v_{21}=\ell_1/t_1
\label{eq:speed}
\end{eqnarray} 
If $S_2$ is in uniform motion relative to $S_1$, then if $S_1$ takes another measurement
$v_{21}'=\ell_1'/t_1'$, we find
\begin{eqnarray}
	v_{21}=v_{21}'
\end{eqnarray} 

In addition to rigid bodies moving uniformly with respect to each other, the orientations of their frames may also be rotated with respect to one another. As we have noted before, the isotropy of space allows the rotation of a body, here $S_2$, relative to the axis system of another, $S_1$, to be written in terms of the product operation between vectors. In section \ref{AaAIII2010} we showed how to derive the plane of the rotation given the initial and final positions of three points of a rigid body. That calculation gave the rotation plane, and the angle in that plane, as a bi-vector.

\subsection{The speed of light and the Clifford algebra $C\ell(1,3)$ \label{sec:speedlight}}

Consider now a short pulse of light emitted at event $E_1=(t_1\that +x_1\xhat +y_1\yhat +z_1\zhat)$ and received at event $E_2=(t_2\that +x_2\xhat +y_2\yhat +z_2\zhat)$, both measured in the frame of a rigid body $S_1$ in inertial motion. Extending the definition, eq(\ref{eq:speed}), of relative speeds of rigid bodies, to the speed of light, $c$, gives for this situation
\begin{eqnarray}
	c=\ell_{21}/(t_2-t_1)
\end{eqnarray}
where $\ell_{21}$ is given by Pythagoras
\begin{eqnarray}
	\ell_{21}=\sqrt{(x_2-x_1)^2+(y_2-y_1)^2+(z_2-z_1)^2}
\end{eqnarray}
Eliminating $\ell_{21}$ and rearranging gives
\begin{eqnarray}
	c^2 (t_2-t_1)^2- (x_2-x_1)^2- (y_2-y_1)^2-(z_2-z_1)^2 = 0
\label{eq:lorinvar}
\end{eqnarray}

It is experimentally observed that in a frame $S_1$, all measurements of $c$ give the same value, independently of the location of the emission and absorption events. Furthermore, if these two events are observed by observers in the frame of another  rigid body $S_2$, then the same result holds, for the same value of $c$, regardless of the position, orientation or speed of the two rigid bodies relative to each other. In other words eq(\ref{eq:lorinvar}) is the generalisation of Pythagoras to spacetime distances between events connected by light.


The speed of light is rather different to the speed of a rigid body. Although Einstein reports \cite{Einstein19wavefront} that he found the thought experiment of imagining that he was traveling with a light wave, as a key step in coming to his Special Theory of Relativity, it is experimentally observed that no rigid body ever travels at the speed of light relative to another rigid body. Thus we cannot define the motion of a rigid body relative to light, only the speed of light relative to a rigid body.  The transformation laws between measurements made using clocks and rigid bodies derived in the next subsection explains this -- explains in the sense that we must modify at least one of the assumptions and at least one of the axioms of our sections if we were ever to observe rigid bodies traveling at the speed of light.

In order to find the algebraic product relationship between the basis vectors, we can repeat the argument from section II \ref{AaAII2010}. We want the free product $(c (t_2-t_1)\that +(x_2-x_1)\xhat+ (y_2-y_1)\yhat+ (z_2-z_1)\zhat)^2$ to reproduce eq(\ref{eq:lorinvar}). After a few simple steps analogous to the 2D and 3D cases, we deduce that we need in addition to the product rules for the spatial unit vectors, eqs(2-6) of section III \ref{AaAIII2010}, the rules
\begin{eqnarray}
	\that^2  &=& 1 \nonumber \\
	\that\xhat &=&  -\xhat\that \nonumber \\
	\that\yhat &=&  -\yhat\that \nonumber \\
	\that\zhat &=&  -\zhat\that 
	\label{eq:4ddefs}
\end{eqnarray}
The rules eqs(2-6) of III, together with eqs(\ref{eq:4ddefs}) define the Clifford algebra $C\ell(1,3)$.

The  Clifford algebra of spacetime $C\ell(1,3)$ has other basis elements, they are constructed from products of the defining elements of eq(\ref{eq:defe}). The extra elements we write as 
\begin{eqnarray}
e_{\mu\nu} &\equiv& e_\mu e_\nu \ =- e_\nu e_\mu \ =- e_{\nu\mu} \nonumber \\
e_{\mu\nu\rho} &\equiv& e_\mu e_\nu e_\rho \ = e_{\nu\rho\mu}  \ {\textrm{ and other cyclic permutations of }}{\mu\nu\rho}   \nonumber \\
							&=& - e_{\nu\mu\rho} \ {\textrm{ and other non-cyclic permutations} } \nonumber \\
   e &\equiv& e_0 e_1 e_2 e_3 \ =\ e_{0123} \ \ {\textrm{ and cyclic permutations}}  \nonumber \\
    	&=& - e_{1023}\ \   {\textrm{ and other non-cyclic permutations}}
\end{eqnarray} 
Using this notation we may readily expand out the square, $a^2$,  of the vector $a$ that represents the distance between two events as
\begin{eqnarray}
	a^2 &=& (a_0e_0 + a_1e_1 +a_2e_2 +a_3e_3)^2 \nonumber \\
			&=& a_0^2e_0^2 + (a_ie_i)^2 + a_0a_ie_{0i} + a_ia_0e_{i0}  \nonumber \\
			&=& a_0^2 - a_1^2 - a_2^2 - a_3^2
\label{eq:4vect2}
\end{eqnarray}
since $e_{0i} =-e_{i0}$. By choosing $e_0=\that$, the parameter $a_0$ in the first of eqs(\ref{eq:4vect}) is related to the time coordinate $t$ by the scale factor $c$, $a_0 = ct$.

Observe that the spacetime Clifford algebra $C\ell(1,3)$ contains the 16 linearly independent elements
\begin{eqnarray}
\textrm{the scalar, }\ \bbI &=& 1 \nonumber \\
\textrm{the 4 vectors, } e_\mu&=& e_0, e_1, e_2, e_3\ = \ \that, \xhat,\yhat,\zhat \nonumber \\
\textrm{the 3 spatial bi-vectors, }  e_{ij} &=& e_{23}, e_{31}, e_{12}\ \ = \  \ihat,\jhat,\khat \nonumber \\
\textrm{the 3 spacetime bi-vectors, } e_{i0} &=& e_{10}, e_{20}, e_{30} \nonumber \\
\textrm{the 4 tri-vectors, } e_\mu e&=& e_{123},e_{023}, e_{031}, e_{012} \nonumber \\
\textrm{the quadri-vector or spacetime pseudoscalar, }\ e &=& e_{0123} 
\label{eq:cl13elem}
\end{eqnarray}
where the six elements $1, e_0, e_{10},  e_{20},  e_{30},$ and $e_{123}$ square to $+1$ and the ten elements $e_1, e_2, e_3,  e_{23}, e_{31},$ $e_{12},$ $e_{023},$ $e_{031}, e_{012}$ and $e$ square to $-1$. The finite group $C\ell^\textrm{group}(1,3)$ consists of 32 elements, these 16 elements and their negatives.

We have used the invariance of the speed of light, both with respect to measurements in the frame of one inertial rigid body, and with respect to measurements in differing inertial rigid bodies, to deduce the Lorentz invariant metric of spacetime. This metric is the generalisation of the Pythagoras result for orthonormal axes \xhat, \yhat\ and \zhat\ to include \that. Pythagoras says that the length squared of the line on a rigid body in 3D is the sum of the squares of the components with respect to orthonormal reference axes, and is invariant under motion of the rigid body. This incapsulates the homogeneity and isotropy of 3D space. The Lorentz metric extends Pythagoras from 3D to the 4D of spacetime by giving a precise meaning to the statement that the time axis of an inertial rigid body is orthogonal to all three space axes. Further, the Lorentz metric uses the speed of light, $c$, as the  constant  relating the length of the space measuring sticks (chosen to be 1 metre) to the time measuring stick (chosen to be 1 second). Although we have not chosen units where $c=1$, we have chosen the  vector \that\ to be unit, $\mod{\that}=1$.

\subsection{Spacetime events and simultaneity \label{sec:spacetime}}

We employ the techniques introduced in section \ref{AaAIII2010} for transforming between reference frames in a relativistic setting and obtain the Lorentz and Poincar\'e transformations as a result. We begin by considering two frames $S_1$ and $S_2$ representing two inertial rigid body observers. For simplicity we assume that the two frames are not rotated with respect to one another and that initially at $t=0$ the origins of the two frames coincide, that is $O_1=O_2$. By this we mean that the spacetime coordinates of the initial event are the same in both frames
\begin{eqnarray}
S_1&=&(0,0,0,0)=0\that+0\xhat+0\yhat+0\zhat\\
S_2&=&(0,0,0,0)'=0\that'+0\xhat'+0\yhat'+0\zhat'
\end{eqnarray}

We consider now how the coordinates change in the two frames as $S_2$ moves with respect to $S_1$ in the $\xhat$ direction with speed $v$. The position of $S_1$ as measured by observers in $S_1$ is 
\begin{eqnarray}
S_1(S_1)=ct\that=s_1
\end{eqnarray}
because observers in $S_1$ see their clocks ticking.
Similarly, the position of $S_2$ as measured by observers in $S_1$ is
\begin{eqnarray}
S_1(S_2)=ct\that+vt\xhat=s_2
\end{eqnarray}
Therefore the transformation from the the first frame $S_1$ to the second frame $S_2$ is given by a spacetime rotation $\mathrm{Rot}(S_1 \rightarrow S_2)$
\begin{eqnarray}
\mathrm{Rot}(S_1 \rightarrow S_2)=\mathrm{Rot}(ct\that\rightarrow ct\that+vr\xhat)\equiv \mathrm{Rot}(s_1\rightarrow s_2)
\end{eqnarray}

Now consider a general spacetime event $P$ measured by $S_1$ and $S_2$. The coordinates of $P$ are given by
\begin{eqnarray}
P&=&ct\that+x\xhat+y\yhat+z\zhat\\
&=& ct'\that'+x'\xhat'+y'\yhat'+z'\zhat'
\end{eqnarray}
respectively. Given the coordinates of $P$ in the first frame $S_1$, the coordinates in the second frame $S_2$ are given by
\begin{eqnarray}
S_2(P)&=&\mathrm{Rot}(S_1 \rightarrow S_2)(S_1(P))\\
&=&(s_1 s_2)^{1/2}S_1(P)(s_1 s_2)^{-1/2}\\
&=&\frac{1}{(\hat{s}_1+\hat{s}_2)^2}\hat{s}_1(\hat{s}_1+\hat{s}_2)S_1(P)(\hat{s}_1+\hat{s}_2)\hat{s}_1
\end{eqnarray}
where we have used the expression (38) of section \ref{AaAII2010} for the square root in terms of the unit vectors $\hat{s}_1=\that$ and $\hat{s}_2=\gamma(\that+\beta\xhat)$ where $\beta=v/c$ and $\gamma=1/\sqrt{1-v^2/c^2}$. This expression can be evaluated directly, but the algebra is simplified by noting that for rotations in the $\that\xhat$ plane, the $\yhat$ and $\zhat$ components are unchanged and using the generalization of equation (29) of section \ref{AaAIII2010} we have

\begin{eqnarray}
S_2(P)&=&\hat{s}_1\hat{s}_2(ct\that+x\xhat)+y\yhat+z\zhat\nonumber\\
&=& \gamma\that(\that+\beta\xhat)(ct\that+x\xhat)+y\yhat+z\zhat\nonumber\\
&=& \gamma(ct\that-\beta ct\xhat+x\xhat-x\beta\that)+y\yhat+z\zhat\nonumber\\
&=& \gamma c(t-vx/c^2)\that+\gamma(x-vt)\xhat+y\yhat+z\zhat
\end{eqnarray}
Consequently
\begin{eqnarray}
t'&=&\gamma(t-vx/c^2)\nonumber\\
x'&=&\gamma(x-vt)\nonumber\\
y'&=&y\nonumber\\
z'&=&z
\end{eqnarray}
These are the standard Lorentz transformations. 

For simplicity we assumed that the origins of $S_1$ and $S_2$ coincided at $t=0$ and that the spatial orientations of the frames are equal. For the case where the orientations are not the same, the above calculations still hold but one has to introduce a rotation (see section \ref{AaAIII2010}) to align the frames. For the case where the frames do not share a common origin (both in space and time), the origins of the two frames will be connected via a translation in spacetime 
\begin{eqnarray}
\trans{S_2}{S_1}O_2 =O_2+\mathbf{T} = O_1
\end{eqnarray}
where $T=t_0\that+x_0\xhat+y_0\yhat+z_0\zhat$ is a vector. So if at $t=0$ (in frame $S_2$), the location of the origin of frame $S_1$ as measured by an observer in frame $S_2$ is given by
\begin{eqnarray}
S_2(O_1)|_{t=o} =x_0\xhat+y_0\yhat + z_0\zhat
\end{eqnarray}
then the transformation between the two frames is given by
\begin{eqnarray}
t'&=&t_0+\gamma(t-vx/c^2)\\
x'&=&x_0+\gamma(x-vt)\\
y'&=&y_0+y\\
z'&=&z_0+z
\end{eqnarray}
These are the Poincar\'e transformations for parallel spatial axes.

\subsection{Concluding remarks - What have we achieved?}\label{sec:conclusion}

In this section we have extended the work of the previous three sections to include time. The invariance properties of clocks corresponds strongly with the homogeneity of space. We associated a scale to the time parameter by developing a time measuring stick $OT$ (known as a `clock') and the unit vector $\that$. We may treat distances in the time direction (time intervals) in similar ways that we treat distance in any one of the three linearly independent space directions. An event $a$ in spacetime can then be labelled in terms of the four basis vectors $(\that,\xhat,\yhat,\zhat)$. The vector space properties of the time dimension have been shown to be analogous to those properties derived from the homogeneity of each of the three dimensions of physical space. Time can be treated as a fourth coordinate, that is physical spacetime corresponds to a four dimensional vector space. The time coordinate is a `coordinate', `parameter' or `direction' that is linearly independent of the three space `coordinates', `parameters' or `directions'.

The relationship of the time axis to the space axes is given by the Lorentz metric, just as the relationship of the space axes to each other is given by the Pythagorean metric. The Lorentz metric follows from the invariance of the speed of light, just as the Pythagorean metric follows from the isotropy of space, being the invariance of rigid bodies under rotation. 

It is a guiding general principle of science that observations are essentially independent of the observer. Einstein's 1905 paper \cite{einstein1905elektrodynamik} uses a more specific form of this general principle, ``That physics is the same for all inertial observers'' (translation as used by \cite{serway2008psa}). In this section we have assumed that all inertial observers measure the speed of light, $c$. All of this is well known and standard. Our argument contains some novelty in the following results.

If the Pythagorean metric is assumed to apply to one rigid body frame, then the assumption of homogeneity means it applies to all rigid body frames. likewise, we need only assume that the speed of light $c$ is invariant in one inertial rigid body frame for we can use the homogeneity of spacetime, and the isotropy of space to deduce that the Clifford algebra $C\ell(1,3)$ gives the transformation laws for all spacetime measurements between all inertial rigid body frames.

We have extended the Clifford algebra $C\ell(0,3)$ to incorporate our knowledge of the relationship of the time parameter to the three spatial parameters. The concept of `distance' in spacetime arising from the constancy of the speed of light enabled us to expand the four-dimensional vector space to an associative algebra of dimension $2^4 = 16$. We found the algebraic product relationship between the basis vectors by repeating the argument from section \ref{AaAII2010}. The rules eqs(2-6) of section \ref{AaAIII2010}, and eqs(\ref{eq:4ddefs}) define the Clifford algebra $C\ell(1,3)$.

We explored this algebra in subsection \ref{sec:spacetime} and employed the techniques introduced in section \ref{AaAIII2010} for transforming measurements of events (and by extension the notion of rigid bodies) between inertial reference frames in a relativistic setting to obtain the Lorentz and Poincar\'e transformations.

This paper considers only inertial rigid body frames. Whereas frames are allowed to be moving at a constant speed and two frames are allowed to be rotated with respect to each other, that is have different orientations, we do not allow a frame to be accelerating or rotating. such a frame would no longer be an inertial frame. Work is needed to use the Clifford algebra to account for any accelerations that are being experienced by observers. This will allow us to deduce the extension of the properties of rigid bodies and clocks to accelerated motion. We further anticipate that this future work will resolve many of the paradoxes, such as the twin paradox, and various open problems (see for example \cite{selleri1998open}) in special relativity.

\section{Algebraic Structure of $C\ell(1,3)$}\label{AaAV2010}
We have argued that the Clifford algebra $C\ell(1,3)$ is the appropriate algebra to describe spacetime. We have shown that the rational numbers $\mathbb{Q}$ are needed as the field over which the algebra is defined. In this section we explore further the algebraic structure of spacetime.

Matrices are a natural and very useful way to study the properties of algebras. In this section we review the matrix representations of Clifford algebras $C\ell(1,3)$ and $C\ell(3,1)$ and some of their lower dimensional subalgebras. For representations of $C\ell(p,q)$ up to $p+q=7$ and for some general results, the reader is referred to Lounesto \cite{lounesto2001caa}.

Although matrix representations are a useful tool for studying the Clifford algebras of space and spacetime, and indeed algebras in general, it must be remembered that the Clifford algebras retain a stronger link to the points, lines etc. of spacetime than matrix representations which are only determined up to a similarity transformation.

There are two important points we must consider when looking for a matrix representation. First the dimension of the algebra is normally no more than the number of independent components of the matrices. Second, it is  important to find connections between the
geometry of the Clifford algebras and the geometry which is present in
their matrix representations. This is a powerful incentive to consider mainly matrix representations over the rational or real numbers.

One claimed weakness \cite{penrose2004rrc} of the Clifford algebra $C\ell(1,3)$ in being able to mathematically describe physical
reality is that the algebra is not a division algebra, meaning that there are elements of the algebra other than zero for which no
inverse can be found. There are in fact very few linear spaces which admit the structure of a division algebra; the algebra of the rationals, the reals, the complex numbers and the quaternion algebra are examples.
The Clifford algebra $C\ell(1,3)$ is not a division algebra as there exist many elements $A$ for which no inverse $A^{-1}$ can be defined.  

It has been shown by van der Mark and Williamson \cite{vandermark2007} that the areas of the algebra where the inverse does not exists, that is where division cannot be
defined, are where certain invariant quantities become zero, for example on the light cone. These areas are referred to as null-hyperplanes because they correspond to null multivectors and
correspond exactly to cases of physical interest. The fact that there does not exist an inverse for every element
is therefore not a weakness but a necessity  because the breakdown of invertibility in these areas matches the behavior of nature.

Later in this section, we confirm some of the results found in \cite{vandermark2007}, but not by means of defining a
new conjugate, but by using the matrix representations of the spacetime Clifford algebra $C\ell(1,3)$. The use of matrix
representations make it a straightforward task to determine which elements of the algebra are or are not invertible. Given
that an element is invertible, it is then straightforward to calculate its inverse. The invertibility or
non-invertibility of multivectors give us physical insight into conserved quantities and limitations of physical systems.

\subsection{$C\ell(1,0)$ and $C\ell(0,1)$}\label{sec:1dim}%
The Clifford algebras $C\ell(1,0)$ and $C\ell(0,1)$ each have two basis elements, $1$ and $e_1$ satisfying, 
\begin{eqnarray}
1^2=1,\quad e_1^2=+1
\end{eqnarray}
for $C\ell(1,0)$ and 
\begin{eqnarray}
1^2=1,\quad e_1^2=-1 
\end{eqnarray}
for $C\ell(0,1)$. Both algebras have two degrees of freedom and can be represented as $2\times 2$ real matrices. 

A representation of $C\ell(1,0)$ requires two $2\times 2$ matrices that square to unity. A suitable basis for this algebra is
\begin{eqnarray}
1 =\left(\begin{array}{cc} 1&0\\0&1 \end{array}\right) \quad
e_1 =\left(\begin{array}{cc} 0&1\\1&0 \end{array}\right) 
\end{eqnarray}
meaning a general element $A\in C\ell(1,0)$ may be written as
\begin{eqnarray}
A=a+be_1=\left(\begin{array}{cc} a&b\\b&a \end{array}\right)
\end{eqnarray}

For $C\ell(0,1)$, $e_1^2=-1$ and so this algebra is isomorphic to the algebra of complex numbers $\mathbb{C}$
\begin{eqnarray}
\mathbb{C}\cong C\ell(0,1)
\end{eqnarray}
An arbitrary element $A\in C\ell(0,1)$ (or equivalently an arbitrary complex number) may be written as a linear combination of the $C\ell(0,1)$ elements 1 and  $e_1$ or equivalently as a linear combination of the $2\times 2$ unimodular basis matrices
\begin{eqnarray}
1 =\left(\begin{array}{cc} 1&0\\0&1 \end{array}\right) \quad
e_1 =\left(\begin{array}{cc} 0&1\\-1&0 \end{array}\right) 
\end{eqnarray}
Thus
\begin{eqnarray}
A=a+be_1=\left(\begin{array}{cc} a&b\\-b&a \end{array}\right)\qquad a,b\in \mathbb{R}
\end{eqnarray}

The geometry of these two Clifford algebras is different from the standard Argand diagram view of complex numbers where a complex number $z$ is a point in a two dimensional plane. In the complex number algebra, $z$ can be rotated in the complex plane. In the one dimensional geometries described by $C\ell(1,0)$ and $C\ell(0,1)$ however, there is no physical rotation operator because space is simply not big enough. Given a vector in a one dimensional space, and a corresponding set of lines in a physical space,  there is no physical operation that will transform the lines into minus themselves (that is, an inversion) even though such a mathematical operator exists (multiply by $-1$). More generally we say that in an $n$-dimensional space, an $n$-vector may have a mathematical inversion, but there is no geometric operation that will turn the corresponding geometric object, an $n$-multivector, into minus itself. We discussed this issue in sections \ref{AaAIII2010} and \ref{AaAIV2010}.

As a final observation, notice that 
\begin{eqnarray}
\mathrm{det}(A)=a^2+b^2\quad\mathrm{if}\;A\in C\ell(0,1) \\
\mathrm{det}(A)=a^2-b^2\quad\mathrm{if}\;A\in C\ell(1,0)
\end{eqnarray}
In $C\ell(0,1)$ only the trivial element with $a=b=0$ does not have an inverse. In $C\ell(1,0)$ there are many elements for which an inverse is not defined (whenever $a=\pm b$). 
\subsection{$C\ell(2,0)$ and $C\ell(0,2)$}\label{sec:2dim}%
We require a set of four linearly independent matrices that satisfy the commutation relations of the basis elements $\left\{1,e_1,e_1,e_{12}\right\}$ of the algebras $C\ell(0,2)$ and $C\ell(2,0)$. We encountered these algebras in section \ref{AaAII2010} where it was shown that the homogeneity and isotropy of 2D space, together with Pythagoras' theorem, gives one of these two algebras depending on the choice of metric.

We begin by considering the algebra of all $2\times 2$ matrices with real entries, $\mathrm{Mat}(2,\mathbb{R})$. One useful basis for this algebra is ${I_2,\ell,m,n}$ with
\begin{eqnarray}
I_2=\left(\begin{array}{cc} 1&0\\0&1 \end{array}\right)\quad \ell=\left(\begin{array}{cc} 1&0\\0&-1 \end{array}\right)\quad
m=\left(\begin{array}{cc} 0&1\\1&0 \end{array}\right)\quad n=\left(\begin{array}{cc} 0&1\\-1&0 \end{array}\right)
\end{eqnarray}
These basis elements satisfy
\begin{eqnarray}
1^2=\ell^2=m^2=1,\;\mathrm{and}\;n^2=-1
\end{eqnarray}
An arbitrary $2\times 2$ matrix can be written in this basis as
\begin{eqnarray}
\left(\begin{array}{cc} a&b\\c&d \end{array}\right)=\frac{a+d}{2}\left(\begin{array}{cc} 1&0\\0&1 \end{array}\right)+\frac{a-d}{2}\left(\begin{array}{cc} 1&0\\0&-1 \end{array}\right) 
+\frac{b+c}{2}\left(\begin{array}{cc} 0&1\\1&0 \end{array}\right)+\frac{b-c}{2}\left(\begin{array}{cc} 0&1\\-1&0 \end{array}\right)
\end{eqnarray}
$I_2,l,m$ and $n$ satisfy the multiplication rules
\begin{eqnarray}
\ell m=-m \ell=n, \quad mn=-nm=\ell, \;\mathrm{and}\;n\ell=-\ell n=m
\end{eqnarray}
which are precisely the rules satisfied by the basis elements $e_1,e_2$ and $e_{12}$ of $C\ell(2,0)$. An arbitrary multivector $A$ in this algebra may therefore be represented as
\begin{eqnarray}
A=a1+be_1+ce_2+de_{12}=\left(\begin{array}{cc} a+b & c+d\\c-d & a-b \end{array}\right)
\end{eqnarray}

A matrix representation of $C\ell(0,2)$ in terms of $\mathrm{Mat}(2,\mathbb{R})$ cannot be found because $\mathrm{Mat}(2,\mathbb{R})$ has only one of its basis elements square to minus unity whereas $C\ell(0,2)$ has three. A set of three matrices that square to minus unity is needed. It is easily proved that $3 \times 3$ real matrices are also too small. 

A representation can be found in terms of the $2 \times 2$ matrices with \textit{complex} entries or a representation in terms of $4 \times 4$ matrices with real entries. 
A suitable set of sixteen $4 \times 4$ matrices is constructed by taking tensor products of the $2 \times 2$ basis elements $I_2,\ell,m,n$.
\begin{eqnarray}
\notag A_1&=&I_2\otimes I_2=\left(\begin{array}{cc} I_2&0\\0&I_2 \end{array}\right)\qquad A_2=I_2\otimes \ell=\left(\begin{array}{cc} I_2&0\\0&-I_2 \end{array}\right)\\
\notag A_3&=&I_2\otimes m=\left(\begin{array}{cc} 0&I_2\\I_2&0 \end{array}\right)\qquad A_4=I_2\otimes n=\left(\begin{array}{cc} 0&I_2\\-I_2&0 \end{array}\right)\\
\notag A_5&=&\ell \otimes I_2=\left(\begin{array}{cc} \ell&0\\0&\ell \end{array}\right)\qquad A_6=\ell\otimes \ell=\left(\begin{array}{cc} \ell&0\\0&-\ell \end{array}\right)\\
\notag A_7&=&\ell\otimes m=\left(\begin{array}{cc} 0&\ell\\ \ell&0 \end{array}\right)\qquad A_8=\ell\otimes n=\left(\begin{array}{cc} 0&\ell\\-\ell&0 \end{array}\right)\\
\notag A_9&=&m\otimes I_2=\left(\begin{array}{cc} m&0\\0&m \end{array}\right)\qquad A_{10}=m\otimes \ell=\left(\begin{array}{cc} m&0\\0&-m \end{array}\right)\\
\notag A_{11}&=&m\otimes m=\left(\begin{array}{cc} 0&m\\m&0 \end{array}\right)\qquad A_{12}=m\otimes n=\left(\begin{array}{cc} 0&m\\-m&0 \end{array}\right)\\
\notag A_{13}&=&n\otimes I_2=\left(\begin{array}{cc} n&0\\0&n \end{array}\right)\qquad A_{14}=n\otimes \ell=\left(\begin{array}{cc} n&0\\0&-n \end{array}\right)\\
 A_{15}&=&n\otimes m=\left(\begin{array}{cc} 0&n\\n&0 \end{array}\right)\qquad A_{16}=n\otimes n=\left(\begin{array}{cc} 0&n\\-n&0 \end{array}\right)\label{set1}
\end{eqnarray}
These matrices satisfy
\begin{eqnarray}
\notag A^2_i&=&+1,\quad\mathrm{for}\quad i=1,2,3,5,6,7,9,10,11,16\\
\notag A^2_i&=&-1,\quad\mathrm{for}\quad i=4,8,12,13,14,15
\end{eqnarray}

One possible representation of $C\ell(0,2)$ is to choose
\begin{eqnarray}
e_1=A_8=\left(\begin{array}{cc} 0&\ell\\-\ell&0 \end{array}\right)\quad e_2=A_{12}=\left(\begin{array}{cc} 0&m\\-m&0 \end{array}\right)\quad e_{12}=-A_{13}=-\left(\begin{array}{cc} n&0\\0&n \end{array}\right)
\end{eqnarray}
so that an arbitrary multivector $A$ in this algebra may be represented as
\begin{eqnarray}
A=a1+be_1+ce_2+de_{12}=\left(\begin{array}{cc} aI_2+dn&bl+cm\\-bl-cm&aI_2-dn \end{array}\right)
\end{eqnarray}
This choice of representation is however not unique.

These matrices also give a representation of the quaternion algebra $\mathbb{H}$, and so the quaternion algebra is isomorphic to $C\ell(0,2)$
\begin{eqnarray}
C\ell(0,2)\cong \mathbb{H}
\end{eqnarray}
This $4\times 4$ real representation of the quaternions $(i,j,k)$ is given by
\begin{eqnarray}
\label{i} i=\left(\begin{array}{cc} 0&I_2\\-I_2&0 \end{array}\right)=\left(\begin{array}{cccc} 0&0&1&0\\0&0&0&1\\-1&0&0&0\\0&-1&0&0 \end{array}\right),
\label{j} j=\left(\begin{array}{cc} 0&n\\n&0 \end{array}\right)=\left(\begin{array}{cccc} 0&0&0&1\\0&0&-1&0\\0&1&0&0\\-1&0&0&0 \end{array}\right),
\label{k} k=\left(\begin{array}{cc} n&0\\0&n \end{array}\right)=\left(\begin{array}{cccc} 0&1&0&0\\-1&0&0&0\\0&0&0&-1\\0&0&1&0 \end{array}\right)
\end{eqnarray}
A $2\times2$ complex representation of the quaternions $i,j,k$ is given by
\begin{eqnarray} 
i=\left(\begin{array}{cc} i&0\\0&-i \end{array}\right),\qquad j=\left(\begin{array}{cc} 0&1\\-1&0 \end{array}\right),\qquad k=\left(\begin{array}{cc} 0&i\\i&0 \end{array}\right)
\end{eqnarray}
where the matrix element $i=\sqrt{-1}$. We apologize for two different uses for $i$ in the same equation!

These different matrix representations demonstrate that we have the embedding
\begin{eqnarray}
\mathrm{Mat}(1,\mathbb{H})\subset\mathrm{Mat}(2,\mathbb{C})\subset \mathrm{Mat}(4,\mathbb{R})
\end{eqnarray}
More generally $\mathrm{Mat}(n,\mathbb{H})\subset \mathrm{Mat}(2n,\mathbb{C})\subset \mathrm{Mat}(4n,\mathbb{R})$ for integer $n$.

If we wish to, we can of course also represent $C\ell(2,0)$ using $4 \times 4$ matrices. For example, a possible representation of $C\ell(2,0)$ in $\mathrm{Mat}(4,\mathbb{R})$ is
\begin{eqnarray}
e_1=\left(\begin{array}{cc} \ell&0\\0&\ell \end{array}\right)\quad e_2=\left(\begin{array}{cc} m&0\\0&m \end{array}\right)\quad e_{12}=\left(\begin{array}{cc} n&0\\0&n \end{array}\right)
\end{eqnarray}
however this representation is just two copies of its representation in $\mathrm{Mat}(2,\mathbb{R})$.

\subsection{$C\ell(3,0)$ and $C\ell(0,3)$}\label{sec:3d}%
The algebras $C\ell(0,3)$ and $C\ell(3,0)$ are both eight dimensional with basis $\left\{1,e_1,e_2,e_3,e_{23},e_{31},e_{12},e_{123}\right\}$. In section III \ref{AaAIII2010} we showed that both these algebras contain some cyclic structure. Both algebras have a four dimensional subalgebra called the even subalgebra spanned by the scalar and the three bivectors $\left\{{1,e_{23},e_{31},e_{12}}\right\}$. These two even subalgebras are isomorphic to each other and also isomorphic to the quaternion algebra with the isomorphism
\begin{eqnarray}
1&\leftrightarrow& 1,\qquad i\leftrightarrow-e_{23},\ \quad
 j\leftrightarrow-e_{31},\ \quad 
 k\leftrightarrow-e_{12}\quad\mathrm{for}\quad C\ell^{+}(3,0) \\
1&\leftrightarrow& 1,\qquad i\leftrightarrow e_{23},\qquad j\leftrightarrow e_{31},\qquad k\leftrightarrow e_{12}\qquad\mathrm{for}\quad C\ell^{+}(0,3)
\end{eqnarray}
We note as we did in section \ref{AaAIII2010} that $C\ell(3,0)$ does not respect the cyclic structure of $e_1 e_2=e_{12}$ that we have for $C\ell(0,3)$.

$C\ell(3,0)$ has four basis elements that square to unity and four that square to minus unity. A representation of the algebra can be found in term of $4\times 4$ real matrices
\begin{eqnarray}
1=A_1=\left(\begin{array}{cc} I_2&0\\0&I_2 \end{array}\right)\quad e_1=A_5=\left(\begin{array}{cc} \ell&0\\0&\ell \end{array}\right)\quad
 e_2=A_{10}=\left(\begin{array}{cc} m&0\\0&-m \end{array}\right)\quad e_3=A_{11}=\left(\begin{array}{cc} 0&m\\m&0 \end{array}\right)
\end{eqnarray}
From these the matrix representations of the other basis elements are readily found to be
\begin{eqnarray}
e_{12}=A_{14}\qquad e_{23}=A_{4}\qquad e_{31}=-A_{15}\qquad e_{123}=A_8
\end{eqnarray}

It is also possible to find a $2\times 2$ complex matrix representation by choosing
\begin{eqnarray}
e_1=\sigma_1,\qquad e_2=\sigma_2,\qquad e_3=\sigma_3.
\end{eqnarray}
where
\begin{eqnarray}
I=\left(\begin{array}{cc} 1&0\\0&1 \end{array}\right)\quad \sigma_1=\left(\begin{array}{cc} 0&1\\1&0 \end{array}\right)\quad \sigma_2=\left(\begin{array}{cc} 0&-i\\i&0\end{array}\right)\quad \sigma_3=\left(\begin{array}{cc} 1&0\\0&-1 \end{array}\right)
\end{eqnarray}
are the Pauli matrices that give a matrix representation of the Pauli algebra
\begin{eqnarray}
[\sigma_a, \sigma_b]=2i\epsilon_{abc}\sigma_c
\end{eqnarray}
From these it is then easy to show that
\begin{eqnarray}
e_{12}&=&\sigma_1\sigma_2=i\sigma_3 \nonumber\\
e_{23}&=&\sigma_2\sigma_3=i\sigma_1 \nonumber\\
e_{31}&=&\sigma_3\sigma_1=i\sigma_2 \nonumber\\
e_{123}&=&\sigma_1\sigma_2\sigma_3=i
\end{eqnarray}

There are no representations of $C\ell(0,3)$ in terms of $4\times 4$ real matrices or $2\times 2$ complex matrices because this algebra has six of its eight basis vector square to minus unity. It is possible to find a representation in terms of $8\times 8$ real matrices, $4\times 4$ complex matrices and also in terms of the $2\times 2$ matrix algebra over the quaternions. This quaternion representation is given by
\begin{eqnarray}
1=I_2\qquad e_1=\left(\begin{array}{cc} 0&i\\i&0 \end{array}\right) \qquad e_2=\left(\begin{array}{cc} 0&j\\j&0 \end{array}\right)\qquad e_3=\left(\begin{array}{cc} 0&k\\k&0 \end{array}\right)
\end{eqnarray}
\begin{eqnarray}
e_{23}=\left(\begin{array}{cc} i&0\\0&i \end{array}\right)\quad e_{31}=\left(\begin{array}{cc} j&0\\0&j \end{array}\right)\quad e_{12}=\left(\begin{array}{cc} k&0\\0&k \end{array}\right)\quad e_{123}=\left(\begin{array}{cc} 0&-1\\-1&0 \end{array}\right)
\end{eqnarray}
Using equations (\ref{i}), these matrices can be rewritten as $8\times 8$ real matrices because $i,j,k$ can be written as $4\times 4$ real matrices.

\subsection{$C\ell(3,1)$ and $C\ell(1,3)$}\label{sec:4d}%
Both $C\ell(1,3)$ and $C\ell(3,1)$ are sixteen
dimensional algebras that are candidates to describe the 4-dimensional geometry of spacetime. The matrix representations of these two algebras are quite distinct as
\begin{eqnarray}
C\ell(1,3):\quad\mathrm{has}\; &10&\;\mathrm{roots\;of\;+1}\;\mathrm{and}\nonumber\\
&6&\;\mathrm{roots\;of\;-1},\;\mathrm{while}\nonumber\\
C\ell(3,1):\quad\mathrm{has}\; &6&\;\mathrm{roots\;of\;+1}\;\mathrm{and}\nonumber\\
&10&\;\mathrm{roots\;of\;-1}\nonumber
\end{eqnarray}

Because $C\ell(3,1)$ has only six roots of $-1$ we can find a representation in terms of $4\times 4$ real matrices. A suitable representation is:

\begin{eqnarray}
1 =\left(\begin{array}{cc} I_2&0\\0&I_2 \end{array}\right) \quad
e_0 &=&\left(\begin{array}{cc} m&0\\0&m \end{array}\right) \nonumber\\
e_1 =\left(\begin{array}{cc} n&0\\0&n \end{array}\right)\quad
e_2 &=&\left(\begin{array}{cc} \ell&0\\0&-\ell \end{array}\right)\quad
e_3 =\left(\begin{array}{cc} 0&\ell\\\ell&0 \end{array}\right)\nonumber\\
e_{10} =\left(\begin{array}{cc} \ell&0\\0&\ell \end{array}\right)\quad
e_{20} &=&\left(\begin{array}{cc} -n&0\\0&n \end{array}\right)\quad
e_{30} =\left(\begin{array}{cc} 0&-n\\-n&0 \end{array}\right)\nonumber\\
e_{23} =\left(\begin{array}{cc} 0&\ell\\-\ell&0 \end{array}\right)\quad
e_{31} &=&\left(\begin{array}{cc} 0&-m\\-m&0 \end{array}\right)\quad
e_{12} =\left(\begin{array}{cc} m&0\\0&-m \end{array}\right)\nonumber\\
e_{023} =\left(\begin{array}{cc} 0&m\\-m&0 \end{array}\right)\quad
e_{031} &=&\left(\begin{array}{cc} 0&I_2\\I_2&0 \end{array}\right)\quad
e_{012} =\left(\begin{array}{cc} -I_2&0\\0&I_2 \end{array}\right)\nonumber\\
e_{123} =\left(\begin{array}{cc} 0&n\\n&0 \end{array}\right)\quad
e_{0123} &=&\left(\begin{array}{cc} 0&-\ell\\ \ell&0 \end{array}\right)\label{set2}
\end{eqnarray}
(Here, $l$, $m$ and $n$ are as defined in subsection 2 and the set (\ref{set2}) is a renaming of the set (\ref{set1})).

Because $C\ell(1,3)$ has ten roots of minus unity,the smallest possible real matrices are $8\times8$. The sixteen dimension algebra may also be represented by $2\times 2$ matrices with quaternion entries, that is, by
$\mathrm{Mat}(2,\mathbb{H})$ as follows
\begin{eqnarray}
1 =\left(\begin{array}{cc} I_2&0\\0&I_2 \end{array}\right) \quad
e_0 &=&\left(\begin{array}{cc} I_2&0\\0&-I_2 \end{array}\right)\nonumber\\
e_1 =\left(\begin{array}{cc} 0&i\\i&0 \end{array}\right)\quad
e_2 &=&\left(\begin{array}{cc} 0&j\\j&0 \end{array}\right)\quad
e_3 =\left(\begin{array}{cc} 0&k\\k&0 \end{array}\right)\nonumber\\
e_{10} =\left(\begin{array}{cc} 0&-i\\i&0 \end{array}\right)\quad
e_{20} &=&\left(\begin{array}{cc} 0&-j\\j&0 \end{array}\right)\quad
e_{30} =\left(\begin{array}{cc} 0&-k\\k&0 \end{array}\right)\nonumber\\
e_{23} =\left(\begin{array}{cc} i&0\\0&i \end{array}\right)\quad
e_{31} &=&\left(\begin{array}{cc} j&0\\0&j \end{array}\right)\quad
e_{12} =\left(\begin{array}{cc} k&0\\0&k \end{array}\right)\nonumber\\
e_{023} =\left(\begin{array}{cc} i&0\\0&-i \end{array}\right)\quad
e_{031} &=&\left(\begin{array}{cc} j&0\\0&-j \end{array}\right)\quad
e_{012} =\left(\begin{array}{cc} k&0\\0&-k \end{array}\right)\nonumber\\
e_{123} =\left(\begin{array}{cc} 0&-I\\-I&0 \end{array}\right)\quad
e_{0123} &=&\left(\begin{array}{cc} 0&-I\\I&0 \end{array}\right)\label{set3}
\end{eqnarray}
and by using equations (\ref{i}), these can be rewritten as $4\times 4$ complex, or $8\times 8$ real matrices and we have an embedding 
\begin{eqnarray}
C\ell(1,3)\cong\mathrm{Mat}(2,\mathbb{H})\subset \mathrm{Mat}(4,\mathbb{C})\subset  \mathrm{Mat}(8,\mathbb{R})\label{matalg}
\end{eqnarray}

Note that this representation explicitly highlights the link of the quaternions with the space-space bi-vectors $e_{ij}$ and the Pauli
spin matrices with the space-time bi-vectors $e_{0i}$. Spacetime rotations can thus be given an acceptable treatment in any of the matrix
algebras of equation (\ref{matalg}).

\subsection{Finding inverses in the spacetime algebra $C\ell(1,3)$}\label{divinverse}%
We now suggest that the easiest method of finding inverses of multivectors in $C\ell(1,3)$ is by making use of the
$2\times 2$ quaternion matrix representation of the algebra. The advantage of this approach is that it avoids the introduction of a new conjugate
operator as in \cite{vandermark2007}. A matrix is invertible if and only if its determinant is non-zero. We can
apply this condition to the matrix representation of the algebra $C\ell(1,3)$ and in this way find what multivectors are
invertible and which are not.

An arbitrary 16-component multivector $A\in C\ell(1,3)$ can be written in terms of the $\mathrm{Mat}(2,\mathbb{H})$ representation (\ref{set3}) as
\begin{eqnarray}
A=\left(\begin{array}{cc} q_{11}&q_{12}\\q_{21}&q_{22}
\end{array}\right)\qquad\mathrm{where}\;q_{ij}\;\mathrm{are\;quaternions}
\end{eqnarray}

Writing a quaternion $q=q_1+q_2i+q_3j+q_4k$ as a $2\times2$ complex matrix, an easy calculation shows that the determinant of a quaternion $q$ is given by $\det(q)=q_1^2-q_2^2-q_3^2-q_4^2=|q|^2$. The determinant of $A$ may be expressed as\footnote{see for example the website
http://en.wikipedia.org/wiki/Determinant}
\begin{eqnarray}
\det(A)&=&|q_{11}|^2|q_{22}-q_{21}q_{11}^{-1}q_{12}|^2\qquad (q_{11}\neq 0)\nonumber\\
&=&|q_{22}|^2|q_{11}-q_{12}q_{22}^{-1}q_{21}|^2\qquad (q_{22}\neq 0)\nonumber\\
&=& |q_{12}|^2\,|q_{21}|^2\qquad (q_{11}=q_{22}=0)
\end{eqnarray}
$A$ is singular if and only if $\det(A)=0$ so that the above equations determines whether a given multivector $A$ has an inverse or not.
Provided an inverse does exist, it is straightforward to write down a formula for the inverse $A$. For example when neither $q_{11}$ nor $q_{22}$ are $0$,
\begin{eqnarray}
\Omega^{-1}=\left(\begin{array}{cc} q_{11}&q_{12}\\q_{21}&q_{22} \end{array}\right)^{-1}=\left(\begin{array}{cc}
w_{11}&w_{12}\\w_{21}&w_{22} \end{array}\right)
\end{eqnarray}
where
\begin{eqnarray}
w_{11}&=&(q_{11}-q_{12}q_{22}^{-1}q_{21})^{-1}\nonumber\\
w_{12}&=&q_{11}^{-1}q_{12}(q_{21}q_{11}^{-1}q_{12}-q_{22})^{-1}\nonumber\\
w_{21}&=&(q_{21}q_{11}^{-1}q_{12}-q_{22})^{-1}q_{21}q_{11}^{-1}\nonumber\\
w_{22}&=&(q_{22}-q_{21}q_{11}^{-1}q_{12})^{-1}
\end{eqnarray}

It is thus straightforward (although perhaps tedious) to find the determinants and, when possible, the inverses of multivectors
in the algebra $C\ell(1,3)$. In the next subsection we highlight the physical significance of when a multivector is not invertible. 

\subsection{The non-invertible elements of $C\ell(1,3)$}\label{invertibility}
Let us now consider some specific multivectors and find when they are singular. In particular we will consider as explicit examples a mono-vector and a bivector. A more complete treatment of what follows can be found in \cite{vandermark2007}, where more general multivectors, including those of mixed grade, are considered.

Consider first a general mono-vector $x=(x_0,\mathbf{x})=x_0e_0+x_1e_1+x_2e_2+x_3e_3$ in $C\ell(1,3)$. In terms of the $\mathrm{Mat}(2,\mathbb{H})$ representation
\begin{eqnarray}
x=\left(\begin{array}{cc} x_0&\mathbf{x}\\\mathbf{x}&-x_0 \end{array}\right)=\left(\begin{array}{cc} x_0&P\\P&-x_0
\end{array}\right)
\end{eqnarray}
where $P=x_1i+x_2j+x_3k$ is a pure quaternion (that is, a quaternion with no real part). From the previous subsection, an easy calculation gives the determinant of this vector as
\begin{eqnarray}
\det(x)=|x_0^2+P^2|^2
\end{eqnarray}
and so $x$ fails to have an inverse if and only if
\begin{eqnarray}\label{invint}
x^2=x_0^2-x_1^2-x_2^2-x_3^3=0
\end{eqnarray}

For the case where $x$ is a position vector in spacetime, $x^2$  is just the invariant interval. From relativity we know that this interval  being zero, corresponds to $x$ being on the lightcone. Therefore the hyperplane
where division is not defined for mono-vectors is precisely in agreement with physical limitations set in place by the speed of
light.

As another example of a mono-vector, consider the differential operator $d$,
\begin{eqnarray}
d=e_0\partial_0-e_1\partial_1-e_2\partial_2-e_3\partial_3
\end{eqnarray}
This operator is singular if
\begin{eqnarray}
\partial_0^{2}-\nabla^2=0
\end{eqnarray}

Similarly, the vector potential $A=(\phi,\mathbf{A})=\phi e_0+A_1e_1+A_2e_2+A_3e_3$ does not have an
inverse when
\begin{eqnarray}
\phi^2=|\mathbf{A}|^2
\end{eqnarray}
Via Lorentz transformation it is always possible to find a frame where $A_2=A_3=0$ in which case $|\mathbf{A}|^2=|A_1|^2$. In this
frame, the potential $A$ does not have an inverse if $\phi=\pm|A_1|$.

Next, consider a bi-vector $F\in C\ell(1,3)$, written as
\begin{eqnarray}
F=\left(\begin{array}{cc} P_1&-P_2\\P_2&P_1 \end{array}\right)
\end{eqnarray}
where $P_1$ and $P_2$ are both pure quaternions.
For a pure quaternion $P$, $P^2=-|P|^2$ and therefore the inverse of $P$ is given by
\begin{eqnarray}
P^{-1}=-\frac{P}{|P|^2}
\end{eqnarray}
Therefore $F$ does not have an inverse when
\begin{eqnarray}\label{conditioninverse}
P_1&=&P_2P_1^{-1}P_2
\end{eqnarray}
Note that
\begin{eqnarray}
P_2P_1&=&P_2^2\frac{P_1}{|P_1|^2}P_2\nonumber\\
&=&-\frac{|P_2|^2}{|P_1|^2}P_1P_2\nonumber\\
&=& \frac{|P_2|^2}{|P_1|^2}P_2P_1
\end{eqnarray}
so that this condition implies that $|P_1|=|P_2|$.

The two pure quaternions can therefore be written as
\begin{eqnarray}
\notag P_1&=&|P_1|\hat{P}_1,\qquad \hat{P}_1^2=1\\
\notag P_2&=&|P_2|\hat{P}_2,\qquad \hat{P}_2^2=1
\end{eqnarray}
Equation (\ref{conditioninverse}) now implies that 
\begin{eqnarray}
P_1P_2=P_2P_1,
\end{eqnarray}
Thus by regarding $P_1$ and $P_2$ as vectors in $3$-space, we have that $P_1\perp P_2$.
In other words, $F$ is singular is equivalent to the conditions
\begin{eqnarray}
|P_1|=|P_2|,\;\mathrm{and}\; P_1\perp P_2
\end{eqnarray}

The electromagnetic field can be written as a bi-vector in $C\ell(1,3)$.
Explicitly,
\begin{eqnarray}
F=E_1e_{01}+E_2e_{02}+E_3e_{03}+B_1e_{23}+B_2e_{31}+B_3e_{12}
\end{eqnarray}
where $E_i$ and $B_i$ are the electric and magnetic field components respectively \cite{butler2005}.

The reader is reminded that the space-space bi-vectors $e_{ij}$ are isomorphic to the pure quaternions. We substitute
\begin{eqnarray}
P_1=\mathbf{B},\qquad P_2=-e\mathbf{E}
\end{eqnarray}
where $\mathbf{E}=(E_1,E_2,E_3)$ and $\mathbf{B}=(B_1,B_2,B_3)$ are the electric and magnetic Heaviside-Gibbs field vectors and
$e=e_{0123}$ is the pseudoscalar. The lack of an inverse then implies that
\begin{eqnarray}
|\mathbf{E}|=|\mathbf{B}|,\quad \mathrm{and}\quad \mathbf{E}\perp \mathbf{B}
\end{eqnarray}
that is, $F$ is the bivector that corresponds to free electromagnetic waves.

\subsection{Concluding remarks - What have we achieved?}\label{sec:conclusion}%
Matrices are a natural and useful way of studying various properties of algebras. One down side of working with matrices is that the matrix representations are not so clearly tied to the geometry.

Although most of this section has used the reals, the complex numbers and the quaternions, we observe that we need only the rational number field for all the calculations. Because the Clifford algebras over the rationals include elements that square to minus unity, we do not need the complex number field.

The representation of $C\ell(1,3)$ in terms of $\mathrm{Mat}(2,\mathbb{H})$ highlights the link between the quaternions and the
bi-vectors $e_{ij}$ and the Pauli spin matrices and the bi-vectors $e_{0i}$. Rotations can be given an acceptable treatment in
any of the appropriate algebras. However, as was shown in sections \ref{AaAIII2010} and \ref{AaAIV2010}, only the algebras $C\ell(0,3)$ and $C\ell(1,3)$ preserve cyclic structure.

The spacetime Clifford algebra $C\ell(1,3)$ is not a division algebra. This has led the author of reference
\cite{penrose2004rrc}, and others, to suggest that the algebra is therefore not a suitable mathematical structure
to model physical reality. The existence of inverses is indeed very important.

We have confirmed the observations made in \cite{vandermark2007} that the areas of the algebra where division is not
defined correspond to the situations on the lightcone. Therefore, the behaviour of the
algebra matches the behaviour of the physical universe. We conclude therefore that the lack of division throughout the entire algebra is not to be regarded as a weakness of the algebra but a necessity since, it matches the behaviour of our physical universe.

\section{Conclusion}
In this paper we have taken up the challenge by \cite{smolin2007tpr,woit2006new,penrose2004rrc} to reinvestigate the assumptions and axioms of physics. We have reviewed the basic assumptions made about physical space, in particular its geometry. From these assumptions we set out to develop the most appropriate mathematical framework within which to describe physical phenomena.

In section \ref{AaAI2010} we showed that the observed translational features of rigid objects in the geometric space of the 2D physical world leads to a set of operations on the points, lines, and areas of rigid objects, and to a vector space. To obtain this result we only had to assume that physical space is homogeneous. We did not have to make the usual assumption that space is continuous. Because all physical rigid bodies are finite, and measurements of translations have both upper and lower limits \lmax\ and \lmin\, not all the operations of the vector space have physical counterparts. No physical situation requires the number $\infty$ to be introduced nor do physical situations require lines whose lengths approach zero by a Cauchy process. 

Section \ref{AaAII2010} reviewed the assumptions that underlie the rotational properties of two dimensional rigid bodies. Making the assumption that physical space is isotropic in addition to being homogeneous allowed us to find operations on lines to describe rotations. We found that it is not necessary to introduce the unit imaginary $i$ to describe rotations by an Argand diagram but rather that rotations in the plane can be better described by the geometric bivector $\xhat\yhat$. 

The concepts of homogeneity and isotropy are readily extended from two dimensions to three. Insisting on maintaining cyclic structure for rotations uniquely led to $C\ell(0,3)$ as the appropriate algebra to describe 3D geometry. This algebra contains three bivectors $\xhat\yhat,\yhat\zhat,\zhat\xhat$ that take the place of the usual $ix,iy,iz$ to describe rotations in the three orthogonal planes. Again no continuity conditions are needed to be made to arrive at these conclusions.  

In section \ref{AaAIV2010} we extended our discussion of space to include time. We show that we may use clocks to deduce that we may regard time as a fourth dimension in a vector space over the rational numbers. The homogeneity of time, clocks tick at the same rate today as they did yesterday, together with the finite speed of light observed to be $c$ in all inertial frames led to the Lorentz metric of spacetime and consequently the spacetime Clifford algebra $C\ell(1,3)$. Time is different from space. Unlike the three spatial axes, the time axis cannot be rotated. Whereas a 3D rigid body may be rotated in a such a way that the orientation of a particular line within this rigid body is reversed, the orientation of the rigid bodies clock cannot be reversed. Clocks and rigid bodies are thus quite distinct. By assuming that the equivalence principle holds for  inertial rigid body frames and that the speed of light $c$ is observed to be the same in all such frames we derived the Lorentz and Poincar\'e transformations.

Section \ref{AaAV2010} addressed the issue raised by Penrose \cite{penrose2004rrc} and others, that Clifford algebras are in general not division algebras. We have shown that the matrix representation of $2\times 2$ matrices over the field of quaternions is a very powerful tool to do manipulations and find inverses in $C\ell(1,3)$. Vectors that do no have inverses are on the lightcone. Other elements of $C\ell(1,3)$ without inverses are generalizations of null vectors. We thereby showed that the existence of non-invertible elements in the algebra is not a limitation of the usefulness to physics of the algebra but rather that it reflects accurately the spacetime properties of physical systems.

We have demonstrated that a careful study of the assumptions and axioms associated with spacetime leads to a somewhat richer structure than the standard Lorentz and Poincar\'e algebras. We have thus gone part of the way to answering the question raised by Smolin \cite{smolin2007tpr}, Woit \cite{woit2006new}, and Penrose \cite{penrose2004rrc}. 
\begin{acknowledgements}
The authors wish to acknowledge John Williamson and Martin van der Mark for their insights and discussions relating to section \ref{AaAV2010} this paper.
\end{acknowledgements}

\bibliography{Niels}  
\bibliographystyle{unsrt}  
\end{document}